\documentclass[aps,prb,twocolumn,superscriptaddress,notitlepage,noshowkeys]{revtex4-1}  
\usepackage{graphicx}  
\usepackage{dcolumn}   
\usepackage{bm}        
\usepackage{amssymb}   
\usepackage{mleftright}
\mleftright
\usepackage[utf8]{inputenc} 
\usepackage[english]{babel} 
\usepackage{amsmath, mathtools}
\usepackage{amsthm}
\usepackage{enumerate}
\usepackage{wrapfig}
\usepackage[textsize=tiny]{todonotes}
\usepackage[many]{tcolorbox}
\usepackage{tensor} 
\usepackage{here} 
\usepackage{hhline}
\usepackage{braket}
\newcommand{\red}[1]{\textcolor{red}{#1}}
\definecolor{light-gray}{gray}{0.85}
\definecolor{nicegreen}{RGB}{60,183,82}
\usepackage{hyperref}
\definecolor{PRBblue}{RGB}{0,0,254}
\definecolor{bwred}{rgb}{0.94,0.5,0.5}
\definecolor{bwblue}{rgb}{0.13,0.67,0.8}
\definecolor{atomictangerine}{rgb}{1.0, 0.6, 0.4}
\hypersetup{
	colorlinks, 
	linkcolor=PRBblue, 
	citecolor=PRBblue, 
	filecolor=PRBblue, 
	urlcolor=PRBblue,
	pdftitle={String-nets for unitary fusion categories}, 
	pdfauthor={Alexander Hahn, and Ramona Wolf}
}
\usepackage{stackengine}

\definecolor{LightGray}{RGB}{200,200,200}

\allowdisplaybreaks

\usepackage{tikz}
\usetikzlibrary{decorations.pathreplacing,shapes.misc}
\usetikzlibrary{arrows,shapes,positioning}
\usetikzlibrary{decorations.markings}
\usetikzlibrary{through,calc}
\usetikzlibrary{patterns}
\usetikzlibrary{plotmarks}
\usetikzlibrary{automata}
\tikzstyle arrowstyle=[scale=1]
\usepackage{extarrows}

\DeclareMathOperator{\supp}{supp}
\DeclareMathOperator{\conf}{conf}
\DeclareMathOperator{\tr}{tr}
\DeclareMathOperator{\im}{im}
\DeclareMathOperator{\id}{id}
\DeclareMathOperator{\vspan}{span}
\DeclareMathOperator{\res}{res}
\DeclareMathOperator{\real}{Re}
\DeclareMathOperator{\imag}{Im}
\DeclareMathOperator{\diff}{diff}
\DeclareMathOperator{\homeo}{Homeo}
\DeclareMathOperator{\Mor}{Mor}
\DeclareMathOperator{\Ob}{Ob}
\DeclareMathOperator{\target}{Target}
\DeclareMathOperator{\source}{Source}
\DeclareMathOperator{\aut}{Aut}
\DeclareMathOperator{\Sch}{Sch}
\DeclareMathOperator{\Conf}{Conf}
\DeclareMathOperator{\Diff}{Diff}
\DeclareMathOperator{\Endstar}{End^*}
\DeclareMathOperator{\End}{End}

\newcommand{\cat}{\mathcal{C}}
\newcommand{\ann}{\mathrm{Ann}}
\renewcommand{\Vec}{\textbf{Vec}}
\newcommand{\Z}{\mathbb{Z}}

\theoremstyle{plain}

\theoremstyle{definition}
\newtheorem{definition}{Definition}[section]

\theoremstyle{remark}

\newcommand{\rom}[1]{\uppercase\expandafter{\romannumeral #1\relax}}

\newcommand{\arho}{{}_\alpha\rho}
\newcommand{\asrho}{{}_{\alpha^*}\!\rho}
\newcommand{\dothis}[1]{{\color{red} #1}}

\newcommand{\itemone}{\begin{tikzpicture}[baseline={([yshift=-3pt]current bounding box.center)}]\node[circle,scale=0.6,draw,red] at (0,0) {$1$};\end{tikzpicture}}
\newcommand{\itemtwo}{\begin{tikzpicture}[baseline={([yshift=-3pt]current bounding box.center)}]\node[circle,scale=0.6,draw,red] at (0,0) {$2$};\end{tikzpicture}}
\newcommand{\itemthree}{\begin{tikzpicture}[baseline={([yshift=-3pt]current bounding box.center)}]\node[circle,scale=0.6,draw,red] at (0,0) {$3$};\end{tikzpicture}}
\newcommand{\itemfour}{\begin{tikzpicture}[baseline={([yshift=-3pt]current bounding box.center)}]\node[circle,scale=0.6,draw,red] at (0,0) {$4$};\end{tikzpicture}}
\newcommand{\itemfive}{\begin{tikzpicture}[baseline={([yshift=-3pt]current bounding box.center)}]\node[circle,scale=0.6,draw,red] at (0,0) {$5$};\end{tikzpicture}}
\newcommand{\itemsix}{\begin{tikzpicture}[baseline={([yshift=-3pt]current bounding box.center)}]\node[circle,scale=0.6,draw,red] at (0,0) {$6$};\end{tikzpicture}}

\tikzset{->-/.style={decoration={
			markings,
			mark=at position .6 with {\arrow[>=stealth]{>}}},postaction={decorate}}}
\tikzset{-<-/.style={decoration={
			markings,
			mark=at position .6 with {\arrow[>=stealth]{<}}},postaction={decorate}}}

\begin{document}
	
	
	\title{Generalized string-nets for unitary fusion categories without tetrahedral symmetry}
	\author{Alexander Hahn}
	\email{alexander.hahn@htp-tel.de}
	\author{Ramona Wolf}
	\email{ramona.wolf@itp.uni-hannover.de}
	\affiliation{Institut für Theoretische Physik, Leibniz Universität Hannover, Appelstraße 2, 30167 Hannover, Germany}
	
	\begin{abstract}
		The Levin-Wen model of string-net condensation explains how topological phases emerge from the microscopic degrees of freedom of a physical system. However, the original construction is not applicable to all unitary fusion category since some additional symmetries for the $F$-symbols are imposed. In particular, the so-called tetrahedral symmetry is not fulfilled by many interesting unitary fusion categories. In this paper, we present a generalized construction of the Levin-Wen model for arbitrary multiplicity-free unitary fusion categories that works without requiring these additional symmetries. We explicitly calculate the matrix elements of the Hamiltonian and, furthermore, show that it has the same properties as the original one.
	\end{abstract}
	
	\maketitle
	
\section{Introduction}

For a long time, it was thought that all phases and continuous phase transitions could be described by Landau's theory of symmetry breaking \cite{Landau1965}. However, with the discovery of the fractional quantum Hall effect in the early 1980s \cite{FQH1,FQH2}, it became clear that there are systems that exhibit a different kind of order---topological order \cite{Wen1990,Wen1995}---going beyond the scope of Landau's theory. Since then, topological phases in condensed matter systems have been extensively studied and are still an active area of research \cite{Haldane1983,Haldane1983a,PTBO10,PBTO12,Wen2017}, with applications ranging from fractional quantum Hall systems \cite{WN90,BW90,Read1990,FK91,ZMP13} over quantum spin systems \cite{WWZ89,RS91,Wen1991,SF00,Wen2002,SP02} to, more recently, topological quantum computation \cite{Ioffe2002,DKLP02,Kitaev2003,Nayak2008,Terhal2015}. Moreover, exploiting powerful tensor network variational methods \cite{Schollwoeck2005,Schollwoeck2011} to numerically study topological phases has led to numerous important insights \cite{Vidal2004,Verstraete2004,Trebst2008,Koenig2010,Haegeman2011,Haegeman2012,Finch2014,Singh2014}. These techniques fall under the name of \emph{density matrix renormalization group methods}.

For a complete understanding of the theory of topological phases similar to Landau's theory of symmetry breaking, one has to face several challenges: for instance, Landau's theory provides low energy effective theories for general ordered phases, namely Ginzburg-Landau field theories \cite{GL50}. In the theory of topological phases, this has been successfully understood  by topological quantum field theories (TQFTs) \cite{Witten1988,Witten1989}. 
Another important aspect that is explained by Landau's theory is a physical picture for the emergence of ordered phases---particle condensation. Additionally, the theory provides a framework to characterize and classify these phases, namely group theory.

The last two issues are addressed in \cite{LW05} for a large class of topological phases, so-called doubled topological phases. In their work, the authors describe a lattice model, in which the emergence of topological phases is explained by so-called string-net condensation. This physical mechanism yields a much richer class of phases than the one that is given by Landau's theory. Their approach exploits the mathematical framework of unitary fusion categories (UFCs) \cite{Kassel95,Etingof2015}, which solves the problem of finding an analogue of group theory for the phase characterization. The idea behind their approach is to describe the universal properties of a string-net condensed phase via the ground state wave function, which is determined by local constraints.

Their construction has several important properties: First of all, it yields a lattice model with an exactly solvable Hamiltonian and a description of its ground state wave function. Moreover, it is interesting from a purely mathematical point of view: given a unitary fusion category fulfilling some additional constraints, one can use the string-net construction to build a unitary modular tensor category (UMTC), since the latter is described by the emerging quasiparticles of the string-net model. This UMTC then serves as a mathematical description of anyonic particles.

Although this is a quite promising approach, especially because of its constructive nature, there is one caveat: It is not possible to use it for \emph{general} unitary fusion categories. More precisely, the $F$-symbols (also called $6j$-symbols) of the category have to fulfill the so-called \emph{tetrahedral symmetry} condition and some unitarity constraint. However, there are unitary fusion categories that do not fulfill these conditions (see, e.g., \cite{Hong09,Osborne2019} and Appendix~\ref{app:H3} of this paper). 

Nevertheless, the construction of \cite{LW05} can be adapted for general unitary fusion categories. While it was already questioned whether these symmetry conditions are a necessary criterion \cite{HW12,LW14}, it was shown in \cite{Hong09}, that \emph{any} unitary fusion category yields an exactly solvable Hamiltonian. One can also take a more category-theoretical point of view to face this problem: It was shown that the string-net space for a category $\mathcal{C}$ in the sense of \cite{LW05} is equal to the state space of the Turaev-Viro TQFT for $\mathcal{C}$ \cite{KMR10,KKR10,Ki11}, which itself is isomorphic to the state space of the Reshetikin-Turaev theory for the Drinfeld double of $\mathcal{C}$ \cite{KB10,TV10,Balsam2010}. While in \cite{LW05} the additional symmetries arise quite naturally both of the latter formalisms get along without any additional constraints. Thereby, they provide many important results regarding string-nets and TQFTs \cite{FRS02,FRS04a,FRS04b,FRS05,FFRS06}.

However, these categorical approaches as well as the one in \cite{Hong09} have a severe drawback: They are not constructive, which means that they lack an explicit formula for the matrix elements of the Hamiltonian. Especially for physicists, who do not necessarily have a solid background in category theory, this lack of a constructive way to define the Hamiltonian is a serious hurdle when using the Levin-Wen approach for arbitrary UFCs. Therefore, it was still noted as an important open question in physics to study generalized string-net models \cite{HW12}. For instance, in \cite{LW14} the authors present a construction of the string-net Hamiltonian without demanding tetrahedral symmetry, but it is only applicable to fusion categories where the objects form an abelian group under the fusion operation. Hence, giving a construction of the Levin-Wen model for general unitary fusion categories is still an unsolved problem.

In this work we answer this question by explicitly constructing the Hamiltonian without imposing any additional constraints on the unitary fusion category. Moreover, we give proofs of the important properties of the Hamiltonian. We also show that our formula for the Hamiltonian can be transformed into the original formula given in \cite{LW05} if we impose the additional constraints (see Appendix~\ref{app:tetrahedral}).

The paper is organized as follows: Section~\ref{sec:model} is the main part of the paper. We begin by revising the original construction of the Hamiltonian from \cite{LW05} and point out where the construction fails for general UFCs. Afterwards, we do the explicit calculation of the matrix elements without imposing any additional symmetries. In the last part of this section we explain the ideas behind the (rather technical) proofs of the properties of the Hamiltonian, which can be found in detail in Appendix~\ref{app:Hamprops}. In Section~\ref{sec:excitations}, we explain how the definition of excitations in the original paper can be generalized to the framework of general UFCs. Finally, we conclude in Section~\ref{sec:conclusion} and also mention some interesting open questions in this area.

In the Appendix, several technical calculations can be found: Appendix~\ref{app:UFCgraphcalc} is dedicated to the graphical calculus of UFCs that we use for computations throughout this work. Appendix~\ref{app:GHcalc} provides technical details that are necessary for the calculation of the Hamiltonian. As mentioned above, the proofs for several properties of the Hamiltonian can be found in Appendix~\ref{app:Hamprops}. In Appendix~\ref{app:tetrahedral}, we show how our construction can be transformed into the original formula if one imposes additional symmetry constraints. In Appendix~\ref{app:H3} we give an explicit example for a UFC which breaks tetrahedral symmetry.

\section{The string-net model}
\label{sec:model}

In this section we describe the string-net model as proposed by Levin and Wen in their seminal paper \cite{LW05} and point out which aspects do not work for general unitary fusion categories. Furthermore, we show how to explicitly compute the matrix elements of a matrix representation for the Hamiltonian of the generalized Levin-Wen model. This formula is applicable for any unitary fusion category.

We begin by recalling how the original model is constructed. In the general string-net picture, we need some data to specify the model:
\begin{enumerate}
	\item \textbf{String types.} We need to specify the types of strings that can appear, and also the total number $N+1$ of different types. We label different string types with integers: $i=0,1,2,\dots,N$. where $i=0$ represents the vacuum string.
	\item \textbf{Branching rules.} It is necessary to specify which string types $i,j,k$ are allowed to meet at a vertex:
	\begin{equation}
		\begin{tikzpicture}[baseline=(current bounding box.center), decoration={markings,mark=at position .5 with {\arrow[>=stealth]{>}}},scale=0.8]
		\draw[{postaction=decorate}] (330:1cm) to node[below] {\small $j$} (0,0);
		\draw[{postaction=decorate}] (210:1cm) to node[below] {\small $i$} (0,0);
		\draw[{postaction=decorate}] (90:1cm) to node[right] {\small $k$} (0,0);
		\end{tikzpicture}
	\end{equation}
	\item \textbf{String orientations.} With every string type $i$ we associate a dual string type $i^*$ that satisfies $(i^*)^*=i$. The string of type $i^*$ corresponds to the type-$i$ string with opposite orientation:
	\begin{equation}
		\begin{tikzpicture}[baseline=(current bounding box.center), decoration={markings,mark=at position .5 with {\arrow[>=stealth]{>}}},scale=1]
		\draw[{postaction=decorate}] (0,0) to node[left] {\small $i$} (0,1);
		\end{tikzpicture}\ =\ 
		\begin{tikzpicture}[baseline=(current bounding box.center), decoration={markings,mark=at position .5 with {\arrow[>=stealth]{>}}},scale=1]
		\draw[{postaction=decorate}] (0,0) to node[right] {\small $i^*$} (0,-1);
		\end{tikzpicture}
	\end{equation}
\end{enumerate}
After specifying this data, we can define the corresponding Hilbert space of the string-net model. The states in the Hilbert space are simply linear combinations of different spatial configurations of string-nets.

Before we continue to discuss ground states of the string-net model, a brief comment on string diagrams is necessary. In this paper, we use the convention that all unoriented string diagrams point upwards, i.e.
\begin{equation}
	\begin{tikzpicture}[baseline=(current bounding box.center), decoration={markings,mark=at position .5 with {\arrow[>=stealth]{>}}},scale=0.8]
	\draw[] (330:1cm) to node[below] {\small $j$} (0,0);
	\draw[] (210:1cm) to node[below] {\small $i$} (0,0);
	\draw[] (90:1cm) to node[right] {\small $k$} (0,0);
	\end{tikzpicture}\ \equiv\ 
	\begin{tikzpicture}[baseline=(current bounding box.center), decoration={markings,mark=at position .5 with {\arrow[>=stealth]{>}}},scale=0.8]
	\draw[{postaction=decorate}] (330:1cm) to node[below] {\small $j$} (0,0);
	\draw[{postaction=decorate}] (210:1cm) to node[below] {\small $i$} (0,0);
	\draw[{postaction=decorate}] (0,0) to node[right] {\small $k$} (90:1cm);
	\end{tikzpicture}.\label{eq:StringOrientation}
\end{equation}
For the vacuum string we usually use dotted lines to distinguish it from the other string types. Furthermore, in contrast to the original paper, we never use horizontal lines. The reason behind this is that without imposing tetrahedral symmetry the meaning of these lines is ambiguous. Hence, throughout this paper we make an effort to translate all relevant diagrams to ones that have no horizontal lines (for example, equations \eqref{eq:groundstate1} -- \eqref{eq:groundstate5}).

In the original paper, the authors reason that the ground state wave function $\Phi$ of a Hamiltonian acting on a string-net state can be uniquely specified by local constraints. These constraints are the following:
\begin{align}
	\Phi\left(
	\begin{tikzpicture}[baseline={([yshift=-3pt]current bounding box.center)}, decoration={markings,mark=at position .5 with {\arrow[>=stealth]{>}}},scale=0.5]
	\draw[{postaction=decorate}] (1,0.5) to node[above] {\small $i$} (3,1.5);
	\draw[fill=LightGray,rounded corners,LightGray] (0,0) rectangle (1,2);
	\draw[fill=LightGray,rounded corners,LightGray] (3,0) rectangle (4,2);
	\end{tikzpicture}\right)&=	\Phi\left(	
	\begin{tikzpicture}[baseline={([yshift=-3pt]current bounding box.center)}, decoration={markings,mark=at position .55 with {\arrow[>=stealth]{>}}},scale=0.5]
	\draw[{postaction=decorate}] (1,0.5) -- (1.5,0.75) to [bend right=80] node [above,pos=0.4] {\small $i$} (2.5,1.25) -- (3,1.5);
	\draw[fill=LightGray,rounded corners,LightGray] (0,0) rectangle (1,2);
	\draw[fill=LightGray,rounded corners,LightGray] (3,0) rectangle (4,2);
	\end{tikzpicture}\right) \label{eq:groundstate1}\\
	\Phi\left(
	\begin{tikzpicture}[baseline={([yshift=-3pt]current bounding box.center)}, decoration={markings,mark=at position .0 with {\arrow[>=stealth]{>}}},scale=0.5]
	\draw[fill=LightGray,rounded corners,LightGray] (0,0) rectangle (1,2);
	\draw[{postaction=decorate}] (2,1) circle (0.5cm);
	\node at (2.9,1) {\small $i$};
	\end{tikzpicture}
	\right)	&=d_i \ \Phi\left(
	\begin{tikzpicture}[baseline={([yshift=-3pt]current bounding box.center)}, decoration={markings,mark=at position .0 with {\arrow[>=stealth]{>}}},scale=0.5]
	\draw[fill=LightGray,rounded corners,LightGray] (0,0) rectangle (1,2);
	\end{tikzpicture}
	\right) \label{eq:groundstate2}\\
	\Phi\left(
	\begin{tikzpicture}[baseline={([yshift=-3pt]current bounding box.center)}, decoration={markings,mark=at position .55 with {\arrow[>=stealth]{>}}},scale=0.5]
	\draw[{postaction=decorate}] (0.8,0.5) to node[above] {\small $i$} (1.5,0.75);
	\draw[{postaction=decorate}] (1.5,0.75) to [bend right=80] node [below,pos=0.4] {\small $l$} (2.5,1.25);
	\draw[{postaction=decorate}] (1.5,0.75) to [bend left=80] node [above,pos=0.4] {\small $k$} (2.5,1.25);
	\draw[{postaction=decorate}] (2.5,1.25) to node[above] {\small $j$} (3.2,1.5);
	\draw[fill=LightGray,rounded corners,LightGray] (-0.2,0) rectangle (0.8,2);
	\draw[fill=LightGray,rounded corners,LightGray] (3.2,0) rectangle (4.2,2);
	\end{tikzpicture}
	\right) &=\delta_{ij} \Phi\left(
	\begin{tikzpicture}[baseline={([yshift=-3pt]current bounding box.center)}, decoration={markings,mark=at position .6 with {\arrow[>=stealth]{>}}},scale=0.5]
	\draw[{postaction=decorate}] (0.8,0.5) to node[above] {\small $i$} (1.5,0.75);
	\draw[{postaction=decorate}] (1.5,0.75) to [bend right=80] node [below,pos=0.4] {\small $l$} (2.5,1.25);
	\draw[{postaction=decorate}] (1.5,0.75) to [bend left=80] node [above,pos=0.4] {\small $k$} (2.5,1.25);
	\draw[{postaction=decorate}] (2.5,1.25) to node[above] {\small $i$} (3.2,1.5);
	\draw[fill=LightGray,rounded corners,LightGray] (-0.2,0) rectangle (0.8,2);
	\draw[fill=LightGray,rounded corners,LightGray] (3.2,0) rectangle (4.2,2);
	\end{tikzpicture}
	\right) \label{eq:groundstate3}\\
	\Phi\left(
	\begin{tikzpicture}[baseline={([yshift=-3pt]current bounding box.center)}, decoration={markings,mark=at position .6 with {\arrow[>=stealth]{>}}},scale=0.5]
	\draw[fill=LightGray,rounded corners,LightGray] (0,-0.2) rectangle (2,0.8);
	\draw[fill=LightGray,rounded corners,LightGray] (0,3.2) rectangle (2,4.2);
	\draw[{postaction=decorate}] (0.5,0.8) to node[left] {\small $i$} (0.5,2.5);
	\draw[{postaction=decorate}] (0.5,2.5) to node[left] {\small $k$} (0.5,3.2);
	\draw[{postaction=decorate}] (1.5,0.8) to node[right] {\small $j$} (1.5,1.5);
	\draw[{postaction=decorate}] (1.5,1.5) to node[right] {\small $l$} (1.5,3.2);
	\draw[{postaction=decorate}] (1.5,1.5) to node[above] {\small $m$} (0.5,2.5);
	\end{tikzpicture}\right) &=\sum_n \left(G_{ij}^{kl}\right)_{mn}\ \Phi\left(
	\begin{tikzpicture}[baseline={([yshift=-3pt]current bounding box.center)}, decoration={markings,mark=at position .6 with {\arrow[>=stealth]{>}}},scale=0.5]
	\draw[fill=LightGray,rounded corners,LightGray] (0,-0.2) rectangle (2,0.8);
	\draw[fill=LightGray,rounded corners,LightGray] (0,3.2) rectangle (2,4.2);
	\draw[{postaction=decorate}] (0.5,0.8) to node[left] {\small $i$} (1,1.5);
	\draw[{postaction=decorate}] (1.5,0.8) to node[right] {\small $j\ $} (1,1.5);
	\draw[{postaction=decorate}] (1,1.5) to node[right] {\small $n$} (1,2.5);
	\draw[{postaction=decorate}] (1,2.5) to node[left] {\small $\ k$} (0.5,3.2);
	\draw[{postaction=decorate}] (1,2.5) to node[right] {\small $l$} (1.5,3.2);
	\end{tikzpicture}
	\right) \label{eq:groundstate4}\\
	\Phi\left(
	\begin{tikzpicture}[baseline={([yshift=-3pt]current bounding box.center)}, decoration={markings,mark=at position .6 with {\arrow[>=stealth]{>}}},scale=0.5]
	\draw[fill=LightGray,rounded corners,LightGray] (0,-0.2) rectangle (2,0.8);
	\draw[fill=LightGray,rounded corners,LightGray] (0,3.2) rectangle (2,4.2);
	\draw[{postaction=decorate}] (0.5,0.8) to node[left] {\small $i$} (0.5,1.5);
	\draw[{postaction=decorate}] (0.5,1.5) to node[left] {\small $k$} (0.5,3.2);
	\draw[{postaction=decorate}] (1.5,0.8) to node[right] {\small $j$} (1.5,2.5);
	\draw[{postaction=decorate}] (1.5,2.5) to node[right] {\small $l$} (1.5,3.2);
	\draw[{postaction=decorate}] (0.5,1.5) to node[above] {\small $m$} (1.5,2.5);
	\end{tikzpicture}\right) &=\sum_n \left(H_{ij}^{kl}\right)_{mn}\ \Phi\left(
	\begin{tikzpicture}[baseline={([yshift=-3pt]current bounding box.center)}, decoration={markings,mark=at position .6 with {\arrow[>=stealth]{>}}},scale=0.5]
	\draw[fill=LightGray,rounded corners,LightGray] (0,-0.2) rectangle (2,0.8);
	\draw[fill=LightGray,rounded corners,LightGray] (0,3.2) rectangle (2,4.2);
	\draw[{postaction=decorate}] (0.5,0.8) to node[left] {\small $i$} (1,1.5);
	\draw[{postaction=decorate}] (1.5,0.8) to node[right] {\small $j\ $} (1,1.5);
	\draw[{postaction=decorate}] (1,1.5) to node[right] {\small $n$} (1,2.5);
	\draw[{postaction=decorate}] (1,2.5) to node[left] {\small $\ k$} (0.5,3.2);
	\draw[{postaction=decorate}] (1,2.5) to node[right] {\small $l$} (1.5,3.2);
	\end{tikzpicture}
	\right) \label{eq:groundstate5}
\end{align}

\begin{figure}[t]
	\centering
	\includegraphics{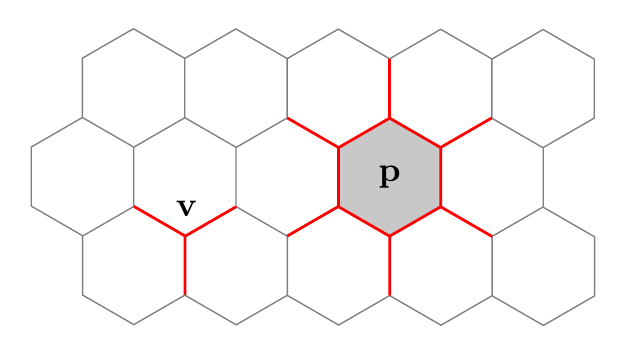}
	\caption{\textbf{Hamiltonian on the honeycomb lattice.} A general Hamiltonian consists of operators that act on vertices $Q_\mathbf{v}$ and ones that act on plaquettes $B_\mathbf{p}$.\label{fig:honeycomb}}
\end{figure}

Note that the gray rectangles always represent the remaining parts of the string-net that are not affected by the local relations. The $d_i$ are complex numbers (called \emph{quantum dimensions}) assigned to the string types and the operators $G_{ij}^{kl}$ and $H_{ij}^{kl}$ are related to the $F$-symbols \eqref{eq:Fsymbol} (or $6j$-symbols) of the underlying unitary fusion categories (see Appendix~\ref{app:GHcalc}):
\begin{align}
	\left(G_{ij}^{kl}\right)_{mn}&=\sqrt{\frac{d_m d_n}{d_j d_k}} \overline{\left(F_n^{iml}\right)}_{kj}\\
	\left(H_{ij}^{kl}\right)_{mn}&=\sqrt{\frac{d_m d_n}{d_i d_l}} \left(F_n^{kmj}\right)_{il}.
\end{align}
However, not all choices of $F$-symbols and quantum dimensions lead to self-consistent constraints \eqref{eq:groundstate1}--\eqref{eq:groundstate5}. More precisely, to yield self-consistent constraints the $F$-symbols have to fulfill the pentagon equation:
\begin{equation}
	\left(F_m^{nkl}\right)_{pl}
	\left(F_m^{ijs}\right)_{nr}
	=
	\sum_q
	\left(F_{p}^{ijk}\right)_{nq}
	\left(F_m^{iql}\right)_{pr}
	\left(F_{r}^{jkl}\right)_{qs}.
	\label{eq:pentagon}
\end{equation}
This is the only consistency condition we request. In contrast, in the original model the so-called tetrahedral symmetry has to be fulfilled, which is defined as follows:
\begin{align}
	\begin{split}
		\left(F_i^{jkl}\right)_{mn}&=\left(F^{kji^*}_{l^*}\right)_{mn^*}\\
		&\hspace{-15pt}=\left(F^{i^*lk}_{j^*}\right)_{m^*n}=\sqrt{\frac{d_m d_n}{d_j d_l}}\left(F_{i^*}^{m^*kn^*}\right)_{j^*l^*}\label{tetrahedralSymmetryUFC}.
	\end{split}
\end{align}	
We discuss this condition in more detail in Appendix~\ref{app:tetrahedral} and also show that, when additionally assuming tetrahedral symmetry together with some other conditions that are required in the original construction, our form of the Hamiltonian can be converted into the original form.

\subsection{The Hamiltonian}

After we have specified the local constraints for the ground state we can construct exactly solvable lattice spin Hamiltonians with exactly these states as ground states. As proposed in the original paper, we consider the honeycomb lattice where the degrees of freedom are on the edges. The Hamiltonian then consists of two types of operators: ones that act on the vertices $\mathbf{v}$, denoted $Q_\mathbf{v}$, and ones that act on plaquettes $\mathbf{p}$, denoted $B_\mathbf{p}$ (see Fig.~\ref{fig:honeycomb}). The exactly solvable Levin-Wen Hamiltonian $H$ on a honeycomb lattice is given by taking the sum of these operators over all vertices and all plaquettes:
\begin{equation}
	H=-\sum_{\mathbf{v}} Q_\mathbf{v} - \sum_{\mathbf{p}} B_\mathbf{p}.\label{Hamiltonian}
\end{equation}
The negative signs ensure that those string-net configurations that obey the branching rules and relations \eqref{eq:groundstate1} -- \eqref{eq:groundstate5} are energetically favored and therefore in the ground state.

The vertex operator $Q_\mathbf{v}$, which is also called the \emph{electric charge operator}, always acts on three degrees of freedom. It ensures that the string-net configurations of the ground state obey the branching rules:
\begin{equation}
	Q_\mathbf{v}\Bigg| \begin{tikzpicture}[baseline=(current bounding box.center), decoration={markings,mark=at position .5 with {\arrow[>=stealth]{>}}},scale=0.5]
	\draw[{postaction=decorate}] (330:1cm) to node[below] {\small $j$} (0,0);
	\draw[{postaction=decorate}] (210:1cm) to node[below] {\small $i$} (0,0);
	\draw[{postaction=decorate}] (90:1cm) to node[right] {\small $k$} (0,0);
	\end{tikzpicture}\Bigg\rangle=\delta_{ij}^{k^*} \Bigg| \begin{tikzpicture}[baseline=(current bounding box.center), decoration={markings,mark=at position .5 with {\arrow[>=stealth]{>}}},scale=0.5]
	\draw[{postaction=decorate}] (330:1cm) to node[below] {\small $j$} (0,0);
	\draw[{postaction=decorate}] (210:1cm) to node[below] {\small $i$} (0,0);
	\draw[{postaction=decorate}] (90:1cm) to node[right] {\small $k$} (0,0);
	\end{tikzpicture}\Bigg\rangle,
\end{equation}
where 
\begin{equation}
	\delta_{ij}^{k} = 
	\begin{cases}
		1, & i\otimes j = k \text{ is an allowed fusion}\\
		0, & \text{otherwise.}
	\end{cases}\label{eq:Nijk}
\end{equation}

The plaquette operator $B_\mathbf{p}$, also called the \emph{magnetic flux operator}, imposes dynamics to the system. It is a linear combination of $N+1$ terms, one term for each string type (plus the vacuum string):
\begin{equation}
	B_\mathbf{p}=\sum_{s=0}^{N} a_s B_\mathbf{p}^s,
\end{equation}
where the coefficients $a_s$ satisfy $a_{s^*}=a_s^*$ but are otherwise arbitrary for now. Each of the individual terms acts on a plaquette $\mathbf{p}$ of the honeycomb lattice by inserting a loop of type $s$ and fusing this loop into the internal links of the plaquette $\mathbf{p}$. Hence, effectively the operator maps a configuration $g,h,i,j,k,l$ of internal edges to linear combination of different configurations $g',h',i',j',k',l'$:
\begin{widetext}
	\begin{equation}
		B_\mathbf{p}^s\Bigg|
		\begin{tikzpicture} [scale=0.9,rotate=90, scale=1,baseline={([yshift=-3pt]current bounding box.center)}, decoration={markings,mark=at position .5 with {\arrow[>=stealth]{<}}}]
		\draw[{postaction=decorate}] (0,2.73205) -- node[above] {\scriptsize{$g$}} (0.5,1.86603); 
		\draw[{postaction=decorate}] (-1,2.73205) -- node[left] {\scriptsize{$l$}} (0,2.73205); 
		\draw[{postaction=decorate}] (-1.5,1.86603) -- node[below] {\scriptsize{$k$}} (-1,2.73205); 
		\draw[{postaction=decorate}] (-1,1) -- node[below] {\scriptsize{$j$}} (-1.5,1.86603); 
		\draw[{postaction=decorate}] (0,1) -- node[right] {\scriptsize{$i$}} (-1,1); 
		\draw[{postaction=decorate}] (0.5,1.86603) -- node[above] {\scriptsize{$h$}} (0,1); 
		\node (e1) at (0.25,3.16506) {\scriptsize{$a$}};
		\draw[{postaction=decorate}] (0,2.73205) -- (e1); 
		\node (e2) at (-1.25,3.16506) {\scriptsize{$f$}};
		\draw[{postaction=decorate}] (-1,2.73205) -- (e2); 
		\node (e3) at (-2,1.86603) {\scriptsize{$e$}};
		\draw[{postaction=decorate}] (-1.5,1.86603) -- (e3); 
		\node (e4) at (-1.25,0.566987) {\scriptsize{$d$}};
		\draw[{postaction=decorate}] (-1,1) -- (e4); 
		\node (e5) at (0.25,0.566987) {\scriptsize{$c$}};
		\draw[{postaction=decorate}] (0,1) -- (e5); 
		\node (e6) at (1,1.86603) {\scriptsize{$b$}};
		\draw[{postaction=decorate}]  (0.5,1.86603) -- (e6); 
		\end{tikzpicture}
		\Bigg\rangle
		=
		\Bigg|\begin{tikzpicture} [scale=0.9,rotate=90, scale=1,baseline={([yshift=-3pt]current bounding box.center)}, decoration={markings,mark=at position .5 with {\arrow[>=stealth]{<}}}]
		\draw[{postaction=decorate}] (0,2.73205) -- node[above] {\scriptsize{$g$}} (0.5,1.86603); 
		\draw[{postaction=decorate}] (-1,2.73205) -- node[left] {\scriptsize{$l$}} (0,2.73205); 
		\draw[{postaction=decorate}] (-1.5,1.86603) -- node[below] {\scriptsize{$k$}} (-1,2.73205); 
		\draw[{postaction=decorate}] (-1,1) -- node[below] {\scriptsize{$j$}} (-1.5,1.86603); 
		\draw[{postaction=decorate}] (0,1) -- node[right] {\scriptsize{$i$}} (-1,1); 
		\draw[{postaction=decorate}] (0.5,1.86603) -- node[above] {\scriptsize{$h$}} (0,1); 
		\node (e1) at (0.25,3.16506) {\scriptsize{$a$}};
		\draw[{postaction=decorate}] (0,2.73205) -- (e1); 
		\node (e2) at (-1.25,3.16506) {\scriptsize{$f$}};
		\draw[{postaction=decorate}] (-1,2.73205) -- (e2); 
		\node (e3) at (-2,1.86603) {\scriptsize{$e$}};
		\draw[{postaction=decorate}] (-1.5,1.86603) -- (e3); 
		\node (e4) at (-1.25,0.566987) {\scriptsize{$d$}};
		\draw[{postaction=decorate}] (-1,1) -- (e4); 
		\node (e5) at (0.25,0.566987) {\scriptsize{$c$}};
		\draw[{postaction=decorate}] (0,1) -- (e5); 
		\node (e6) at (1,1.86603) {\scriptsize{$b$}};
		\draw[{postaction=decorate}]  (0.5,1.86603) -- (e6); 
		\draw[{postaction=decorate}] (-0.8,2.43205) -- node[right] {\scriptsize{$s$}} (-0.2,2.43205); 
		\draw[{postaction=decorate}] (-0.2,1.3) -- node[left] {\scriptsize{$s$}} (-0.8,1.3); 
		\draw[{postaction=decorate}] (0.13,1.86603) -- (-0.2,1.3); 
		\draw[{postaction=decorate}] (-0.2,2.43205) -- (0.13,1.86603); 
		\draw[{postaction=decorate}] (-1.13,1.86603) -- (-0.8,2.43205); 
		\draw[{postaction=decorate}] (-0.8,1.3) -- (-1.13,1.86603); 
		\end{tikzpicture}
		\Bigg\rangle
		=\sum_{\substack{g',h',i',\\j',k',l'}} B_{\mathbf{p},ghijkl}^{s,g'h'i'j'k'l'}(abcdef)\ \Bigg|
		\begin{tikzpicture} [scale=0.9,rotate=90, scale=1,baseline={([yshift=-3pt]current bounding box.center)}, decoration={markings,mark=at position .5 with {\arrow[>=stealth]{<}}}]
		\draw[{postaction=decorate}] (0,2.73205) -- node[above] {\scriptsize{$g'$}} (0.5,1.86603); 
		\draw[{postaction=decorate}] (-1,2.73205) -- node[left] {\scriptsize{$l'$}} (0,2.73205); 
		\draw[{postaction=decorate}] (-1.5,1.86603) -- node[below] {\scriptsize{$k'$}} (-1,2.73205); 
		\draw[{postaction=decorate}] (-1,1) -- node[below] {\scriptsize{$j'$}} (-1.5,1.86603); 
		\draw[{postaction=decorate}] (0,1) -- node[right] {\scriptsize{$i'$}} (-1,1); 
		\draw[{postaction=decorate}] (0.5,1.86603) -- node[above] {\scriptsize{$h'$}} (0,1); 
		\node (e1) at (0.25,3.16506) {\scriptsize{$a$}};
		\draw[{postaction=decorate}] (0,2.73205) -- (e1); 
		\node (e2) at (-1.25,3.16506) {\scriptsize{$f$}};
		\draw[{postaction=decorate}] (-1,2.73205) -- (e2); 
		\node (e3) at (-2,1.86603) {\scriptsize{$e$}};
		\draw[{postaction=decorate}] (-1.5,1.86603) -- (e3); 
		\node (e4) at (-1.25,0.566987) {\scriptsize{$d$}};
		\draw[{postaction=decorate}] (-1,1) -- (e4); 
		\node (e5) at (0.25,0.566987) {\scriptsize{$c$}};
		\draw[{postaction=decorate}] (0,1) -- (e5); 
		\node (e6) at (1,1.86603) {\scriptsize{$b$}};
		\draw[{postaction=decorate}]  (0.5,1.86603) -- (e6); 
		\end{tikzpicture}
		\Bigg\rangle.\label{BpsDef}
	\end{equation}
\end{widetext}
The operator $B_\mathbf{p}^s$ acts on the twelve links of the plaquette and hence has a representation as a $(N+1)^{12}\times (N+1)^{12}$ matrix. However, since the external legs $a,\dots,f$ are not changed by this operator, the matrix has a block-diagonal structure involving $(N+1)^6$ blocks of dimension $(N+1)^6\times (N+1)^6$, where each block is labeled by a fixed configuration of external legs.

We now evaluate the action of the operator $B_\mathbf{p}^s$ on a fixed plaquette $\mathbf{p}$ step by step to get an equation for the matrix elements $B_{\mathbf{p},ghijkl}^{s,g'h'i'j'k'l'}(abcdef)$. The first step in this calculation is fusing the $s$-type loop string to the internal links of the plaquette. To achieve this, we apply the completeness relation \eqref{eq:completeness} at every internal link.
\begin{widetext}
	\begin{equation}
		\begin{split}
			\begin{tikzpicture} [rotate=90, scale=1,baseline=(current bounding box.center), decoration={markings,mark=at position .5 with {\arrow[>=stealth]{<}}}, scale=1]
			\draw[{postaction=decorate}] (0,2.73205) -- node[above] {\scriptsize{$g$}} (0.5,1.86603); 
			\draw[{postaction=decorate}] (-1,2.73205) -- node[left] {\scriptsize{$l$}} (0,2.73205); 
			\draw[{postaction=decorate}] (-1.5,1.86603) -- node[below] {\scriptsize{$k$}} (-1,2.73205); 
			\draw[{postaction=decorate}] (-1,1) -- node[below] {\scriptsize{$j$}} (-1.5,1.86603); 
			\draw[{postaction=decorate}] (0,1) -- node[right] {\scriptsize{$i$}} (-1,1); 
			\draw[{postaction=decorate}] (0.5,1.86603) -- node[above] {\scriptsize{$h$}} (0,1); 
			\node (e1) at (0.25,3.16506) {\scriptsize{$a$}};
			\draw[{postaction=decorate}] (0,2.73205) -- (e1); 
			\node (e2) at (-1.25,3.16506) {\scriptsize{$f$}};
			\draw[{postaction=decorate}] (-1,2.73205) -- (e2); 
			\node (e3) at (-2,1.86603) {\scriptsize{$e$}};
			\draw[{postaction=decorate}] (-1.5,1.86603) -- (e3); 
			\node (e4) at (-1.25,0.566987) {\scriptsize{$d$}};
			\draw[{postaction=decorate}] (-1,1) -- (e4); 
			\node (e5) at (0.25,0.566987) {\scriptsize{$c$}};
			\draw[{postaction=decorate}] (0,1) -- (e5); 
			\node (e6) at (1,1.86603) {\scriptsize{$b$}};
			\draw[{postaction=decorate}]  (0.5,1.86603) -- (e6); 
			\draw[{postaction=decorate}] (-0.8,2.43205) -- node[right] {\scriptsize{$s$}} (-0.2,2.43205); 
			\draw[{postaction=decorate}] (-0.2,1.3) -- node[left] {\scriptsize{$s$}} (-0.8,1.3); 
			\draw[{postaction=decorate}] (0.13,1.86603) -- (-0.2,1.3); 
			\draw[{postaction=decorate}] (-0.2,2.43205) -- (0.13,1.86603); 
			\draw[{postaction=decorate}] (-1.13,1.86603) -- (-0.8,2.43205); 
			\draw[{postaction=decorate}] (-0.8,1.3) -- (-1.13,1.86603); 
			\draw[dotted] (0.13,1.86603) -- (0.30492,2.16603);
			\draw[dotted] (-1.13,1.86603) -- (-1.30492,1.56603);
			\end{tikzpicture}
			&=
			\sum_{\substack{g',h',i',\\j',k',l'}}
			\sqrt{\frac{d_{g'^*}}{d_{g^*} d_{s^*}}} \sqrt{\frac{d_{h'}}{d_{h} d_{s}}} \sqrt{\frac{d_{i'}}{d_{i} d_{s}}} \sqrt{\frac{d_{j'}}{d_{j} d_{s}}} \sqrt{\frac{d_{k'^*}}{d_{k^*} d_{s^*}}} \sqrt{\frac{d_{l'^*}}{d_{l^*} d_{s^*}}}\\
			&\begin{tikzpicture} [rotate=90, baseline=(current bounding box.center), decoration={markings,mark=at position .5 with {\arrow[>=stealth]{<}}}, scale=4]
			\node (a) at (0.115471,2.93205) {\small{$a$}};
			\coordinate (aend) at (0,2.73205);
			\draw[{postaction=decorate}] (aend) -- (a); 
			\node (f) at (-1.13453, 2.96506) {\small{$f$}};
			\coordinate (fend) at (-1,2.73205);
			\draw[{postaction=decorate}] (fend) -- (f); 
			\node (e) at (-1.7,1.86603) {\small{$e$}};
			\coordinate (eend) at (-1.5,1.86603);
			\draw[{postaction=decorate}] (eend) -- (e); 
			\node (d) at (-1.13453, 0.766988) {\small{$d$}};
			\coordinate (dend) at (-1,1);
			\draw[{postaction=decorate}] (dend) -- (d); 
			\node (c) at (0.115471,0.799999) {\small{$c$}};
			\coordinate (cend) at (0,1);
			\draw[{postaction=decorate}] (cend) -- (c); 
			\node (b) at (0.7,1.86603) {\small{$b$}};
			\coordinate (bend) at (0.5,1.86603);
			\draw[{postaction=decorate}] (bend)  -- (b); 
			\coordinate (cap) at (0.25,1.86603);
			\coordinate (cup) at (-1.25,1.86603);
			\draw[dotted] (cap) -- (0.420322, 2.00404);
			\draw[dotted] (cup) -- (-1.42032, 1.72802);
			\coordinate (lsup) at (-0.55,2.58205);
			\coordinate (lsdown) at (-0.45,2.58205);
			\draw[{postaction=decorate}] (lsup) -- node[left] {\small{$l'$}} (lsdown);
			\coordinate (slcdown) at (-0.7,2.43205);
			\coordinate (lcdown) at (-0.7,2.73205);
			\coordinate (slcup) at (-0.3,2.43205);
			\coordinate (lcup) at (-0.3,2.73205);
			\draw[{postaction=decorate}] (slcdown) -- (lsup);
			\draw[{postaction=decorate}] (lcdown) -- (lsup);
			\draw[{postaction=decorate}] (lsdown) -- (slcup);
			\draw[{postaction=decorate}] (lsdown) -- (lcup);
			\draw[{postaction=decorate}] (lcup) -- node[left] {\small{$l$}} (0,2.73205);
			\draw[{postaction=decorate}] (-1,2.73205) -- node[left] {\small{$l$}} (lcdown);
			\draw (slcdown) -- (-0.8,2.43205);
			\draw (slcup) -- (-0.2,2.43205);
			\coordinate (isup) at (-0.55,1.15);
			\coordinate (isdown) at (-0.45,1.15);
			\draw[{postaction=decorate}] (isdown) -- node[right] {\small{$i'$}} (isup);
			\coordinate (sicdown) at (-0.7,1);
			\coordinate (icdown) at (-0.7,1.3);
			\coordinate (sicup) at (-0.3,1);
			\coordinate (icup) at (-0.3,1.3);
			\draw[{postaction=decorate}] (isup) -- (sicdown);
			\draw[{postaction=decorate}] (isup) -- (icdown);
			\draw[{postaction=decorate}] (sicup) -- (isdown);
			\draw[{postaction=decorate}] (icup) -- (isdown);
			\draw[{postaction=decorate}] (0,1) -- node[right] {\small{$i$}} (sicup);
			\draw[{postaction=decorate}] (sicdown) -- node[right] {\small{$i$}} (-1,1);	
			\draw (icup) -- (-0.2,1.3);
			\draw (icdown) -- (-0.8,1.3);
			\coordinate (sgdown) at (0.11412, 2.19204);
			\coordinate (sgup) at (0.0765461, 2.25604);
			\draw[{postaction=decorate}] (sgup) -- node[above left] {\small{$g'$}} (sgdown);
			\coordinate (gcup) at (0.340645,2.14204);
			\draw[{postaction=decorate}] (sgdown) -- (gcup);
			\coordinate (gscdown) at (-0.126534, 2.30604);
			\draw[{postaction=decorate}] (gscdown) -- (sgup);
			\coordinate (gcdown) at (0.151662, 2.46937);
			\coordinate (gscup) at (0.16441, 1.97953);
			\draw[{postaction=decorate}] (sgdown) -- (gscup);
			\coordinate (ag) at (0.0288677,2.68205);
			\draw[{postaction=decorate}] (aend) -- (ag);
			\draw[{postaction=decorate}] (ag) -- node[above] {\small{$g$}} (sgup);
			\draw[{postaction=decorate}] (gcup) -- node[above] {\small{$g$}} (0.5,1.86603);
			\draw[{postaction=decorate}] (-0.2,2.43205) -- node[below] {\small{$s$}} (gscdown);
			\draw[{postaction=decorate}] (gscup) -- (cap);
			\coordinate (shdown) at (0.0765468, 1.47602);
			\coordinate (shup) at (0.114118, 1.54001);
			\draw[{postaction=decorate}] (shup) -- node[above right] {\small{$h'$}} (shdown);
			\coordinate (hcup) at (0.340641,1.59001);
			\draw[{postaction=decorate}] (hcup) -- (shup);
			\coordinate (shcdown) at (-0.126529,1.42602);
			\draw[{postaction=decorate}] (shdown) -- (shcdown);
			\coordinate (hsdown) at (0.1516597983468383,1.26268);
			\coordinate (hsup) at (0.164404, 1.75254);
			\draw[{postaction=decorate}] (hsup) -- (shup);
			\draw[{postaction=decorate}] (cap) -- (hsup);
			\draw[{postaction=decorate}] (shcdown) -- node[below] {\small{$s$}} (-0.2,1.3);
			\coordinate (ch) at (0.0288674,1.05);
			\draw[{postaction=decorate}] (ch) -- (cend);
			\draw[{postaction=decorate}] (shdown) -- node[above] {\small{$h$}} (ch);
			\draw[{postaction=decorate}] (0.5,1.86603) -- node[above] {\small{$h$}} (hcup);
			\coordinate (sjdown) at (-1.11412, 1.54001);
			\coordinate (sjup) at (-1.07655, 1.47602);
			\draw[{postaction=decorate}] (sjup) -- node[below right] {\small{$j'$}} (sjdown);
			\coordinate (jscdown) at (-1.34064, 1.59001);
			\draw[{postaction=decorate}] (sjdown) -- (jscdown);
			\coordinate (jcup) at (-0.873471, 1.42602);
			\draw[{postaction=decorate}] (jcup) -- (sjup);
			\coordinate (jcdown) at (-1.16441, 1.75254);
			\draw[{postaction=decorate}] (sjdown) -- (jcdown);
			\coordinate (jscup) at (-1.15166, 1.26268);
			\draw[{postaction=decorate}] (jscdown) -- node[below] {\small{$j$}} (-1.5,1.86603);
			\coordinate (dj) at (-1.02887,1.05);
			\draw[{postaction=decorate}] (dend) -- (dj);
			\draw[{postaction=decorate}] (dj) -- node[below] {\small{$j$}} (sjup);
			\draw[{postaction=decorate}] (jcdown) -- (cup);
			\draw[{postaction=decorate}] (-0.8,1.3) -- node[above] {\small{$s$}} (jcup);
			\coordinate (skup) at (-1.11412, 2.19204);
			\coordinate (skdown) at (-1.07655, 2.25604);
			\draw[{postaction=decorate}] (skup) -- node[below left] {\small{$k'$}} (skdown);
			\coordinate (kscup) at (-0.873466, 2.30604);
			\draw[{postaction=decorate}] (skdown) -- (kscup);
			\coordinate (kcdown) at (-1.34064, 2.14204);
			\draw[{postaction=decorate}] (kcdown) -- (skup);
			\coordinate (kcup) at (-1.15166, 2.46937);
			\coordinate (kscdown) at (-1.16441, 1.97953);
			\draw[{postaction=decorate}] (kscdown) -- (skup);
			\draw[{postaction=decorate}] (cup) -- (kscdown);
			\draw[{postaction=decorate}] (kscup) -- node[above] {\small{$s$}} (-0.8,2.43205);
			\draw[{postaction=decorate}] (-1.5,1.86603) -- node[below] {\small{$k$}} (kcdown);
			\coordinate (fk) at (-1.02887,2.68205);
			\draw[{postaction=decorate}] (skdown) -- (fk);
			\draw[{postaction=decorate}] (fk) -- node[below] {\small{$k$}} (fend);
			\node (sgh) at (0.17,1.86603) {\small{$s$}};
			\node (sjk) at (-1.17,1.86603) {\small{$s$}};
			\node[circle,scale=0.7,draw,red] (vertex1) at (0.375,1.86603) {$1$};
			\node[circle,scale=0.7,draw,red] (vertex2) at (-0.0925,1.15) {$2$};
			\node[circle,scale=0.7,draw,red] (vertex3) at (-0.9075,1.15) {$3$};
			\node[circle,scale=0.7,draw,red] (vertex4) at (-1.365,1.86603) {$4$};
			\node[circle,scale=0.7,draw,red] (vertex5) at (-0.9075,2.58205) {$5$};
			\node[circle,scale=0.7,draw,red] (vertex5) at (-0.0925,2.58205) {$6$};
			\end{tikzpicture}.
		\end{split}\label{eq:bigdiagram}
	\end{equation}
\end{widetext}
We can now evaluate each of the six corners individually. To simplify the diagrams we make use of the graphical calculus of the underlying unitary fusion category. This is explained in more detail in Appendix~\ref{app:UFCgraphcalc}. If necessary, we first rearrange the diagram such that it is easier to see which operation can be applied.
\begin{itemize}
	\item[\itemone] To be able to apply an $F$-move to this picture, we have added a vacuum line \footnote{Note that we are always allowed to add and remove vacuum lines, since they do not change the meaning of the diagram. Mathematically, the loop added by the $B_\mathbf{p}^s$	operator is a composition of an evaluation and a co-evaluation morphism. Therefore, a vacuum line at the top corner as well as at the bottom corner of the loop is already implicit.}. We can then use the modified $F$-moves \eqref{eq:mirror2} and \eqref{eq:mirror} and the bigon relation \eqref{eq:UFCbigon} twice to to simplify the diagram in the following way:
	\begin{align*}
		&\begin{tikzpicture}[rotate=90, baseline=(current bounding box.center), decoration={markings,mark=at position .5 with {\arrow[>=stealth]{<}}}, scale=2]
		\node (b) at (0.8,1.86603) {\scriptsize{$b$}};
		\coordinate (bend) at (0.5,1.86603);
		\draw[{postaction=decorate}] (bend)  -- (b); 
		\coordinate (cap) at (0.25,1.86603);
		\draw[dotted] (cap) -- (0.420322, 2.00404);
		\node (sgh) at (0.15,1.86603) {\scriptsize{$s$}};
		\coordinate (sgdown) at (0.143474, 2.14204);
		\coordinate (sgup) at (0.0471918, 2.30604);
		\draw[{postaction=decorate}] (sgup) -- node[above] {\scriptsize{$g'$}\ } (sgdown);
		\coordinate (gcup) at (0.340645,2.14204);
		\draw[{postaction=decorate}] (sgdown) -- (gcup);
		\coordinate (gcdown) at (0.151662, 2.46937);
		\coordinate (gscup) at (0.16441, 1.97953);
		\draw[{postaction=decorate}] (sgdown) -- (gscup);
		\draw[{postaction=decorate}] (gcup) -- node[above] {\scriptsize{$g$}} (0.5,1.86603);
		\draw[{postaction=decorate}] (gscup) -- (cap);
		\coordinate (shdown) at (0.0471897, 1.42602);
		\coordinate (shup) at (0.143475, 1.59001);
		\draw[{postaction=decorate}] (shup) -- node[above] {\ \scriptsize{$h'$}} (shdown);
		\coordinate (hcup) at (0.340641,1.59001);
		\draw[{postaction=decorate}] (hcup) -- (shup);
		\coordinate (hsdown) at (0.1516597983468383,1.26268);
		\coordinate (hsup) at (0.164404, 1.75254);
		\draw[{postaction=decorate}] (hsup) -- (shup);
		\draw[{postaction=decorate}] (cap) -- (hsup);
		\draw[{postaction=decorate}] (0.5,1.86603) -- node[above] {\scriptsize{$h$}} (hcup);
		\end{tikzpicture}
		=\sum_\alpha \left(F^{g^*s^*s}_{g^*}\right)_{0\alpha^*}^\mathsf{T}
		\begin{tikzpicture}[baseline=(current bounding box.center), decoration={markings,mark=at position .5 with {\arrow[>=stealth]{>}}}]
		\node (b) at (0,0.5) {\scriptsize{$b$}};
		\coordinate (bend) at (0,0);
		\draw[{postaction=decorate}] (b)  -- (bend);
		\coordinate (gs) at (-1,-1.5); 
		\node (gstext) at (-1,-1.7) {\scriptsize{$g'$}};
		\coordinate (g) at (-0.2, -0.3);
		\draw[{postaction=decorate}] (bend) -- node[left] {\scriptsize{$g$}} (g);
		\coordinate (alpha) at (-0.4, -0.6);
		\draw[{postaction=decorate}] (g) -- node[left] {\scriptsize{$\alpha$}} (alpha);
		\coordinate (sg) at (-0.4, -1.05);
		\coordinate (gbigon) at (-0.815385, -0.773077);
		\draw (alpha) -- (sg);
		\coordinate (gsstart) at (-0.8, -1.25);
		\draw[{postaction=decorate}] (sg) -- node[below right] {\scriptsize{$s$}} (gsstart);
		\draw[{postaction=decorate}] (gsstart) -- (gs);
		\draw (alpha) -- (gbigon);
		\draw[{postaction=decorate}] (gbigon) -- node[above left] {\scriptsize{$g$}} (gsstart);
		\coordinate (sh) at (0.25,-1.05);
		\draw[{postaction=decorate}] (sh) -- node[below left] {\scriptsize{$s$}} (g);
		\coordinate (h) at (0.571429,-0.857143);
		\draw[{postaction=decorate}] (h) -- node[above right] {\scriptsize{$h$}} (bend);
		\coordinate (hc) at (0.571429, -1.15714);
		\draw (hc) -- (h);
		\draw (hc) -- (sh);
		\coordinate (hs) at (0.8,-1.5);
		\node (gstext) at (0.8,-1.7) {\scriptsize{$h'$}};
		\draw[{postaction=decorate}] (hs) -- (hc);
		\end{tikzpicture}\\
		&=\sum_\alpha \left(F^{g^*s^*s}_{g^*}\right)_{\alpha^*0}\delta_{g'^*,\alpha^*} \sqrt{\frac{d_{g^*} d_{s^*}}{d_{g'^*}}}
		\begin{tikzpicture}[baseline=(current bounding box.center), decoration={markings,mark=at position .5 with {\arrow[>=stealth]{>}}}]
		\node (b) at (0,0.5) {\scriptsize{$b$}};
		\coordinate (bend) at (0,0);
		\draw[{postaction=decorate}] (b)  -- (bend);
		\coordinate (gs) at (-1,-1.5); 
		\node (gstext) at (-1,-1.7) {\scriptsize{$g'$}};
		\coordinate (g) at (-0.2, -0.3);
		\draw[{postaction=decorate}] (bend) -- node[left] {\scriptsize{$g$}} (g);
		\draw[{postaction=decorate}] (g) -- (gs);
		\coordinate (sh) at (0.25,-1.05);
		\draw[{postaction=decorate}] (sh) -- node[below left] {\scriptsize{$s$}} (g);
		\coordinate (h) at (0.571429,-0.857143);
		\draw[{postaction=decorate}] (h) -- node[above right] {\scriptsize{$h$}} (bend);
		\coordinate (hc) at (0.571429, -1.15714);
		\draw (hc) -- (h);
		\draw (hc) -- (sh);
		\coordinate (hs) at (0.8,-1.5);
		\node (gstext) at (0.8,-1.7) {\scriptsize{$h'$}};
		\draw[{postaction=decorate}] (hs) -- (hc);
		\end{tikzpicture}\\
		&=\sum_\beta \sqrt{\frac{d_{g^*} d_{s^*}}{d_{g'^*}}} \left(F^{g^*s^*s}_{g^*}\right)_{g'^*0} \overline{\left(F_{b^*}^{g'^*sh}\right)}_{g^*\beta}
		\begin{tikzpicture}[baseline=(current bounding box.center), decoration={markings,mark=at position .5 with {\arrow[>=stealth]{>}}}]
		\node (b) at (0,0.5) {\scriptsize{$b$}};
		\coordinate (bend) at (0,0);
		\draw[{postaction=decorate}] (b)  -- (bend);
		\coordinate (gs) at (-1,-1.5); 
		\draw[{postaction=decorate}] (bend) -- (gs);
		\node (gstext) at (-1,-1.7) {\scriptsize{$g'$}};	
		\coordinate (hs) at (1,-1.5);
		\node (gstext) at (1,-1.7) {\scriptsize{$h'$}};
		%
		\coordinate (beta) at (0.4, -0.6);
		\draw[{postaction=decorate}] (beta) -- node[right] {\scriptsize{$\beta$}} (bend);
		\coordinate (sh) at (0.4, -1.05);
		\coordinate (hbigon) at (0.815385, -0.773077);
		\draw (beta) -- (sh);
		\coordinate (hsend) at (0.8, -1.25);
		\draw[{postaction=decorate}] (hsend) -- node[below left] {\scriptsize{$s$}} (sh);
		\draw[{postaction=decorate}] (hs) -- (hsend);
		\draw (beta) -- (hbigon);
		\draw[{postaction=decorate}] (hsend) -- node[above right] {\scriptsize{$g$}} (hbigon);
		\end{tikzpicture}\\
		&=\sum_\beta \sqrt{\frac{d_{g^*} d_{s^*}}{d_{g'^*}}} \left(F^{g^*s^*s}_{g^*}\right)_{g'^*0} \overline{\left(F_{b^*}^{g'^*sh}\right)}_{g^*\beta} \delta_{\beta,h'} \sqrt{\frac{d_s d_h}{d_{h'}}}\!\!
		\begin{tikzpicture}[baseline=(current bounding box.center), decoration={markings,mark=at position .5 with {\arrow[>=stealth]{>}}}]
		\node (btext) at (0,0.5) {\scriptsize{$b$}};
		\coordinate (b) at (0,0.3);
		\coordinate (bend) at (0,0);
		\draw[{postaction=decorate}] (b)  -- (bend);
		\node(gstext) at (-0.25,-0.5) {\scriptsize{$g'$}};
		\coordinate (gs) at (-0.3,-0.3);
		\draw[{postaction=decorate}] (bend)  -- (gs);
		\node (hstext) at (0.25,-0.5) {\scriptsize{$h'$}};
		\coordinate (hs) at (0.3,-0.3);
		\draw[{postaction=decorate}] (hs)  -- (bend);
		\end{tikzpicture}\\
		&=\sqrt{\frac{d_{g^*} d_{s^*}}{d_{g'^*}}} \sqrt{\frac{d_s d_h}{d_{h'}}} \left(F^{g^*s^*s}_{g^*}\right)_{g'^*0} \overline{\left(F_{b^*}^{g'^*sh}\right)}_{g^*h'}
		\begin{tikzpicture}[baseline=(current bounding box.center), decoration={markings,mark=at position .5 with {\arrow[>=stealth]{>}}}]
		\node (btext) at (0,0.5) {\scriptsize{$b$}};
		\coordinate (b) at (0,0.3);
		\coordinate (bend) at (0,0);
		\draw[{postaction=decorate}] (b)  -- (bend);
		\node(gstext) at (-0.3,-0.5) {\scriptsize{$g'$}};
		\coordinate (gs) at (-0.3,-0.3);
		\draw[{postaction=decorate}] (bend)  -- (gs);
		\node (hstext) at (0.3,-0.5) {\scriptsize{$h'$}};
		\coordinate (hs) at (0.3,-0.3);
		\draw[{postaction=decorate}] (hs)  -- (bend);
		\end{tikzpicture}
	\end{align*}
	\item[\itemtwo] Here, we can directly apply a $G$-move to simplify the diagram:
	\begin{align*}
		\begin{tikzpicture}[rotate=90, baseline=(current bounding box.center), decoration={markings,mark=at position .5 with {\arrow[>=stealth]{<}}}, scale=2]
		\node (c) at (0.115471,0.799999) {\scriptsize{$c$}};
		\coordinate (cend) at (0,1);
		\draw[{postaction=decorate}] (cend) -- (c);
		\coordinate (shdown) at (0.0765468, 1.47602);
		\coordinate (shup) at (0.114118, 1.54001);
		\draw[{postaction=decorate}] (shup) -- node[above left] {\scriptsize{$h'$}} (shdown);
		\coordinate (shcdown) at (-0.126529,1.42602);
		\draw[{postaction=decorate}] (shdown) -- (shcdown);
		\coordinate (hsdown) at (0.1516597983468383,1.26268);
		\draw[{postaction=decorate}] (shcdown) -- node[below] {\scriptsize{$s$}} (-0.2,1.3);
		\coordinate (ch) at (0.0288674,1.05);
		\draw[{postaction=decorate}] (ch) -- (cend);
		\draw[{postaction=decorate}] (shdown) -- node[above] {\scriptsize{$h$}} (ch);
		\coordinate (isup) at (-0.55,1.15);
		\coordinate (isdown) at (-0.45,1.15);
		\draw[{postaction=decorate}] (isdown) -- node[below] {\scriptsize{$i'$}} (isup);
		\coordinate (sicup) at (-0.3,1);
		\coordinate (icup) at (-0.3,1.3);
		\draw[{postaction=decorate}] (sicup) -- (isdown);
		\draw[{postaction=decorate}] (icup) -- (isdown);
		\draw[{postaction=decorate}] (0,1) -- node[right] {\scriptsize{$i$}} (sicup);
		\draw (icup) -- (-0.2,1.3);
		\end{tikzpicture}
		&=
		\begin{tikzpicture}[decoration={markings,mark=at position .5 with {\arrow[>=stealth]{>}}},baseline=(current bounding box.center)]
		\node (is) at (0,0) {\scriptsize{$i'$}};
		\coordinate (isend) at (0,0.5);
		\draw[{postaction=decorate}] (is) -- (isend);
		\coordinate (s) at (-0.5,0.8);
		\coordinate (i) at (0.5,0.8);
		\draw (isend) -- (s);
		\draw (isend) -- (i);
		\coordinate (ih) at (0.5,1);
		\coordinate (sh) at (-0.5,1.6);
		\draw[{postaction=decorate}] (s) -- node[left] {\scriptsize{$s$}}(sh);
		\draw[{postaction=decorate}] (i) -- node[right] {\scriptsize{$i$}} (ih);
		\draw[{postaction=decorate}] (ih) -- node[above] {\scriptsize{$h$}} (sh);
		\node (hs) at (-0.5,2.1) {\scriptsize{$h'$}};
		\node (c) at (0.5,2.1) {\scriptsize{$c$}};
		\draw[{postaction=decorate}] (sh) -- (hs);
		\draw[{postaction=decorate}] (c) -- (ih);
		\end{tikzpicture}\\
		&=\sum_\alpha \left(G_{si}^{h'c^*}\right)_{h\alpha}
		\begin{tikzpicture}[decoration={markings,mark=at position .5 with {\arrow[>=stealth]{>}}},baseline=(current bounding box.center)]
		\node (is) at (0,0) {\scriptsize{$i'$}};
		\coordinate (isend) at (0,0.5);
		\draw[{postaction=decorate}] (is) -- (isend);
		\coordinate (s) at (-0.3,0.8);
		\coordinate (i) at (0.3,0.8);
		\draw[{postaction=decorate}] (isend) -- node[left] {\scriptsize{$s$}} (s);
		\draw[{postaction=decorate}] (isend) -- node[right] {\scriptsize{$i$}} (i);
		\coordinate (adown) at (0,1.1);
		\draw (s) -- (adown);
		\draw (i) -- (adown);
		\coordinate (aup) at (0,1.5);
		\draw[{postaction=decorate}] (adown) -- node[right] {\scriptsize{$\alpha$}} (aup);
		\node (hs) at (-0.5,2.1) {\scriptsize{$h'$}};
		\node (c) at (0.5,2.1) {\scriptsize{$c$}};
		\draw[{postaction=decorate}] (aup) -- (hs);
		\draw[{postaction=decorate}] (c) -- (aup);
		\end{tikzpicture}\\
		&=\sum_\alpha \left(G_{si}^{h'c^*}\right)_{h\alpha} \sqrt{\frac{d_s d_i}{d_{i'}}} \delta_{\alpha,i'}
		\begin{tikzpicture}[baseline=(current bounding box.center), decoration={markings,mark=at position .5 with {\arrow[>=stealth]{>}}},rotate=180]
		\node (btext) at (0,0.5) {\scriptsize{$i'$}};
		\coordinate (b) at (0,0.3);
		\coordinate (bend) at (0,0);
		\draw[{postaction=decorate}] (b)  -- (bend);
		\node(gstext) at (-0.3,-0.5) {\scriptsize{$c$}};
		\coordinate (gs) at (-0.3,-0.3);
		\draw[{postaction=decorate}] (gs)  -- (bend);
		\node (hstext) at (0.3,-0.5) {\scriptsize{$h'$}};
		\coordinate (hs) at (0.3,-0.3);
		\draw[{postaction=decorate}] (bend)  -- (hs);
		\end{tikzpicture}\\
		&=\sqrt{\frac{d_s d_i}{d_{i'}}} \left(G_{si}^{h'c^*}\right)_{hi'} 
		\begin{tikzpicture}[baseline=(current bounding box.center), decoration={markings,mark=at position .5 with {\arrow[>=stealth]{>}}},rotate=180]
		\node (btext) at (0,0.5) {\scriptsize{$i'$}};
		\coordinate (b) at (0,0.3);
		\coordinate (bend) at (0,0);
		\draw[{postaction=decorate}] (b)  -- (bend);
		\node(gstext) at (-0.3,-0.5) {\scriptsize{$c$}};
		\coordinate (gs) at (-0.3,-0.3);
		\draw[{postaction=decorate}] (gs)  -- (bend);
		\node (hstext) at (0.3,-0.5) {\scriptsize{$h'$}};
		\coordinate (hs) at (0.3,-0.3);
		\draw[{postaction=decorate}] (bend)  -- (hs);
		\end{tikzpicture}\\
		&=\sqrt{\frac{d_s d_h}{d_{h'}}} \overline{\left(F_{i'}^{shc^*}\right)}_{ih'}
		\begin{tikzpicture}[baseline=(current bounding box.center), decoration={markings,mark=at position .5 with {\arrow[>=stealth]{>}}},rotate=180]
		\node (btext) at (0,0.5) {\scriptsize{$i'$}};
		\coordinate (b) at (0,0.3);
		\coordinate (bend) at (0,0);
		\draw[{postaction=decorate}] (b)  -- (bend);
		\node(gstext) at (-0.3,-0.5) {\scriptsize{$c$}};
		\coordinate (gs) at (-0.3,-0.3);
		\draw[{postaction=decorate}] (gs)  -- (bend);
		\node (hstext) at (0.3,-0.5) {\scriptsize{$h'$}};
		\coordinate (hs) at (0.3,-0.3);
		\draw[{postaction=decorate}] (bend)  -- (hs);
		\end{tikzpicture}
	\end{align*}
	\item[\itemthree] This diagram requires us to apply a $H$-move:
	\begin{align*}
		\begin{tikzpicture}[rotate=90, baseline=(current bounding box.center), decoration={markings,mark=at position .5 with {\arrow[>=stealth]{<}}}, scale=2]
		\node (d) at (-1.13453, 0.766988) {\scriptsize{$d$}};
		\coordinate (dend) at (-1,1);
		\draw[{postaction=decorate}] (dend) -- (d);
		\coordinate (sjdown) at (-1.11412, 1.54001);
		\coordinate (sjup) at (-1.07655, 1.47602);
		\draw[{postaction=decorate}] (sjup) -- node[below left] {\scriptsize{$j'$}} (sjdown);
		\coordinate (jcup) at (-0.873471, 1.42602);
		\draw[{postaction=decorate}] (jcup) -- (sjup);
		\coordinate (jscup) at (-1.15166, 1.26268);
		\coordinate (dj) at (-1.02887,1.05);
		\draw[{postaction=decorate}] (dend) -- (dj);
		\draw[{postaction=decorate}] (dj) -- node[below] {\scriptsize{$j$}} (sjup);
		\draw[{postaction=decorate}] (-0.8,1.3) -- node[above] {\scriptsize{$s$}} (jcup);
		\coordinate (isup) at (-0.55,1.15);
		\coordinate (isdown) at (-0.45,1.15);
		\draw[{postaction=decorate}] (isdown) -- node[above] {\scriptsize{$i'$}} (isup);
		\coordinate (sicdown) at (-0.7,1);
		\coordinate (icdown) at (-0.7,1.3);
		\draw[{postaction=decorate}] (isup) -- (sicdown);
		\draw[{postaction=decorate}] (isup) -- (icdown);
		\draw[{postaction=decorate}] (sicdown) -- node[right] {\scriptsize{$i$}} (-1,1);	
		\draw (icdown) -- (-0.8,1.3);
		\end{tikzpicture}
		&=
		\begin{tikzpicture}[decoration={markings,mark=at position .5 with {\arrow[>=stealth]{>}}},baseline=(current bounding box.center),rotate=180]
		\node (is) at (0,0) {\scriptsize{$i'$}};
		\coordinate (isend) at (0,0.5);
		\draw[{postaction=decorate}] (isend) -- (is);
		\coordinate (i) at (-0.5,0.8);
		\coordinate (s) at (0.5,0.8);
		\draw (i) -- (isend);
		\draw (s) -- (isend);
		\coordinate (sj) at (0.5,1.6);
		\coordinate (ij) at (-0.5,1);
		\draw[{postaction=decorate}] (ij) -- node[right] {\scriptsize{$i$}}(i);
		\draw[{postaction=decorate}] (sj) -- node[left] {\scriptsize{$s$}} (s);
		\draw[{postaction=decorate}] (sj) -- node[below] {\scriptsize{$j$}} (ij);
		\node (d) at (-0.5,2.1) {\scriptsize{$d$}};
		\node (js) at (0.5,2.1) {\scriptsize{$j'$}};
		\draw[{postaction=decorate}] (d) -- (ij);
		\draw[{postaction=decorate}] (js) -- (sj);
		\end{tikzpicture}\\
		&=\sum_\alpha \left(H_{j'd}^{si}\right)_{j\alpha}
		\begin{tikzpicture}[decoration={markings,mark=at position .5 with {\arrow[>=stealth]{>}}},baseline=(current bounding box.center),rotate=180]
		\node (is) at (0,0) {\scriptsize{$i'$}};
		\coordinate (isend) at (0,0.5);
		\draw[{postaction=decorate}] (isend) -- (is);
		\coordinate (s) at (-0.3,0.8);
		\coordinate (i) at (0.3,0.8);
		\draw[{postaction=decorate}] (s) -- node[right] {\scriptsize{$i$}} (isend);
		\draw[{postaction=decorate}] (i) -- node[left] {\scriptsize{$s$}} (isend);
		\coordinate (adown) at (0,1.1);
		\draw (s) -- (adown);
		\draw (i) -- (adown);
		\coordinate (aup) at (0,1.5);
		\draw[{postaction=decorate}] (aup) -- node[right] {\scriptsize{$\alpha$}} (adown);
		\node (hs) at (-0.5,2.1) {\scriptsize{$d$}};
		\node (c) at (0.5,2.1) {\scriptsize{$j'$}};
		\draw[{postaction=decorate}] (hs) -- (aup);
		\draw[{postaction=decorate}] (c) -- (aup);
		\end{tikzpicture}\\
		&=\sum_\alpha \left(H_{j'd}^{si}\right)_{j\alpha} \sqrt{\frac{d_{s} d_{i}}{d_{i'}}} \delta_{\alpha,i'}
		\begin{tikzpicture}[baseline=(current bounding box.center), decoration={markings,mark=at position .5 with {\arrow[>=stealth]{>}}}]
		\node (btext) at (0,0.5) {\scriptsize{$i'$}};
		\coordinate (b) at (0,0.3);
		\coordinate (bend) at (0,0);
		\draw[{postaction=decorate}] (bend)  -- (b);
		\node(gstext) at (-0.3,-0.5) {\scriptsize{$j'$}};
		\coordinate (gs) at (-0.3,-0.3);
		\draw[{postaction=decorate}] (gs)  -- (bend);
		\node (hstext) at (0.3,-0.5) {\scriptsize{$d$}};
		\coordinate (hs) at (0.3,-0.3);
		\draw[{postaction=decorate}] (hs)  -- (bend);
		\end{tikzpicture}\\
		&=\sqrt{\frac{d_{s} d_{i}}{d_{i'}}} \left(H_{j'd}^{si}\right)_{ji'}
		\begin{tikzpicture}[baseline=(current bounding box.center), decoration={markings,mark=at position .5 with {\arrow[>=stealth]{>}}}]
		\node (btext) at (0,0.5) {\scriptsize{$i'$}};
		\coordinate (b) at (0,0.3);
		\coordinate (bend) at (0,0);
		\draw[{postaction=decorate}] (bend)  -- (b);
		\node(gstext) at (-0.3,-0.5) {\scriptsize{$j'$}};
		\coordinate (gs) at (-0.3,-0.3);
		\draw[{postaction=decorate}] (gs)  -- (bend);
		\node (hstext) at (0.3,-0.5) {\scriptsize{$d$}};
		\coordinate (hs) at (0.3,-0.3);
		\draw[{postaction=decorate}] (hs)  -- (bend);
		\end{tikzpicture}\\
		&=\sqrt{\frac{d_{s} d_{j}}{d_{j'}}} \left(F_{i'}^{sjd}\right)_{j'i}
		\begin{tikzpicture}[baseline=(current bounding box.center), decoration={markings,mark=at position .5 with {\arrow[>=stealth]{>}}}]
		\node (btext) at (0,0.5) {\scriptsize{$i'$}};
		\coordinate (b) at (0,0.3);
		\coordinate (bend) at (0,0);
		\draw[{postaction=decorate}] (bend)  -- (b);
		\node(gstext) at (-0.3,-0.5) {\scriptsize{$j'$}};
		\coordinate (gs) at (-0.3,-0.3);
		\draw[{postaction=decorate}] (gs)  -- (bend);
		\node (hstext) at (0.3,-0.5) {\scriptsize{$d$}};
		\coordinate (hs) at (0.3,-0.3);
		\draw[{postaction=decorate}] (hs)  -- (bend);
		\end{tikzpicture}
	\end{align*}
	\item[\itemfour] This diagram works analogously to the first one by adding a vacuum line. This time, we apply the $F$-moves \eqref{eq:Fsymbol} and \eqref{eq:hermit} together with the bigon relation \eqref{eq:UFCbigon}:
	\begin{align*}
		&\begin{tikzpicture}[rotate=270, baseline=(current bounding box.center), decoration={markings,mark=at position .5 with {\arrow[>=stealth]{<}}}, scale=2]
		\node (b) at (0.8,1.86603) {\scriptsize{$e$}};
		\coordinate (bend) at (0.5,1.86603);
		\draw[{postaction=decorate}] (bend)  -- (b); 
		\coordinate (cap) at (0.25,1.86603);
		\draw[dotted] (cap) -- (0.420322, 2.00404);
		\node (sgh) at (0.15,1.86603) {\scriptsize{$s$}};
		\coordinate (sgdown) at (0.143474, 2.14204);
		\coordinate (sgup) at (0.0471918, 2.30604);
		\draw[{postaction=decorate}] (sgup) -- node[above] {\scriptsize{$j'$}} (sgdown);
		\coordinate (gcup) at (0.340645,2.14204);
		\draw[{postaction=decorate}] (sgdown) -- (gcup);
		\coordinate (gcdown) at (0.151662, 2.46937);
		\coordinate (gscup) at (0.16441, 1.97953);
		\draw[{postaction=decorate}] (sgdown) -- (gscup);
		\draw[{postaction=decorate}] (gcup) -- node[below] {\scriptsize{$j$}} (0.5,1.86603);
		\draw[{postaction=decorate}] (gscup) -- (cap);
		\coordinate (shdown) at (0.0471897, 1.42602);
		\coordinate (shup) at (0.143475, 1.59001);
		\draw[{postaction=decorate}] (shup) -- node[above] {\scriptsize{$k'$}} (shdown);
		\coordinate (hcup) at (0.340641,1.59001);
		\draw[{postaction=decorate}] (hcup) -- (shup);
		\coordinate (hsdown) at (0.1516597983468383,1.26268);
		\coordinate (hsup) at (0.164404, 1.75254);
		\draw[{postaction=decorate}] (hsup) -- (shup);
		\draw[{postaction=decorate}] (cap) -- (hsup);
		\draw[{postaction=decorate}] (0.5,1.86603) -- node[below] {\scriptsize{$k$}} (hcup);
		\end{tikzpicture}
		=\sum_\alpha \left(F^{s^*sj}_{j}\right)_{0\alpha}
		\begin{tikzpicture}[baseline=(current bounding box.center), decoration={markings,mark=at position .5 with {\arrow[>=stealth]{>}}},rotate=180]
		\node (b) at (0,0.5) {\scriptsize{$e$}};
		\coordinate (bend) at (0,0);
		\draw[{postaction=decorate}] (b)  -- (bend);
		\coordinate (gs) at (-1,-1.5); 
		\node (gstext) at (-1,-1.7) {\scriptsize{$j'$}};
		\coordinate (g) at (-0.2, -0.3);
		\draw[{postaction=decorate}] (bend) -- node[right] {\scriptsize{$j$}} (g);
		\coordinate (alpha) at (-0.4, -0.6);
		\draw[{postaction=decorate}] (g) -- node[right] {\scriptsize{$\alpha$}} (alpha);
		\coordinate (sg) at (-0.4, -1.05);
		\coordinate (gbigon) at (-0.815385, -0.773077);
		\draw (alpha) -- (sg);
		\coordinate (gsstart) at (-0.8, -1.25);
		\draw[{postaction=decorate}] (sg) -- node[above left] {\scriptsize{$s$}} (gsstart);
		\draw[{postaction=decorate}] (gsstart) -- (gs);
		\draw (alpha) -- (gbigon);
		\draw[{postaction=decorate}] (gbigon) -- node[below right] {\scriptsize{$j$}} (gsstart);
		\coordinate (sh) at (0.25,-1.05);
		\draw[{postaction=decorate}] (sh) -- node[above right] {\scriptsize{$s$}} (g);
		\coordinate (h) at (0.571429,-0.857143);
		\draw[{postaction=decorate}] (h) -- node[below left] {\scriptsize{$k$}} (bend);
		\coordinate (hc) at (0.571429, -1.15714);
		\draw (hc) -- (h);
		\draw (hc) -- (sh);
		\coordinate (hs) at (0.8,-1.5);
		\node (gstext) at (0.8,-1.7) {\scriptsize{$k'$}};
		\draw[{postaction=decorate}] (hs) -- (hc);
		\end{tikzpicture}\\
		&=\sum_\alpha \left(F^{s^*sj}_{j}\right)_{0\alpha} \delta_{j',\alpha} \sqrt{\frac{d_{j} d_{s}}{d_{j'}}}
		\begin{tikzpicture}[baseline=(current bounding box.center), decoration={markings,mark=at position .5 with {\arrow[>=stealth]{>}}},rotate=180]
		\node (b) at (0,0.5) {\scriptsize{$e$}};
		\coordinate (bend) at (0,0);
		\draw[{postaction=decorate}] (b)  -- (bend);
		\coordinate (gs) at (-1,-1.5); 
		\node (gstext) at (-1,-1.7) {\scriptsize{$j'$}};
		\coordinate (g) at (-0.2, -0.3);
		\draw[{postaction=decorate}] (bend) -- node[below right] {\scriptsize{$j$}} (g);
		\draw[{postaction=decorate}] (g) -- (gs);
		\coordinate (sh) at (0.25,-1.05);
		\draw[{postaction=decorate}] (sh) -- node[above right] {\scriptsize{$s$}} (g);
		\coordinate (h) at (0.571429,-0.857143);
		\draw[{postaction=decorate}] (h) -- node[below left] {\scriptsize{$k$}} (bend);
		\coordinate (hc) at (0.571429, -1.15714);
		\draw (hc) -- (h);
		\draw (hc) -- (sh);
		\coordinate (hs) at (0.8,-1.5);
		\node (gstext) at (0.8,-1.7) {\scriptsize{$k'$}};
		\draw[{postaction=decorate}] (hs) -- (hc);
		\end{tikzpicture}\\
		&=\sum_\beta \sqrt{\frac{d_{j} d_{s}}{d_{j'}}} \left(F^{s^*sj}_{j}\right)_{0j'} \left(F_{e}^{k^*s^*j'}\right)_{j\beta^*}^\dagger
		\begin{tikzpicture}[baseline=(current bounding box.center), decoration={markings,mark=at position .5 with {\arrow[>=stealth]{>}}},rotate=180]
		\node (b) at (0,0.5) {\scriptsize{$e$}};
		\coordinate (bend) at (0,0);
		\draw[{postaction=decorate}] (b)  -- (bend);
		\coordinate (gs) at (-1,-1.5); 
		\draw[{postaction=decorate}] (bend) -- (gs);
		\node (gstext) at (-1,-1.7) {\scriptsize{$j'$}};	
		\coordinate (hs) at (1,-1.5);
		\node (gstext) at (1,-1.7) {\scriptsize{$k'$}};
		%
		\coordinate (beta) at (0.4, -0.6);
		\draw[{postaction=decorate}] (beta) -- node[below left] {\scriptsize{$\beta$}} (bend);
		\coordinate (sh) at (0.4, -1.05);
		\coordinate (hbigon) at (0.815385, -0.773077);
		\draw (beta) -- (sh);
		\coordinate (hsend) at (0.8, -1.25);
		\draw[{postaction=decorate}] (hsend) -- node[above right] {\scriptsize{$s$}} (sh);
		\draw[{postaction=decorate}] (hs) -- (hsend);
		\draw (beta) -- (hbigon);
		\draw[{postaction=decorate}] (hsend) -- node[below left] {\scriptsize{$j$}} (hbigon);
		\end{tikzpicture}\\
		&=\sum_\beta \sqrt{\frac{d_j d_s}{d_{j'}}} \left(F_j^{s^*sj}\right)_{0j'} \overline{\left(F_e^{k^*s^*j'}\right)}_{\beta^*j} \delta_{\beta^*,k'^*} \sqrt{\frac{d_{k^*} d_{s^*}}{d_{k'^*}}}
		\begin{tikzpicture}[baseline=(current bounding box.center), decoration={markings,mark=at position .5 with {\arrow[>=stealth]{>}}},rotate=180]
		\node (btext) at (0,0.5) {\scriptsize{$e$}};
		\coordinate (b) at (0,0.3);
		\coordinate (bend) at (0,0);
		\draw[{postaction=decorate}] (b)  -- (bend);
		\node(gstext) at (-0.25,-0.5) {\scriptsize{$j'$}};
		\coordinate (gs) at (-0.3,-0.3);
		\draw[{postaction=decorate}] (bend)  -- (gs);
		\node (hstext) at (0.25,-0.5) {\scriptsize{$k'$}};
		\coordinate (hs) at (0.3,-0.3);
		\draw[{postaction=decorate}] (hs)  -- (bend);
		\end{tikzpicture}\\
		&=\sqrt{\frac{d_j d_s}{d_{j'}}} \sqrt{\frac{d_{k^*} d_{s^*}}{d_{k'^*}}} \left(F_j^{s^*sj}\right)_{0j'} \overline{\left(F_e^{k^*s^*j'}\right)}_{k'^*j}
		\begin{tikzpicture}[baseline=(current bounding box.center), decoration={markings,mark=at position .5 with {\arrow[>=stealth]{>}}},rotate=180]
		\node (btext) at (0,0.5) {\scriptsize{$e$}};
		\coordinate (b) at (0,0.3);
		\coordinate (bend) at (0,0);
		\draw[{postaction=decorate}] (b)  -- (bend);
		\node(gstext) at (-0.25,-0.5) {\scriptsize{$j'$}};
		\coordinate (gs) at (-0.3,-0.3);
		\draw[{postaction=decorate}] (bend)  -- (gs);
		\node (hstext) at (0.25,-0.5) {\scriptsize{$k'$}};
		\coordinate (hs) at (0.3,-0.3);
		\draw[{postaction=decorate}] (hs)  -- (bend);
		\end{tikzpicture}
	\end{align*}
	\item[\itemfive] Here, we apply a $G$-move to the diagram:
	\begin{align*}
		\begin{tikzpicture}[rotate=270, baseline=(current bounding box.center), decoration={markings,mark=at position .5 with {\arrow[>=stealth]{<}}}, scale=2]
		\node (c) at (0.115471,0.799999) {\scriptsize{$f$}};
		\coordinate (cend) at (0,1);
		\draw[{postaction=decorate}] (cend) -- (c);
		\coordinate (shdown) at (0.0765468, 1.47602);
		\coordinate (shup) at (0.114118, 1.54001);
		\draw[{postaction=decorate}] (shup) -- node[below right] {\scriptsize{$k'$}} (shdown);
		\coordinate (shcdown) at (-0.126529,1.42602);
		\draw[{postaction=decorate}] (shdown) -- (shcdown);
		\coordinate (hsdown) at (0.1516597983468383,1.26268);
		\draw[{postaction=decorate}] (shcdown) -- node[above] {\scriptsize{$s$}} (-0.2,1.3);
		\coordinate (ch) at (0.0288674,1.05);
		\draw[{postaction=decorate}] (ch) -- (cend);
		\draw[{postaction=decorate}] (shdown) -- node[below] {\scriptsize{$k$}} (ch);
		\coordinate (isup) at (-0.55,1.15);
		\coordinate (isdown) at (-0.45,1.15);
		\draw[{postaction=decorate}] (isdown) -- node[above] {\scriptsize{$l'$}} (isup);
		\coordinate (sicup) at (-0.3,1);
		\coordinate (icup) at (-0.3,1.3);
		\draw[{postaction=decorate}] (sicup) -- (isdown);
		\draw[{postaction=decorate}] (icup) -- (isdown);
		\draw[{postaction=decorate}] (0,1) -- node[left] {\scriptsize{$l$}} (sicup);
		\draw (icup) -- (-0.2,1.3);
		\end{tikzpicture}
		&=
		\begin{tikzpicture}[decoration={markings,mark=at position .5 with {\arrow[>=stealth]{>}}},baseline=(current bounding box.center),rotate=180]
		\node (is) at (0,0) {\scriptsize{$l'$}};
		\coordinate (isend) at (0,0.5);
		\draw[{postaction=decorate}] (is) -- (isend);
		\coordinate (s) at (-0.5,0.8);
		\coordinate (i) at (0.5,0.8);
		\draw (isend) -- (s);
		\draw (isend) -- (i);
		\coordinate (ih) at (0.5,1);
		\coordinate (sh) at (-0.5,1.6);
		\draw[{postaction=decorate}] (s) -- node[right] {\scriptsize{$s$}}(sh);
		\draw[{postaction=decorate}] (i) -- node[left] {\scriptsize{$l$}} (ih);
		\draw[{postaction=decorate}] (ih) -- node[below] {\scriptsize{$k$}} (sh);
		\node (hs) at (-0.5,2.1) {\scriptsize{$k'$}};
		\node (c) at (0.5,2.1) {\scriptsize{$f$}};
		\draw[{postaction=decorate}] (sh) -- (hs);
		\draw[{postaction=decorate}] (c) -- (ih);
		\end{tikzpicture}\\
		&=\sum_\alpha \left(G_{fk'^*}^{l^*s^*}\right)_{k^*\alpha^*}
		\begin{tikzpicture}[decoration={markings,mark=at position .5 with {\arrow[>=stealth]{>}}},baseline=(current bounding box.center),rotate=180]
		\node (is) at (0,0) {\scriptsize{$l'$}};
		\coordinate (isend) at (0,0.5);
		\draw[{postaction=decorate}] (is) -- (isend);
		\coordinate (s) at (-0.3,0.8);
		\coordinate (i) at (0.3,0.8);
		\draw[{postaction=decorate}] (isend) -- node[right] {\scriptsize{$s$}} (s);
		\draw[{postaction=decorate}] (isend) -- node[left] {\scriptsize{$l$}} (i);
		\coordinate (adown) at (0,1.1);
		\draw (s) -- (adown);
		\draw (i) -- (adown);
		\coordinate (aup) at (0,1.5);
		\draw[{postaction=decorate}] (adown) -- node[right] {\scriptsize{$\alpha$}} (aup);
		\node (hs) at (-0.5,2.1) {\scriptsize{$k'$}};
		\node (c) at (0.5,2.1) {\scriptsize{$f$}};
		\draw[{postaction=decorate}] (aup) -- (hs);
		\draw[{postaction=decorate}] (c) -- (aup);
		\end{tikzpicture}\\
		&=\sum_\alpha \left(G_{fk'^*}^{l^*s^*}\right)_{k^*\alpha^*} \sqrt{\frac{d_{s^*} d_{l^*}}{d_{l'^*}}} \delta_{\alpha^*,l'^*}\!
		\begin{tikzpicture}[baseline=(current bounding box.center), decoration={markings,mark=at position .5 with {\arrow[>=stealth]{>}}}]
		\node (btext) at (0,0.5) {\scriptsize{$l'$}};
		\coordinate (b) at (0,0.3);
		\coordinate (bend) at (0,0);
		\draw[{postaction=decorate}] (b)  -- (bend);
		\node(gstext) at (-0.3,-0.5) {\scriptsize{$f$}};
		\coordinate (gs) at (-0.3,-0.3);
		\draw[{postaction=decorate}] (gs)  -- (bend);
		\node (hstext) at (0.3,-0.5) {\scriptsize{$k'$}};
		\coordinate (hs) at (0.3,-0.3);
		\draw[{postaction=decorate}] (bend)  -- (hs);
		\end{tikzpicture}\\
		&=\sqrt{\frac{d_{s^*} d_{l^*}}{d_{l'^*}}} \left(G_{fk'^*}^{l^*s^*}\right)_{k^*l'^*}
		\begin{tikzpicture}[baseline=(current bounding box.center), decoration={markings,mark=at position .5 with {\arrow[>=stealth]{>}}}]
		\node (btext) at (0,0.5) {\scriptsize{$l'$}};
		\coordinate (b) at (0,0.3);
		\coordinate (bend) at (0,0);
		\draw[{postaction=decorate}] (b)  -- (bend);
		\node(gstext) at (-0.3,-0.5) {\scriptsize{$f$}};
		\coordinate (gs) at (-0.3,-0.3);
		\draw[{postaction=decorate}] (gs)  -- (bend);
		\node (hstext) at (0.3,-0.5) {\scriptsize{$k'$}};
		\coordinate (hs) at (0.3,-0.3);
		\draw[{postaction=decorate}] (bend)  -- (hs);
		\end{tikzpicture}\\
		&\phantom{=}\\
		&=\sqrt{\frac{d_{s^*} d_{k^*}}{d_{k'^*}}} \overline{\left(F_{l'^*}^{fk^*s^*}\right)}_{l^*k'^*}
		\begin{tikzpicture}[baseline=(current bounding box.center), decoration={markings,mark=at position .5 with {\arrow[>=stealth]{>}}}]
		\node (btext) at (0,0.5) {\scriptsize{$l'$}};
		\coordinate (b) at (0,0.3);
		\coordinate (bend) at (0,0);
		\draw[{postaction=decorate}] (b)  -- (bend);
		\node(gstext) at (-0.3,-0.5) {\scriptsize{$f$}};
		\coordinate (gs) at (-0.3,-0.3);
		\draw[{postaction=decorate}] (gs)  -- (bend);
		\node (hstext) at (0.3,-0.5) {\scriptsize{$k'$}};
		\coordinate (hs) at (0.3,-0.3);
		\draw[{postaction=decorate}] (bend)  -- (hs);
		\end{tikzpicture}
	\end{align*}
	\item[\itemsix] Analogously to the third diagram we use a $H$-move to simplify this one:
	\begin{align*}
		\begin{tikzpicture}[rotate=270, baseline=(current bounding box.center), decoration={markings,mark=at position .5 with {\arrow[>=stealth]{<}}}, scale=2]
		\node (d) at (-1.13453, 0.766988) {\scriptsize{$a$}};
		\coordinate (dend) at (-1,1);
		\draw[{postaction=decorate}] (dend) -- (d);
		\coordinate (sjdown) at (-1.11412, 1.54001);
		\coordinate (sjup) at (-1.07655, 1.47602);
		\draw[{postaction=decorate}] (sjup) -- node[above right] {\scriptsize{$g'$}} (sjdown);
		\coordinate (jcup) at (-0.873471, 1.42602);
		\draw[{postaction=decorate}] (jcup) -- (sjup);
		\coordinate (jscup) at (-1.15166, 1.26268);
		\coordinate (dj) at (-1.02887,1.05);
		\draw[{postaction=decorate}] (dend) -- (dj);
		\draw[{postaction=decorate}] (dj) -- node[above] {\scriptsize{$g$}} (sjup);
		\draw[{postaction=decorate}] (-0.8,1.3) -- node[below] {\scriptsize{$s$}} (jcup);
		\coordinate (isup) at (-0.55,1.15);
		\coordinate (isdown) at (-0.45,1.15);
		\draw[{postaction=decorate}] (isdown) -- node[below] {\scriptsize{$l'$}} (isup);
		\coordinate (sicdown) at (-0.7,1);
		\coordinate (icdown) at (-0.7,1.3);
		\draw[{postaction=decorate}] (isup) -- (sicdown);
		\draw[{postaction=decorate}] (isup) -- (icdown);
		\draw[{postaction=decorate}] (sicdown) -- node[left] {\scriptsize{$l$}} (-1,1);	
		\draw (icdown) -- (-0.8,1.3);
		\end{tikzpicture}
		&=
		\begin{tikzpicture}[decoration={markings,mark=at position .5 with {\arrow[>=stealth]{>}}},baseline=(current bounding box.center)]
		\node (is) at (0,0) {\scriptsize{$l'$}};
		\coordinate (isend) at (0,0.5);
		\draw[{postaction=decorate}] (isend) -- (is);
		\coordinate (i) at (-0.5,0.8);
		\coordinate (s) at (0.5,0.8);
		\draw (i) -- (isend);
		\draw (s) -- (isend);
		\coordinate (sj) at (0.5,1.6);
		\coordinate (ij) at (-0.5,1);
		\draw[{postaction=decorate}] (ij) -- node[left] {\scriptsize{$l$}}(i);
		\draw[{postaction=decorate}] (sj) -- node[right] {\scriptsize{$s$}} (s);
		\draw[{postaction=decorate}] (sj) -- node[above] {\scriptsize{$g$}} (ij);
		\node (d) at (-0.5,2.1) {\scriptsize{$a$}};
		\node (js) at (0.5,2.1) {\scriptsize{$g'$}};
		\draw[{postaction=decorate}] (d) -- (ij);
		\draw[{postaction=decorate}] (js) -- (sj);
		\end{tikzpicture}\\
		&=\sum_\alpha \left(H_{l^*s^*}^{a^*g'^*}\right)_{g^*\alpha^*}
		\begin{tikzpicture}[decoration={markings,mark=at position .5 with {\arrow[>=stealth]{>}}},baseline=(current bounding box.center)]
		\node (is) at (0,0) {\scriptsize{$l'$}};
		\coordinate (isend) at (0,0.5);
		\draw[{postaction=decorate}] (isend) -- (is);
		\coordinate (s) at (-0.3,0.8);
		\coordinate (i) at (0.3,0.8);
		\draw[{postaction=decorate}] (s) -- node[left] {\scriptsize{$l$}} (isend);
		\draw[{postaction=decorate}] (i) -- node[right] {\scriptsize{$s$}} (isend);
		\coordinate (adown) at (0,1.1);
		\draw (s) -- (adown);
		\draw (i) -- (adown);
		\coordinate (aup) at (0,1.5);
		\draw[{postaction=decorate}] (aup) -- node[right] {\scriptsize{$\alpha$}} (adown);
		\node (hs) at (-0.5,2.1) {\scriptsize{$a$}};
		\node (c) at (0.5,2.1) {\scriptsize{$g'$}};
		\draw[{postaction=decorate}] (hs) -- (aup);
		\draw[{postaction=decorate}] (c) -- (aup);
		\end{tikzpicture}\\
		&=\sum_\alpha \left(H_{l^*s^*}^{a^*g'^*}\right)_{g^*\alpha^*} \sqrt{\frac{d_{s^*} d_{l^*}}{d_{l'^*}}} \delta_{\alpha^*,l'^*}
		\begin{tikzpicture}[baseline=(current bounding box.center), decoration={markings,mark=at position .5 with {\arrow[>=stealth]{>}}},rotate=180]
		\node (btext) at (0,0.5) {\scriptsize{$l'$}};
		\coordinate (b) at (0,0.3);
		\coordinate (bend) at (0,0);
		\draw[{postaction=decorate}] (bend)  -- (b);
		\node(gstext) at (-0.3,-0.5) {\scriptsize{$g'$}};
		\coordinate (gs) at (-0.3,-0.3);
		\draw[{postaction=decorate}] (gs)  -- (bend);
		\node (hstext) at (0.3,-0.5) {\scriptsize{$a$}};
		\coordinate (hs) at (0.3,-0.3);
		\draw[{postaction=decorate}] (hs)  -- (bend);
		\end{tikzpicture}\\
		&=\sqrt{\frac{d_{s^*} d_{l^*}}{d_{l'^*}}} \left(H_{l^*s^*}^{a^*g'^*}\right)_{g^* l'^*} 
		\begin{tikzpicture}[baseline=(current bounding box.center), decoration={markings,mark=at position .5 with {\arrow[>=stealth]{>}}},rotate=180]
		\node (btext) at (0,0.5) {\scriptsize{$l'$}};
		\coordinate (b) at (0,0.3);
		\coordinate (bend) at (0,0);
		\draw[{postaction=decorate}] (bend)  -- (b);
		\node(gstext) at (-0.3,-0.5) {\scriptsize{$g'$}};
		\coordinate (gs) at (-0.3,-0.3);
		\draw[{postaction=decorate}] (gs)  -- (bend);
		\node (hstext) at (0.3,-0.5) {\scriptsize{$a$}};
		\coordinate (hs) at (0.3,-0.3);
		\draw[{postaction=decorate}] (hs)  -- (bend);
		\end{tikzpicture}\\
		&=\sqrt{\frac{d_{s^*} d_{g^*}}{d_{g'^*}}} \left(F_{l'^*}^{a^*g^*s^*}\right)_{l^*g'^*}
		\begin{tikzpicture}[baseline=(current bounding box.center), decoration={markings,mark=at position .5 with {\arrow[>=stealth]{>}}},rotate=180]
		\node (btext) at (0,0.5) {\scriptsize{$l'$}};
		\coordinate (b) at (0,0.3);
		\coordinate (bend) at (0,0);
		\draw[{postaction=decorate}] (bend)  -- (b);
		\node(gstext) at (-0.3,-0.5) {\scriptsize{$g'$}};
		\coordinate (gs) at (-0.3,-0.3);
		\draw[{postaction=decorate}] (gs)  -- (bend);
		\node (hstext) at (0.3,-0.5) {\scriptsize{$a$}};
		\coordinate (hs) at (0.3,-0.3);
		\draw[{postaction=decorate}] (hs)  -- (bend);
		\end{tikzpicture}
	\end{align*}
\end{itemize}
After we have simplified all six corners of the diagram, we can now insert the results into \eqref{eq:bigdiagram} to determine how the diagram with the inserted loop can be expressed as a linear combination of diagrams without a loop:
\begin{widetext}
	\begin{align}
		\begin{tikzpicture} [rotate=90, scale=1,baseline=(current bounding box.center), decoration={markings,mark=at position .5 with {\arrow[>=stealth]{<}}}, scale=1]
		\draw[{postaction=decorate}] (0,2.73205) -- node[above] {\scriptsize{$g$}} (0.5,1.86603); 
		\draw[{postaction=decorate}] (-1,2.73205) -- node[left] {\scriptsize{$l$}} (0,2.73205); 
		\draw[{postaction=decorate}] (-1.5,1.86603) -- node[below] {\scriptsize{$k$}} (-1,2.73205); 
		\draw[{postaction=decorate}] (-1,1) -- node[below] {\scriptsize{$j$}} (-1.5,1.86603); 
		\draw[{postaction=decorate}] (0,1) -- node[right] {\scriptsize{$i$}} (-1,1); 
		\draw[{postaction=decorate}] (0.5,1.86603) -- node[above] {\scriptsize{$h$}} (0,1); 
		\node (e1) at (0.25,3.16506) {\scriptsize{$a$}};
		\draw[{postaction=decorate}] (0,2.73205) -- (e1); 
		\node (e2) at (-1.25,3.16506) {\scriptsize{$f$}};
		\draw[{postaction=decorate}] (-1,2.73205) -- (e2); 
		\node (e3) at (-2,1.86603) {\scriptsize{$e$}};
		\draw[{postaction=decorate}] (-1.5,1.86603) -- (e3); 
		\node (e4) at (-1.25,0.566987) {\scriptsize{$d$}};
		\draw[{postaction=decorate}] (-1,1) -- (e4); 
		\node (e5) at (0.25,0.566987) {\scriptsize{$c$}};
		\draw[{postaction=decorate}] (0,1) -- (e5); 
		\node (e6) at (1,1.86603) {\scriptsize{$b$}};
		\draw[{postaction=decorate}]  (0.5,1.86603) -- (e6); 
		\draw[{postaction=decorate}] (-0.8,2.43205) -- node[right] {\scriptsize{$s$}} (-0.2,2.43205); 
		\draw[{postaction=decorate}] (-0.2,1.3) -- node[left] {\scriptsize{$s$}} (-0.8,1.3); 
		\draw[{postaction=decorate}] (0.13,1.86603) -- (-0.2,1.3); 
		\draw[{postaction=decorate}] (-0.2,2.43205) -- (0.13,1.86603); 
		\draw[{postaction=decorate}] (-1.13,1.86603) -- (-0.8,2.43205); 
		\draw[{postaction=decorate}] (-0.8,1.3) -- (-1.13,1.86603); 
		\draw[dotted] (0.13,1.86603) -- (0.30492,2.16603);
		\draw[dotted] (-1.13,1.86603) -- (-1.30492,1.56603);
		\end{tikzpicture}
		&=
		\begin{alignedat}[t]{2}
			\sum_{\substack{g',h',i',\\j',k',l'}}&\sqrt{d_s d_{s^*}} \sqrt{\frac{d_{g^*} d_h d_{i'} d_j d_{k^*} d_{l'^*}}{d_{g'^*} d_{h'} d_i d_{j'} d_{k'^*} d_{l'^*}}} \left(F_{g^*}^{g^*s^*s}\right)_{g'^*0} \overline{\left(F_{b^*}^{g'^*sh}\right)}_{g^*h'}\overline{\left(F_{i'}^{shc^*}\right)}_{h'i} \left(F_{i'}^{sjd}\right)_{j'i}\\&\left(F_{j}^{s^*sj}\right)_{0j'} \overline{\left(F_e^{k^*s^*j'}\right)}_{k'^*j} \left(F_{l'^*}^{a^*g^*s^*}\right)_{l^*g'^*} \overline{\left(F_{l'^*}^{fk^*s^*}\right)}_{l^*k'^*}
			\begin{tikzpicture} [rotate=90, scale=1,baseline=(current bounding box.center), decoration={markings,mark=at position .5 with {\arrow[>=stealth]{<}}}]
			\draw[{postaction=decorate}] (0,2.73205) -- node[above] {\scriptsize{$g'$}} (0.5,1.86603); 
			\draw[{postaction=decorate}] (-1,2.73205) -- node[left] {\scriptsize{$l'$}} (0,2.73205); 
			\draw[{postaction=decorate}] (-1.5,1.86603) -- node[below] {\scriptsize{$k'$}} (-1,2.73205); 
			\draw[{postaction=decorate}] (-1,1) -- node[below] {\scriptsize{$j'$}} (-1.5,1.86603); 
			\draw[{postaction=decorate}] (0,1) -- node[right] {\scriptsize{$i'$}} (-1,1); 
			\draw[{postaction=decorate}] (0.5,1.86603) -- node[above] {\scriptsize{$h'$}} (0,1); 
			\node (e1) at (0.25,3.16506) {\scriptsize{$a$}};
			\draw[{postaction=decorate}] (0,2.73205) -- (e1); 
			\node (e2) at (-1.25,3.16506) {\scriptsize{$f$}};
			\draw[{postaction=decorate}] (-1,2.73205) -- (e2); 
			\node (e3) at (-2,1.86603) {\scriptsize{$e$}};
			\draw[{postaction=decorate}] (-1.5,1.86603) -- (e3); 
			\node (e4) at (-1.25,0.566987) {\scriptsize{$d$}};
			\draw[{postaction=decorate}] (-1,1) -- (e4); 
			\node (e5) at (0.25,0.566987) {\scriptsize{$c$}};
			\draw[{postaction=decorate}] (0,1) -- (e5); 
			\node (e6) at (1,1.86603) {\scriptsize{$b$}};
			\draw[{postaction=decorate}]  (0.5,1.86603) -- (e6); 
			\end{tikzpicture}.
		\end{alignedat}
		\\
		\intertext{Combined with Eq.~\eqref{BpsDef} we can conclude}
		B_{\mathbf{p},ghijkl}^{s,g'h'i'j'k'l'}(abcdef)
		&=
		\begin{alignedat}[t]{2}
			&
			\sqrt{d_s d_{s^*}} \sqrt{\frac{d_{g^*} d_h d_{i'} d_j d_{k^*} d_{l'^*}}{d_{g'^*} d_{h'} d_i d_{j'} d_{k'^*} d_{l'^*}}} \left(F_{g^*}^{g^*s^*s}\right)_{g'^*0} \overline{\left(F_{b^*}^{g'^*sh}\right)}_{g^*h'}\overline{\left(F_{i'}^{shc^*}\right)}_{h'i}\\& \left(F_{i'}^{sjd}\right)_{j'i} \left(F_{j}^{s^*sj}\right)_{0j'} \overline{\left(F_e^{k^*s^*j'}\right)}_{k'^*j} \left(F_{l'^*}^{a^*g^*s^*}\right)_{l^*g'^*} \overline{\left(F_{l'^*}^{fk^*s^*}\right)}_{l^*k'^*}.
		\end{alignedat}\label{Bps}
	\end{align}
\end{widetext}
\noindent
Just as in the original Levin-Wen model we make the choice
\begin{equation}
	a_s=\frac{d_s}{D^2},
\end{equation}
\noindent
where 
\begin{equation}
	D=\sqrt{\sum_{i=0}^N d_i^2}
\end{equation}
is the total quantum dimension.

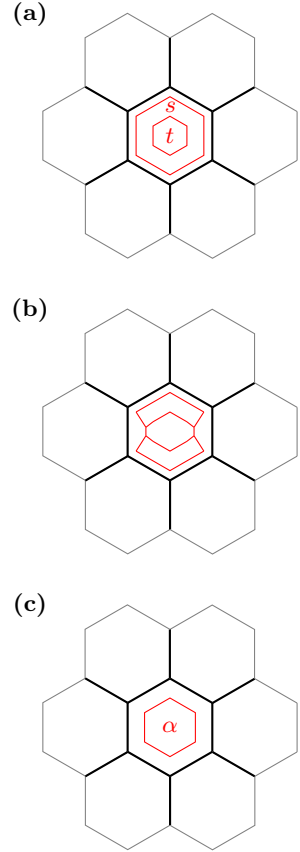
\begin{figure}[t]
	\centering
	\begin{tikzpicture}[scale=0.65]
	\clip(-3,-1.5) rectangle (4.5,4.5);
	\node at (-2,4) {\textbf{(a)}};
	\begin{scope}
	\rotatebox{30}{
		\draw[gray] (0:1cm) -- (60:1cm) -- (120:1cm) -- (180:1cm) -- (240:1cm) -- (300:1cm) -- (0:1cm);
		\draw[thick] (300:1cm) -- (0:1cm);
		\draw[thick] (60:1) -- (120:1cm);
		\begin{scope}[xshift=0,yshift=1.74cm]
		\draw[gray] (300:1cm) -- (0:1cm) -- (60:1cm) -- (120:1cm) -- (180:1cm) -- (240:1cm);
		\draw[thick] (0:1cm) -- (60:1cm);
		\end{scope}
		\begin{scope}[xshift=1.5cm,yshift=0.87cm]
		\draw[red] (0:0.8) -- (60:0.8) -- (120:0.8) -- (180:0.8) -- (240:0.8) -- (300:0.8) -- cycle;
		\draw[red] (0:0.4) -- (60:0.4) -- (120:0.4) -- (180:0.4) -- (240:0.4) -- (300:0.4) -- cycle;
		\node[red] at (0.3,0.5) {\rotatebox{-30}{$s$}};
		\node[red] at (0,0) {\rotatebox{-30}{$t$}};
		\draw[thick] (60:1) -- (120:1);
		\draw[thick] (120:1) -- (180:1);
		\draw[thick] (180:1) -- (240:1);
		\draw[thick] (240:1) -- (300:1);
		\draw[thick] (0:1) -- (60:1);
		\draw[thick] (300:1) -- (0:1);
		\end{scope}
		\begin{scope}[xshift=1.5cm,yshift=-0.87cm]
		\draw[gray] (180:1cm) -- (240:1cm) -- (300:1cm) -- (0:1cm) -- (60:1cm);
		\end{scope}
		\begin{scope}[xshift=3cm,yshift=0cm]
		\draw[thick] (60:1) -- (120:1);
		\draw[thick] (180:1) -- (240:1);
		\draw[gray] (240:1) -- (300:1);
		\draw[gray] (300:1) -- (0:1);
		\draw[gray] (0:1) -- (60:1);
		\end{scope}
		\begin{scope}[xshift=3cm,yshift=1.74cm]
		\draw[gray] (300:1cm) -- (0:1cm) -- (60:1cm) -- (120:1cm) -- (180:1cm);
		\draw[thick] (120:1cm) -- (180:1cm);
		\end{scope}
		\begin{scope}[xshift=1.5cm,yshift=2.61cm]
		\draw[gray] (0:1cm) -- (60:1cm) -- (120:1cm) -- (180:1cm);
		\end{scope}
	}
	\end{scope}
	\end{tikzpicture}
	\begin{tikzpicture}[scale=0.65]
	\clip(-3,-1.5) rectangle (4.5,4.5);
	\node at (-2,4) {\textbf{(b)}};
	\begin{scope}
	\rotatebox{30}{
		\draw[gray] (0:1cm) -- (60:1cm) -- (120:1cm) -- (180:1cm) -- (240:1cm) -- (300:1cm) -- (0:1cm);
		\draw[thick] (300:1cm) -- (0:1cm);
		\draw[thick] (60:1) -- (120:1cm);
		\begin{scope}[xshift=0,yshift=1.74cm]
		\draw[gray] (300:1cm) -- (0:1cm) -- (60:1cm) -- (120:1cm) -- (180:1cm) -- (240:1cm);
		\draw[thick] (0:1cm) -- (60:1cm);
		\end{scope}
		\begin{scope}[xshift=1.5cm,yshift=0.87cm]
		\draw[red] (0:0.8) -- (60:0.8) -- (120:0.8);
		\draw[red] (180:0.8) -- (240:0.8) -- (300:0.8);
		\draw[red] (0:0.4) -- (60:0.4) -- (120:0.4);
		\draw[red] (180:0.4) -- (240:0.4) -- (300:0.4);
		\draw[red] (120:0.8) -- (140:0.5);
		\draw[red] (120:0.4) -- (140:0.5);
		\draw[red] (180:0.8) -- (160:0.5);
		\draw[red] (180:0.4) -- (160:0.5);
		\draw[red] (140:0.5) -- (160:0.5);
		\draw[red] (300:0.8) -- (320:0.5);
		\draw[red] (300:0.4) -- (320:0.5);
		\draw[red] (0:0.8) -- (340:0.5);
		\draw[red] (0:0.4) -- (340:0.5);
		\draw[red] (320:0.5) -- (340:0.5);
		\draw[thick] (60:1) -- (120:1);
		\draw[thick] (120:1) -- (180:1);
		\draw[thick] (180:1) -- (240:1);
		\draw[thick] (240:1) -- (300:1);
		\draw[thick] (0:1) -- (60:1);
		\draw[thick] (300:1) -- (0:1);
		\end{scope}
		\begin{scope}[xshift=1.5cm,yshift=-0.87cm]
		\draw[gray] (180:1cm) -- (240:1cm) -- (300:1cm) -- (0:1cm) -- (60:1cm);
		\end{scope}
		\begin{scope}[xshift=3cm,yshift=0cm]
		\draw[thick] (60:1) -- (120:1);
		\draw[thick] (180:1) -- (240:1);
		\draw[gray] (240:1) -- (300:1);
		\draw[gray] (300:1) -- (0:1);
		\draw[gray] (0:1) -- (60:1);
		\end{scope}
		\begin{scope}[xshift=3cm,yshift=1.74cm]
		\draw[gray] (300:1cm) -- (0:1cm) -- (60:1cm) -- (120:1cm) -- (180:1cm);
		\draw[thick] (120:1cm) -- (180:1cm);
		\end{scope}
		\begin{scope}[xshift=1.5cm,yshift=2.61cm]
		\draw[gray] (0:1cm) -- (60:1cm) -- (120:1cm) -- (180:1cm);
		\end{scope}
	}
	\end{scope}
	\end{tikzpicture}
	\begin{tikzpicture}[scale=0.65]
	\clip(-3,-1.5) rectangle (4.5,4.5);
	\node at (-2,4) {\textbf{(c)}};
	\begin{scope}
	\rotatebox{30}{
		\draw[gray] (0:1cm) -- (60:1cm) -- (120:1cm) -- (180:1cm) -- (240:1cm) -- (300:1cm) -- (0:1cm);
		\draw[thick] (300:1cm) -- (0:1cm);
		\draw[thick] (60:1) -- (120:1cm);
		\begin{scope}[xshift=0,yshift=1.74cm]
		\draw[gray] (300:1cm) -- (0:1cm) -- (60:1cm) -- (120:1cm) -- (180:1cm) -- (240:1cm);
		\draw[thick] (0:1cm) -- (60:1cm);
		\end{scope}
		\begin{scope}[xshift=1.5cm,yshift=0.87cm]
		\draw[red] (0:0.6) -- (60:0.6) -- (120:0.6) -- (180:0.6) -- (240:0.6) -- (300:0.6) -- cycle;
		\node[red] at (0,0) {\rotatebox{-30}{$\alpha$}};
		\draw[thick] (60:1) -- (120:1);
		\draw[thick] (120:1) -- (180:1);
		\draw[thick] (180:1) -- (240:1);
		\draw[thick] (240:1) -- (300:1);
		\draw[thick] (0:1) -- (60:1);
		\draw[thick] (300:1) -- (0:1);
		\end{scope}
		\begin{scope}[xshift=1.5cm,yshift=-0.87cm]
		\draw[gray] (180:1cm) -- (240:1cm) -- (300:1cm) -- (0:1cm) -- (60:1cm);
		\end{scope}
		\begin{scope}[xshift=3cm,yshift=0cm]
		\draw[thick] (60:1) -- (120:1);
		\draw[thick] (180:1) -- (240:1);
		\draw[gray] (240:1) -- (300:1);
		\draw[gray] (300:1) -- (0:1);
		\draw[gray] (0:1) -- (60:1);
		\end{scope}
		\begin{scope}[xshift=3cm,yshift=1.74cm]
		\draw[gray] (300:1cm) -- (0:1cm) -- (60:1cm) -- (120:1cm) -- (180:1cm);
		\draw[thick] (120:1cm) -- (180:1cm);
		\end{scope}
		\begin{scope}[xshift=1.5cm,yshift=2.61cm]
		\draw[gray] (0:1cm) -- (60:1cm) -- (120:1cm) -- (180:1cm);
		\end{scope}
	}
	\end{scope}
	\end{tikzpicture}
	\caption{\label{fig:proj}\textbf{Projector.} \textbf{(a)} We apply the operators $B_\mathbf{p}$ twice to the same plaquette, which insert two loops, one of type $s$ and one of type $t$ (keep in mind that we sum over $s$ and $t$ which is not depicted in the picture). \textbf{(b)} These two loops are then fused together. \textbf{(c)} The result is (a sum over) a single loop of type $\alpha$.}
\end{figure}

\subsection{Properties of the Hamiltonian}

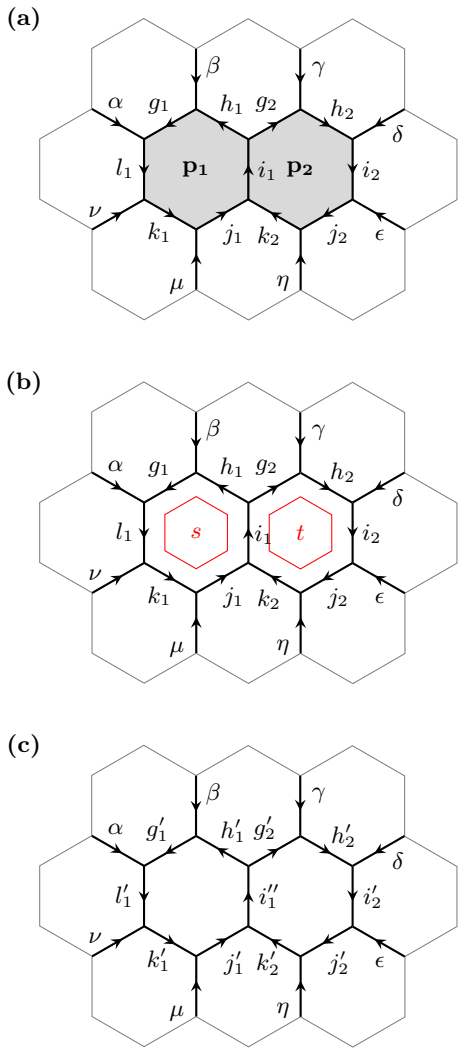
\begin{figure}[t]
	\centering
	\begin{tikzpicture}[scale=0.8]
	\clip(-3,-1.5) rectangle (6.5,4.5);
	\node at (-2,4) {\textbf{(a)}};
	\begin{scope}
	\rotatebox{30}{
		\draw[gray] (0:1cm) -- (60:1cm) -- (120:1cm) -- (180:1cm) -- (240:1cm) -- (300:1cm) -- (0:1cm);
		\draw[thick,->-] (300:1cm) -- (0:1cm);
		\draw[thick,-<-] (60:1cm) -- (120:1cm);
		\begin{scope}[xshift=0,yshift=1.74cm]
		\draw[gray] (300:1cm) -- (0:1cm) -- (60:1cm) -- (120:1cm) -- (180:1cm) -- (240:1cm);
		\draw[thick,-<-] (0:1cm) -- (60:1cm);
		\end{scope}
		\begin{scope}[xshift=1.5cm,yshift=0.87cm]
		\draw[fill=LightGray!70] (0:1cm) -- (60:1cm) -- (120:1cm) -- (180:1cm) -- (240:1cm) -- (300:1cm) -- cycle;
		\node at (0,0) {\rotatebox{-30}{$\mathbf{p_1}$}};
		\draw[->-,thick] (60:1) to node[above] {\rotatebox{-30}{\small$g_1$}} (120:1);
		\draw[->-,thick] (120:1) to node[left,pos=0.3] {\rotatebox{-30}{\small$l_1$}} (180:1);
		\draw[->-,thick] (180:1) to node[left,pos=0.7] {\rotatebox{-30}{\small$k_1$}} (240:1);
		\draw[->-,thick] (240:1) to node[below,pos=0.5] {\rotatebox{-30}{\small$j_1$}} (300:1);
		\draw[->-,thick] (0:1) to node[right,pos=0.7] {\rotatebox{-30}{\small$h_1\ $}} (60:1);
		\end{scope}
		\begin{scope}[xshift=3cm,yshift=-1.74cm]
		\draw[gray] (180:1cm) -- (240:1cm) -- (300:1cm) -- (0:1cm);
		\end{scope}
		\begin{scope}[xshift=1.5cm,yshift=-0.87cm]
		\draw[gray] (180:1cm) -- (240:1cm) -- (300:1cm) -- (0:1cm) -- (60:1cm);
		\draw[thick,->-] (300:1cm) -- (0:1cm);
		\end{scope}
		\begin{scope}[xshift=3cm,yshift=0cm]
		\draw[fill=LightGray!70] (240:1cm) -- (300:1cm) -- (0:1cm) -- (60:1cm) -- (120:1cm) -- (180:1cm) -- cycle;
		\node at (0,0) {\rotatebox{-30}{$\mathbf{p_2}$}};
		\draw[-<-,thick] (60:1) to node[above,pos=0.5] {\rotatebox{-30}{\small$\ g_2$}} (120:1);
		\draw[-<-,thick] (120:1) to node[right,pos=0.7] {\rotatebox{-30}{\small$i_1$}} (180:1);
		\draw[-<-,thick] (180:1) to node[left,pos=0.8] {\rotatebox{-30}{\small$k_2$}} (240:1);
		\draw[-<-,thick] (240:1) to node[below,pos=0.5] {\rotatebox{-30}{\small$j_2$}} (300:1);
		\draw[-<-,thick] (300:1) to node[right,pos=0.3] {\rotatebox{-30}{\small$i_2$}} (0:1);
		\draw[-<-,thick] (0:1) to node[right,pos=0.6] {\rotatebox{-30}{\small$h_2$}} (60:1);
		\end{scope}
		\begin{scope}[xshift=3cm,yshift=1.74cm]
		\draw[gray] (300:1cm) -- (0:1cm) -- (60:1cm) -- (120:1cm) -- (180:1cm);
		\draw[thick,-<-] (300:1cm) -- (0:1cm);
		\draw[thick,->-] (120:1cm) -- (180:1cm);
		\end{scope}
		\begin{scope}[xshift=1.5cm,yshift=2.61cm]
		\draw[gray] (0:1cm) -- (60:1cm) -- (120:1cm) -- (180:1cm);
		\end{scope}
		\begin{scope}[xshift=4.5cm,yshift=0.87cm]
		\draw[gray] (240:1cm) -- (300:1cm) -- (0:1cm) -- (60:1cm) -- (120:1cm);
		\draw[thick,-<-] (240:1cm) -- (300:1cm);
		\end{scope}
		\begin{scope}[xshift=4.5cm,yshift=-0.87cm]
		\draw[gray] (180:1cm) -- (240:1cm) -- (300:1cm) -- (0:1cm) -- (60:1cm);
		\draw[thick,-<-] (180:1cm) -- (240:1cm);
		\end{scope}
		\node at (0.9,2.5) {\rotatebox{-30}{\small$\alpha$}};
		\node at (2.6,2.2) {\rotatebox{-30}{\small$\beta$}};
		\node at (4.1,1.3) {\rotatebox{-30}{\small$\gamma$}};
		\node at (4.7,-0.3) {\rotatebox{-30}{\small$\delta$}};
		\node at (3.6,-1.6) {\rotatebox{-30}{\small$\epsilon$}};
		\node at (1.8,-1.5) {\rotatebox{-30}{\small$\eta$}};
		\node at (0.25,-0.65) {\rotatebox{-30}{\small$\mu$}};
		\node at (-0.3,1.1) {\rotatebox{-30}{\small$\nu$}};
	}
	\end{scope}
	\end{tikzpicture}
	\begin{tikzpicture}[scale=0.8]
	\clip(-3,-1.5) rectangle (6.5,4.5);
	\node at (-2,4) {\textbf{(b)}};
	\begin{scope}
	\rotatebox{30}{
		\draw[gray] (0:1cm) -- (60:1cm) -- (120:1cm) -- (180:1cm) -- (240:1cm) -- (300:1cm) -- (0:1cm);
		\draw[thick,->-] (300:1cm) -- (0:1cm);
		\draw[thick,-<-] (60:1) -- (120:1cm);
		\begin{scope}[xshift=0,yshift=1.74cm]
		\draw[gray] (300:1cm) -- (0:1cm) -- (60:1cm) -- (120:1cm) -- (180:1cm) -- (240:1cm);
		\draw[thick,-<-] (0:1cm) -- (60:1cm);
		\end{scope}
		\begin{scope}[xshift=1.5cm,yshift=0.87cm]
		\draw[red] (0:0.6) -- (60:0.6) -- (120:0.6) -- (180:0.6) -- (240:0.6) -- (300:0.6) -- cycle;
		\node[red] at (0,0) {\rotatebox{-30}{$s$}};
		\draw[->-,thick] (60:1) to node[above] {\rotatebox{-30}{\small$g_1$}} (120:1);
		\draw[->-,thick] (120:1) to node[left,pos=0.3] {\rotatebox{-30}{\small$l_1$}} (180:1);
		\draw[->-,thick] (180:1) to node[left,pos=0.7] {\rotatebox{-30}{\small$k_1$}} (240:1);
		\draw[->-,thick] (240:1) to node[below,pos=0.5] {\rotatebox{-30}{\small$j_1$}} (300:1);
		\draw[->-,thick] (0:1) to node[right,pos=0.7] {\rotatebox{-30}{\small$h_1\ $}} (60:1);
		\end{scope}
		\begin{scope}[xshift=3cm,yshift=-1.74cm]
		\draw[gray] (180:1cm) -- (240:1cm) -- (300:1cm) -- (0:1cm);
		\end{scope}
		\begin{scope}[xshift=1.5cm,yshift=-0.87cm]
		\draw[gray] (180:1cm) -- (240:1cm) -- (300:1cm) -- (0:1cm) -- (60:1cm);
		\draw[thick,->-] (300:1cm) -- (0:1cm);
		\end{scope}
		\begin{scope}[xshift=3cm,yshift=0cm]
		\draw[red] (0:0.6) -- (60:0.6) -- (120:0.6) -- (180:0.6) -- (240:0.6) -- (300:0.6) -- cycle;
		\node[red] at (0,0) {\rotatebox{-30}{$t$}};
		\draw[-<-,thick] (60:1) to node[above,pos=0.5] {\rotatebox{-30}{\small$\ g_2$}} (120:1);
		\draw[-<-,thick] (120:1) to node[right,pos=0.7] {\rotatebox{-30}{\!\small$i_1$}} (180:1);
		\draw[-<-,thick] (180:1) to node[left,pos=0.8] {\rotatebox{-30}{\small$k_2$}} (240:1);
		\draw[-<-,thick] (240:1) to node[below,pos=0.5] {\rotatebox{-30}{\small$j_2$}} (300:1);
		\draw[-<-,thick] (300:1) to node[right,pos=0.3] {\rotatebox{-30}{\small$i_2$}} (0:1);
		\draw[-<-,thick] (0:1) to node[right,pos=0.6] {\rotatebox{-30}{\small$h_2$}} (60:1);
		\end{scope}
		\begin{scope}[xshift=3cm,yshift=1.74cm]
		\draw[gray] (300:1cm) -- (0:1cm) -- (60:1cm) -- (120:1cm) -- (180:1cm);
		\draw[thick,-<-] (300:1cm) -- (0:1cm);
		\draw[thick,->-] (120:1cm) -- (180:1cm);
		\end{scope}
		\begin{scope}[xshift=1.5cm,yshift=2.61cm]
		\draw[gray] (0:1cm) -- (60:1cm) -- (120:1cm) -- (180:1cm);
		\end{scope}
		\begin{scope}[xshift=4.5cm,yshift=0.87cm]
		\draw[gray] (240:1cm) -- (300:1cm) -- (0:1cm) -- (60:1cm) -- (120:1cm);
		\draw[thick,-<-] (240:1cm) -- (300:1cm);
		\end{scope}
		\begin{scope}[xshift=4.5cm,yshift=-0.87cm]
		\draw[gray] (180:1cm) -- (240:1cm) -- (300:1cm) -- (0:1cm) -- (60:1cm);
		\draw[thick,-<-] (180:1cm) -- (240:1cm);
		\end{scope}
		\node at (0.9,2.5) {\rotatebox{-30}{\small$\alpha$}};
		\node at (2.6,2.2) {\rotatebox{-30}{\small$\beta$}};
		\node at (4.1,1.3) {\rotatebox{-30}{\small$\gamma$}};
		\node at (4.7,-0.3) {\rotatebox{-30}{\small$\delta$}};
		\node at (3.6,-1.6) {\rotatebox{-30}{\small$\epsilon$}};
		\node at (1.8,-1.5) {\rotatebox{-30}{\small$\eta$}};
		\node at (0.25,-0.65) {\rotatebox{-30}{\small$\mu$}};
		\node at (-0.3,1.1) {\rotatebox{-30}{\small$\nu$}};
	}
	\end{scope}
	\end{tikzpicture}
	\begin{tikzpicture}[scale=0.8]
	\clip(-3,-1.5) rectangle (6.5,4.5);
	\node at (-2,4) {\textbf{(c)}};
	\begin{scope}
	\rotatebox{30}{
		\draw[gray] (0:1cm) -- (60:1cm) -- (120:1cm) -- (180:1cm) -- (240:1cm) -- (300:1cm) -- (0:1cm);
		\draw[thick,->-] (300:1cm) -- (0:1cm);
		\draw[thick,-<-] (60:1) -- (120:1cm);
		\begin{scope}[xshift=0,yshift=1.74cm]
		\draw[gray] (300:1cm) -- (0:1cm) -- (60:1cm) -- (120:1cm) -- (180:1cm) -- (240:1cm);
		\draw[thick,-<-] (0:1cm) -- (60:1cm);
		\end{scope}
		\begin{scope}[xshift=1.5cm,yshift=0.87cm]
		\draw[->-,thick] (60:1) to node[above] {\rotatebox{-30}{\small$g_1'$}} (120:1);
		\draw[->-,thick] (120:1) to node[left,pos=0.3] {\rotatebox{-30}{\small$l_1'$}} (180:1);
		\draw[->-,thick] (180:1) to node[left,pos=0.7] {\rotatebox{-30}{\small$k_1'$}} (240:1);
		\draw[->-,thick] (240:1) to node[below,pos=0.5] {\rotatebox{-30}{\small$j_1'$}} (300:1);
		\draw[->-,thick] (0:1) to node[right,pos=0.7] {\rotatebox{-30}{\small$h_1'\ $}} (60:1);
		\end{scope}
		\begin{scope}[xshift=3cm,yshift=-1.74cm]
		\draw[gray] (180:1cm) -- (240:1cm) -- (300:1cm) -- (0:1cm);
		\end{scope}
		\begin{scope}[xshift=1.5cm,yshift=-0.87cm]
		\draw[gray] (180:1cm) -- (240:1cm) -- (300:1cm) -- (0:1cm) -- (60:1cm);
		\draw[thick,->-] (300:1cm) -- (0:1cm);
		\end{scope}
		\begin{scope}[xshift=3cm,yshift=0cm]
		\draw[-<-,thick] (60:1) to node[above,pos=0.5] {\rotatebox{-30}{\small$\ g_2'$}} (120:1);
		\draw[-<-,thick] (120:1) to node[right,pos=0.7] {\rotatebox{-30}{\small$i_1''$}} (180:1);
		\draw[-<-,thick] (180:1) to node[left,pos=0.8] {\rotatebox{-30}{\small$k_2'$}} (240:1);
		\draw[-<-,thick] (240:1) to node[below,pos=0.5] {\rotatebox{-30}{\small$j_2'$}} (300:1);
		\draw[-<-,thick] (300:1) to node[right,pos=0.3] {\rotatebox{-30}{\small$i_2'$}} (0:1);
		\draw[-<-,thick] (0:1) to node[right,pos=0.6] {\rotatebox{-30}{\small$h_2'$}} (60:1);
		\end{scope}
		\begin{scope}[xshift=3cm,yshift=1.74cm]
		\draw[gray] (300:1cm) -- (0:1cm) -- (60:1cm) -- (120:1cm) -- (180:1cm);
		\draw[thick,-<-] (300:1cm) -- (0:1cm);
		\draw[thick,->-] (120:1cm) -- (180:1cm);
		\end{scope}
		\begin{scope}[xshift=1.5cm,yshift=2.61cm]
		\draw[gray] (0:1cm) -- (60:1cm) -- (120:1cm) -- (180:1cm);
		\end{scope}
		\begin{scope}[xshift=4.5cm,yshift=0.87cm]
		\draw[gray] (240:1cm) -- (300:1cm) -- (0:1cm) -- (60:1cm) -- (120:1cm);
		\draw[thick,-<-] (240:1cm) -- (300:1cm);
		\end{scope}
		\begin{scope}[xshift=4.5cm,yshift=-0.87cm]
		\draw[gray] (180:1cm) -- (240:1cm) -- (300:1cm) -- (0:1cm) -- (60:1cm);
		\draw[thick,-<-] (180:1cm) -- (240:1cm);
		\end{scope}
		\node at (0.9,2.5) {\rotatebox{-30}{\small$\alpha$}};
		\node at (2.6,2.2) {\rotatebox{-30}{\small$\beta$}};
		\node at (4.1,1.3) {\rotatebox{-30}{\small$\gamma$}};
		\node at (4.7,-0.3) {\rotatebox{-30}{\small$\delta$}};
		\node at (3.6,-1.6) {\rotatebox{-30}{\small$\epsilon$}};
		\node at (1.8,-1.5) {\rotatebox{-30}{\small$\eta$}};
		\node at (0.25,-0.65) {\rotatebox{-30}{\small$\mu$}};
		\node at (-0.3,1.1) {\rotatebox{-30}{\small$\nu$}};
	}
	\end{scope}
	\end{tikzpicture}
	\caption{\label{fig:commut}\textbf{Commutativity.} \textbf{(a)} In this scenario, the operators $B_\mathbf{p_1}$ and $B_\mathbf{p_2}$ are applied to neighboring plaquettes. \textbf{(b)} The operator $B^s_\mathbf{p_1}$ puts a loop of type $s$ into plaquette $\mathbf{p_1}$, and $B^t_\mathbf{p_2}$ puts a loop of type $t$ into $\mathbf{p_2}$ (although not depicted in the picture, we take the sum over $s$ and $t$). \textbf{(c)} After fusing in both loops, the result is a linear combination of string-net configurations which is independent of the order in which the loops have been fused in.}
\end{figure}

Like the original Hamiltonian in \cite{LW05}, this Hamiltonian has some interesting properties, given that the pentagon equation \eqref{eq:pentagon} is fulfilled.

\begin{figure}[t]
	\centering
	\begin{tikzpicture}[scale=0.65,decoration={markings,mark=at position .65 with {\arrow[>=stealth]{>}}}]
	\clip(-4,-2.5) rectangle (5.75,6);
	\begin{scope}
	\rotatebox{30}{
		\begin{scope}[xshift=-1.5cm,yshift=0.87cm]
		\draw[gray] (60:1cm) -- (120:1cm) -- (180:1cm) -- (240:1cm) -- (300:1cm);
		\draw[{postaction=decorate},thick] (240:1cm) -- (300:1cm);
		\draw[{postaction=decorate}, thick] (120:1cm) -- (60:1cm);
		\end{scope}
		\draw[gray] (0:1cm) -- (60:1cm) -- (120:1cm) -- (180:1cm) -- (240:1cm) -- (300:1cm) -- (0:1cm);
		\draw[{postaction=decorate}, ultra thick, red] (120:1cm) -- (180:1cm);
		\draw[{postaction=decorate}, ultra thick, red] (180:1cm) -- (240:1cm);
		\draw[{postaction=decorate}, ultra thick, red] (240:1cm) -- (300:1cm);
		\draw[{postaction=decorate}, thick] (0:1cm) -- (300:1cm);
		\draw[{postaction=decorate}, thick] (60:1cm) -- (120:1cm);
		\begin{scope}[xshift=1.5cm,yshift=-0.87cm]
		\draw[gray] (180:1cm) -- (240:1cm) -- (300:1cm) -- (0:1cm) -- (60:1cm);
		\draw[{postaction=decorate}, ultra thick, red] (180:1cm) -- (240:1cm);
		\draw[{postaction=decorate}, ultra thick, red] (240:1cm) -- (300:1cm);
		\draw[{postaction=decorate}, ultra thick, red] (300:1cm) -- (0:1cm);
		\draw[{postaction=decorate}, thick] (60:1cm) -- (0:1cm);
		\end{scope}
		\begin{scope}[xshift=3cm,yshift=-1.74cm]
		\draw[gray] (180:1cm) -- (240:1cm) -- (300:1cm) -- (0:1cm);
		\draw[{postaction=decorate}, thick] (240:1cm) -- (180:1cm);
		\end{scope}
		\begin{scope}[xshift=-1.5cm,yshift=2.61cm]
		\draw[gray] (0:1cm) -- (60:1cm) -- (120:1cm) -- (180:1cm) -- (240:1cm);
		\draw[{postaction=decorate}, thick] (60:1cm) -- (0:1cm);
		\end{scope}
		\begin{scope}[xshift=0,yshift=1.74cm]
		\draw[gray] (300:1cm) -- (0:1cm) -- (60:1cm) -- (120:1cm) -- (180:1cm) -- (240:1cm);
		\draw[{postaction=decorate}, ultra thick, red] (60:1cm) -- (120:1cm);
		\draw[{postaction=decorate}, ultra thick, red] (120:1cm) -- (180:1cm);
		\draw[{postaction=decorate}, ultra thick, red] (180:1cm) -- (240:1cm);
		\draw[{postaction=decorate}, thick] (0:1cm) -- (60:1cm);
		\end{scope}
		\begin{scope}[xshift=1.5cm,yshift=0.87cm]
		\draw[gray]  (240:1cm) -- (300:1cm) -- (0:1cm) -- (60:1cm) -- (120:1cm);
		\end{scope}
		\begin{scope}[xshift=3cm,yshift=0cm]
		\draw[gray] (240:1cm) -- (300:1cm)-- (0:1cm) -- (60:1cm) -- (120:1cm);
		\draw[{postaction=decorate}, ultra thick, red] (240:1cm) -- (300:1cm);
		\draw[{postaction=decorate}, ultra thick, red] (300:1cm)-- (0:1cm);
		\draw[{postaction=decorate}, ultra thick, red] (0:1cm) -- (60:1cm);
		\draw[{postaction=decorate}, thick] (120:1cm) -- (60:1cm);
		\end{scope}
		\begin{scope}[xshift=4.5cm,yshift=-0.87cm]
		\draw[gray] (180:1cm) -- (240:1cm) -- (300:1cm) -- (0:1cm) -- (60:1cm);
		\draw[{postaction=decorate}, thick] (240:1cm) -- (180:1cm);
		\end{scope}
		\begin{scope}[xshift=0,yshift=3.48cm]
		\draw[gray] (0:1cm) -- (60:1cm) -- (120:1cm) -- (180:1cm);
		\draw[{postaction=decorate}, thick] (60:1cm) -- node[above, xshift=0.15cm,yshift=-0.15cm] {\rotatebox[]{-30}{\small $e_2$}} (0:1cm);
		\end{scope}
		\begin{scope}[xshift=1.5cm,yshift=2.61cm]
		\draw[gray] (0:1cm) -- (60:1cm) -- (120:1cm) -- (180:1cm);
		\draw[{postaction=decorate}, ultra thick, red] (0:1cm) -- (60:1cm);
		\draw[{postaction=decorate}, ultra thick, red] (60:1cm) -- node[below, yshift=0.1cm] {\rotatebox[]{-30}{\small $i_1$}} (120:1cm);
		\draw[{postaction=decorate}, ultra thick, red] (120:1cm) -- node[right, yshift=-0.15cm, xshift=-0.15cm] {\rotatebox[]{-30}{\small $i_2$}} (180:1cm);
		\end{scope}
		\begin{scope}[xshift=3cm,yshift=1.74cm]
		\draw[gray] (300:1cm) -- (0:1cm) -- (60:1cm) -- (120:1cm) -- (180:1cm);
		\draw[{postaction=decorate}, ultra thick, red] (300:1cm) -- (0:1cm);
		\draw[{postaction=decorate}, ultra thick, red] (0:1cm) -- (60:1cm);
		\draw[{postaction=decorate}, ultra thick, red] (60:1cm) -- (120:1cm);
		\draw[{postaction=decorate}, thick] (180:1cm) -- (120:1cm);
		\end{scope}
		\begin{scope}[xshift=4.5cm,yshift=0.87cm]
		\draw[gray] (240:1cm) -- (300:1cm) -- (0:1cm) -- (60:1cm) -- (120:1cm);
		\draw[{postaction=decorate}, thick] (300:1cm) -- (240:1cm);
		\draw[{postaction=decorate}, thick] (60:1cm) -- (120:1cm);
		\end{scope}
		\begin{scope}[xshift=1.5cm,yshift=-2.61cm]
		\draw[gray] (120:1cm) -- (180:1cm) -- (240:1cm) -- (300:1cm) -- (0:1cm);
		\draw[{postaction=decorate}, thick] (180:1cm) -- (120:1cm);
		\end{scope}
		\begin{scope}[xshift=0cm,yshift=-1.74cm]
		\draw[gray] (120:1cm) -- (180:1cm) -- (240:1cm) -- (300:1cm);
		\draw[{postaction=decorate}, thick] (180:1cm) -- (120:1cm);
		\end{scope}
		\begin{scope}[xshift=-1.5cm,yshift=-0.87cm]
		\draw[gray] (120:1cm) -- (180:1cm) -- (240:1cm) -- (300:1cm);
		\end{scope}
		\begin{scope}[xshift=-3cm,yshift=0cm]
		\draw[gray] (60:1cm) -- (120:1cm) -- (180:1cm) -- (240:1cm) -- (300:1cm);
		\end{scope}
		\begin{scope}[xshift=3cm,yshift=-3.48cm]
		\draw[gray] (180:1cm) -- (240:1cm) -- (300:1cm) -- (0:1cm) --(60:1cm);
		\end{scope}
		\begin{scope}[xshift=3cm,yshift=3.48cm]
		\draw[gray] (300:1cm) -- (0:1cm) --(60:1cm) -- (120:1cm) -- (180:1cm);
		\draw[{postaction=decorate}, thick] (0:1cm) -- (300:1cm);
		\draw[{postaction=decorate}, thick] (120:1cm) -- node[right, xshift=-0.05cm, yshift=-0.15cm] {\rotatebox[]{-30}{\small $e_1$}} (180:1cm);
		\end{scope}
		\begin{scope}[xshift=1.5cm,yshift=4.35cm]
		\draw[gray] (0:1cm) -- (60:1cm) -- (120:1cm) -- (180:1cm);
		\end{scope}
		\begin{scope}[xshift=0cm,yshift=5.22cm]
		\draw[gray] (0:1cm) -- (60:1cm) -- (120:1cm) -- (180:1cm) -- (240:1cm);
		\end{scope}
		\begin{scope}[xshift=4.5cm,yshift=2.61cm]
		\draw[gray] (300:1cm) -- (0:1cm) -- (60:1cm) -- (120:1cm);
		\end{scope}
		\begin{scope}[xshift=6cm,yshift=1.74cm]
		\draw[gray] (240:1cm) -- (300:1cm) -- (0:1cm) -- (60:1cm) -- (120:1cm);
		\end{scope}
		\begin{scope}[xshift=-0.435cm,yshift=0.87cm]
		\draw[dashed] (0,0) circle (0.75cm);
		\node (v5) at (-1,0) {\rotatebox[]{-30}{\small $\mathbf{v}_6$}};
		\end{scope}
	}
	\end{scope}
	\end{tikzpicture}
	\caption{\textbf{Closed string operator on the honeycomb lattice.} A closed string operator $W^s(P)$ only acts non-trivially on spin states along the closed path $P$, depicted as thick red line. It creates a type-$s$ string and fuses it into each vertex $\mathbf{v}_1,\mathbf{v}_2,...\mathbf{v}_N$ along $P$. This action only changes the two sites $i_{k-1},i_k$ of $\mathbf{v}_k$, whereas the third site $e_k$ of the vertex is unaffected. In total, $W^s(P)$ transforms the initial spin state $i_1,i_2,...,i_N$ to the final spin state $i'_1,i'_2,...,i'_N$. The labelling convention is demonstratively shown in the picture for $i_1,i_2,e_1,e_2$ and $\mathbf{v}_6$ of an example for a $W^s(P)$ operator.\label{fig:excitations}}
\end{figure}
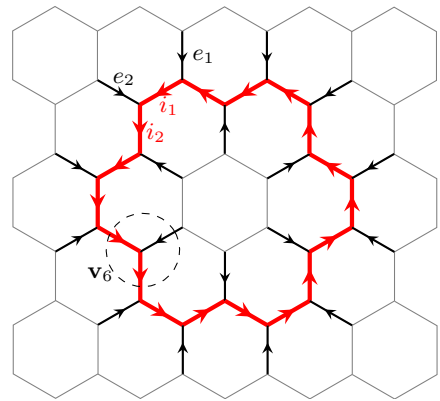

\begin{enumerate}
	\item First, the Hamiltonian is hermitian, which is a crucial property for a Hamiltonian that describes a physical system. Recall that, when acting with the magnetic flux operator $B_\mathbf{p}$ on a plaquette $\mathbf{p}$, we do the following transformation:
	\begin{equation}
		\begin{tikzpicture}[scale=0.65,baseline={([yshift=-0.1cm]current bounding box.center)}]
		\draw[->-] (90:1) -- (150:1);
		\draw[->-] (150:1) -- (210:1);
		\draw[->-] (210:1) -- (270:1);
		\draw[->-] (270:1) -- (330:1);
		\draw[->-] (330:1) -- (30:1);
		\draw[->-] (30:1) -- (90:1);
		\foreach \a in {30,90,150,210,270,330} \draw[-<-] (\a:1) -- (\a:1.5);
		\node at (0,0) {$\vec{p}$};
		\end{tikzpicture}\mapsto\sum_{\vec{p}\,'}C(\vec{p},\vec{p}\,')
		\begin{tikzpicture}[scale=0.65,baseline={([yshift=-0.1cm]current bounding box.center)}]
		\draw[->-] (90:1) -- (150:1);
		\draw[->-] (150:1) -- (210:1);
		\draw[->-] (210:1) -- (270:1);
		\draw[->-] (270:1) -- (330:1);
		\draw[->-] (330:1) -- (30:1);
		\draw[->-] (30:1) -- (90:1);
		\foreach \a in {30,90,150,210,270,330} \draw[-<-] (\a:1) -- (\a:1.5);
		\node at (0,0) {$\vec{p}\,'$};
		\end{tikzpicture},
	\end{equation}
	where the vectors $\vec{p}$ and $\vec{p}\,'$ represent the entirety of the inner indices of the plaquette before and after the application of $B_\mathbf{p}$, respectively, and $C(\vec{p},\vec{p}\,')$ stands for the coefficient in the linear combination (see~\eqref{Bps}). In order to show that the operator $B_\mathbf{p}$ is hermitian it is sufficient to show that $C(\vec{p},\vec{p}\,')=\overline{C(\vec{p}\,',\vec{p})}$ which is a consequence of the mirror symmetry of the $F$-symbols~\eqref{eq:mirror}. The detailed proof can be found in Sec.~\ref{sec:Hermicity}.
	\item Second, the $B_\mathbf{p}$ and $Q_\mathbf{v}$ are projector-valued operators. As depicted in Fig.~\ref{fig:proj}, applying the magnetic flux operator twice corresponds to inserting a loop of type $s$ and a loop of type $t$ into the plaquette and summing over $s,t$. In Sec.~\ref{sec:Projector} we show that this is equivalent to the scenario where we insert only one loop of type $\alpha$ (and sum over $\alpha$).
	\item Furthermore, to build an exactly solvable Hamiltonian, it is necessary that the individual operators $Q_\mathbf{v}$ and $B_\mathbf{p}$ all commute with each other. It is clear that all operators commute when they are applied on plaquettes that have no edges in common. Furthermore, it is easy to see that the electric charge operator $Q_\mathbf{v}$ commutes with everything since it only ensures that valid string-net configurations are energetically favorable. Also, $B_\mathbf{p}$ commutes with itself when it is applied to the same plaquette twice which follows from the property of $B_\mathbf{p}$ being a projector. Hence, it remains to be shown that $B_\mathbf{p}$ commutes with itself on neighboring plaquettes. The main idea here is that, no matter in which order the operators are applied to the plaquettes, the resulting linear combination of string-net configurations is always the same (see Fig.~\ref{fig:commut}). The detailed proof can be found in Sec.~\ref{sec:Commutativity}.
\end{enumerate}

\phantom{\dots}

\section{Excitations}
\label{sec:excitations}

The excited states of the original Levin-Wen Hamiltonian correspond to topologically nontrivial quasiparticles, also referred as to anyons. They are associated to closed string operators $W^s(P)$, which create a type-$s$ string along the closed path $P$ on the honeycomb lattice. In contrast to $B_\mathbf{p}$ (or $Q_\mathbf{v}$) the $W^s(P)$ are no longer just local operators acting on a single plaquette (or vertex). We follow the approach in \cite{LW05} to propose a recipe for computing simple $W^s(P)$ for arbitrary closed paths $P$. Assuming that $W^s(P)$ only changes the spin states along the path $P=\mathbf{v}_1,...,\mathbf{v}_N$ makes it possible to define its action vertex-wise. As in \cite{LW05}, we can then perform the following ansatz for the matrix elements of a simple type-$s$ closed string operator $W^s(P)$ transforming the initial state $i_1,...,i_N$ to the final state $i'_1,...,i'_N$:
\begin{equation}
	W_{i_1 i_2...i_N}^{s,i'_1 i'_2...i'_N}(e_1 e_2...e_N)=\left(\prod_{k=1}^{N} F_k^s\right)\left(\prod_{k=1}^{N} \omega_k\right)\label{eq:StringOperator}.
\end{equation}

This operator only changes two of the three spin states at each vertex $\mathbf{v}_i$ along $P$, see Fig.~\ref{fig:excitations}. The remaining external leg is denoted $e_i$. $F_k^s$ is a combination of $F$-symbols acting on the vertex $\mathbf{v}_k$ which fuses the type-$s$ string generated by $W^s(P)$ into the respective vertex. Hence, we can think of $F_k^s$ as just one of the \itemone \,--\,\itemsix \,\,actions of $B_\mathbf{p}^s$. Then, we have to differ six different cases of vertices $\mathbf{v}_k$:
\begin{widetext}
	\begin{align}
		F_k^s=\begin{cases}
			\sqrt{\frac{d_{s^*} d_s d_{i^*_k} d_{i_{k-1}}}{d_{i'^*_k} d_{i'_{k-1}}}} \left(F_{i^*_k}^{i^*_k s^* s}\right)_{i'^*_k 0} \overline{\left(F_{e^*_k}^{i'^*_k s i_{k-1}}\right)}_{i^*_k i'_{k-1}},
			&\text{if $P$ runs as }\begin{tikzpicture}[decoration={markings,mark=at position .5 with {\arrow[>=stealth]{>}}},baseline=(current bounding box.center),scale=0.5]
			\node (lo) at (330:1.6cm) {\small $i_{k-1}$};
			\draw[{postaction=decorate}, ultra thick] (330:1cm) to (0,0);
			\node (ro) at (210:1.5cm) {\small $i_k$};
			\draw[{postaction=decorate}, ultra thick] (0,0) to (210:1cm);
			\node (u) at (90:1.5cm) {\small $e_{k}$};
			\draw[{postaction=decorate}] (90:1cm) to (0,0);
			\end{tikzpicture}\text{ at $\mathbf{v}_k$ (\itemone)}\\
			\sqrt{\frac{d_s d_{i_k}}{d_{i'_k}}}\, \overline{\left(F_{i'_{k-1}}^{si_k e^*_k}\right)}_{i'_k i_{k-1}},
			&\text{if $P$ runs as }\begin{tikzpicture}[decoration={markings,mark=at position .5 with {\arrow[>=stealth]{>}}},baseline=(current bounding box.center),scale=0.5,rotate=180]
			\node (lo) at (330:1.5cm) {\small $i_k$};
			\draw[{postaction=decorate}, ultra thick] (0,0) to (330:1cm);
			\node (ro) at (210:1.5cm) {\small $e_k$};
			\draw[{postaction=decorate}] (210:1cm) to (0,0);
			\node (u) at (90:1.5cm) {\small $i_{k-1}$};
			\draw[{postaction=decorate},ultra thick] (90:1cm) to (0,0);
			\end{tikzpicture}\text{ at $\mathbf{v}_k$ (\itemtwo)}\\
			\sqrt{\frac{d_s d_{i_{k-1}}}{d_{i'_{k-1}}}} \left(F_{i'_k}^{s i_{k-1} e_k}\right)_{i'_{k-1} i_k}
			&\text{if $P$ runs as }\begin{tikzpicture}[decoration={markings,mark=at position .5 with {\arrow[>=stealth]{>}}},baseline=(current bounding box.center),scale=0.5]
			\node (lo) at (330:1.5cm) {\small $e_k$};
			\draw[{postaction=decorate}] (330:1cm) to (0,0);
			\node (ro) at (210:1.6cm) {\small $i_{k-1}$};
			\draw[{postaction=decorate},ultra thick] (210:1cm) to (0,0);
			\node (u) at (90:1.5cm) {\small $i_k$};
			\draw[{postaction=decorate},ultra thick] (0,0) to (90:1cm);
			\end{tikzpicture}\text{ at $\mathbf{v}_k$ (\itemthree)}\\
			\sqrt{\frac{d_{s^*} d_s d_{i^*_{k-1}} d_{i_k}}{d_{i'^*_{k-1}} d_{i'_k}}} \left(F_{i_k}^{s^*si_k}\right)_{0i'_k} \overline{\left(F_{e_k}^{i^*_{k-1}s^*i'_k}\right)}_{i'_{k-1} i_k},
			&\text{if $P$ runs as }\begin{tikzpicture}[decoration={markings,mark=at position .5 with {\arrow[>=stealth]{>}}},baseline=(current bounding box.center),scale=0.5,rotate=180]
			\node (lo) at (330:1.6cm) {\small $i_{k-1}$};
			\draw[{postaction=decorate}, ultra thick] (330:1cm) to (0,0);
			\node (ro) at (210:1.5cm) {\small $i_k$};
			\draw[{postaction=decorate}, ultra thick] (0,0) to (210:1cm);
			\node (u) at (90:1.5cm) {\small $e_{k}$};
			\draw[{postaction=decorate}] (90:1cm) to (0,0);
			\end{tikzpicture}\text{ at $\mathbf{v}_k$ (\itemfour)}\\
			\sqrt{\frac{d_{s^*} d_{i^*_k}}{d_{i'^*_k}}} \,\overline{\left(F_{i'^*_{k-1}}^{e_k i^*_k s^*}\right)}_{i^*_{k-1} i'^*_k},
			&\text{if $P$ runs as }\begin{tikzpicture}[decoration={markings,mark=at position .5 with {\arrow[>=stealth]{>}}},baseline=(current bounding box.center),scale=0.5]
			\node (lo) at (330:1.5cm) {\small $i_k$};
			\draw[{postaction=decorate}, ultra thick] (0,0) to (330:1cm);
			\node (ro) at (210:1.5cm) {\small $e_k$};
			\draw[{postaction=decorate}] (210:1cm) to (0,0);
			\node (u) at (90:1.5cm) {\small $i_{k-1}$};
			\draw[{postaction=decorate},ultra thick] (90:1cm) to (0,0);
			\end{tikzpicture}\text{ at $\mathbf{v}_k$ (\itemfive)}\\
			\sqrt{\frac{d_{s^*} d_{i^*_{k-1}}}{d_{i'^*_{k-1}}}} \left(F_{i'^*_{k}}^{e^*_k i^*_{k-1}s^*}\right)_{i^*_{k} i'^*_{k-1}},
			&\text{if $P$ runs as }\begin{tikzpicture}[decoration={markings,mark=at position .5 with {\arrow[>=stealth]{>}}},baseline=(current bounding box.center),scale=0.5,rotate=180]
			\node (lo) at (330:1.5cm) {\small $e_k$};
			\draw[{postaction=decorate}] (330:1cm) to (0,0);
			\node (ro) at (210:1.6cm) {\small $i_{k-1}$};
			\draw[{postaction=decorate},ultra thick] (210:1cm) to (0,0);
			\node (u) at (90:1.5cm) {\small $i_k$};
			\draw[{postaction=decorate},ultra thick] (0,0) to (90:1cm);
			\end{tikzpicture}\text{ at $\mathbf{v}_k$ (\itemsix)}
		\end{cases}
	\end{align}
\end{widetext}
In order to compute the $\omega_k$, the authors in \cite{LW05} use the fact that $W^s(P)$ should commute with the Hamiltonian if $P$ is a closed path. This is only the case if $\omega_k$ satisfies a certain constraint (see Eq.~(18) in \cite{LW05}), which in turn gives all simple type-$s$ closed string operators. Unfortunately, we found that in our general approach the derivation of this constraint appears to be highly complicated and not as obvious as in the original model. Therefore, we leave it as an open problem. 

\section{Conclusion and Outlook}
\label{sec:conclusion}

In this paper, we have generalized the Levin-Wen model \cite{LW05} to arbitrary multiplicity-free unitary fusion categories and calculated its matrix elements explicitly. The original Hamiltonian requires several additional symmetries of the underlying UFC which are not mandatory in our construction. The case of fusion rules with multiplicities can easily be obtained by inserting an extra label at each vertex of the honeycomb lattice. However, this only adds further degrees of freedom at each lattice site and does not bring any additional physical insights. Furthermore, current explicit experimental realizations of these particles are limited to very few simple models, and so fusion rules with multiplicity are unlikely to have practical relevance in the near future.

An interesting question which is not answered in this paper, is that of excited states which we briefly addressed in Section~\ref{sec:excitations}. Although it is possible to apply the ansatz of \cite{LW05} in our general case, we believe another approach may be less difficult. In \cite{Buerschaper2009, Sahinoglu2014}, the Levin-Wen model is built as a tensor network of matrix product operators. This idea gives a much simpler framework to calculate the quasiparticle excitations. Such a construction is probably also possible for our model. However, the authors also require tetrahedral symmetry to be fulfilled in their construction, hence one has to carefully check which aspects of the tensor network approach differ for the general case. We hope to address this question in the near future.

From a mathematical point of view, our construction is not the most general one possible. The excitations of a Levin-Wen model correspond to irreducible representations of the Drinfeld double of the underlying category \cite{LW05}. In order to exhibit anyonic quasiparticles, the quantum double needs to have the structure of a modular tensor category \cite{TURAEV1992}. To achieve this, we could relax the conditions of the underlying category to be a $\mathbb{F}$-linear, abelian, spherical, rigid, monoidal category with $\mathrm{End}(0)\simeq\mathbb{F}$ for a closed number field $\mathbb{F}$ \cite{Mueger2003}. However, calculating the matrix elements of such a Levin-Wen model would purely be of theoretical interest since the physically relevant cases are covered by UFCs.
	
	\section*{Acknowledgements}
	We thank Thomas Cope, and Le Phuc Thinh for helpful discussions. We especially thank Tobias J.\@ Osborne for introducing us to the topic and for many insightful comments and discussions. We also thank Jürgen Fuchs for drawing our attention to the tetrahedral symmetry condition and, thereby, motivated us to take a closer look at the Levin-Wen model against that background. This work was supported, in part, by the Deutsche Forschungsgemeinschaft (DFG, German Research Foundation) through SFB 1227 (DQ-mat), the RTG 1991, and funded by the DFG under Germany’s Excellence Strategy -- EXC-2123 QuantumFrontiers -- 390837967.
	
	\onecolumngrid
	
	\begin{appendix}
		
		\section{Unitary fusion categories and their graphical calculus}
		\label{app:UFCgraphcalc}
		
		The underlying mathematical framework of the Levin-Wen constriction is the concept of unitary fusion categories which is a very rich mathematical field itself. We do not want to explain the whole theory here, but refer to \cite{Etingof2015} for a rigorous mathematical treatment of the topic. Here, we rather want to explain the graphical calculus that comes with a unitary fusion category and which we used throughout the paper. This section is intended to be a reference text which can be consulted to understand the individual steps in each of the calculations done in this paper.
		
		Before we continue, a short note on string diagrams: Recall that we use the convention that all unoriented string diagrams point upwards~\eqref{eq:StringOrientation}. Especially in this section we mostly omit the arrows since the equations hold for any chosen orientation.
		
		\begin{definition}
			A \textbf{fusion category} over $\mathbb{C}$ is a $\mathbb{C}$-linear rigid semisimple monoidal category with finitely many simple objects (up to isomorphism) and finite-dimensional morphism spaces such that the identity object is simple.
		\end{definition}
		
		It is not crucial for understanding the paper and the calculations to fully understand the definition of a fusion category, it is rather given here for completeness. Nevertheless, we will highlight some aspects of it to demonstrate how the physical model emerges from the mathematical definition.
		
		First of all, a fusion category is a \emph{monoidal category}, which means that it is equipped with a tensor product. This tensor product is an operation between the simple objects \footnote{The tensor product is defined not only for simple objects but for all objects of the category. We restrict ourselves to simple objects here because it makes it easier to draw the connection to the string types of the lattice model.} of the category ($\equiv$ the string types of the physical model). It determines the action of the vertex operator $Q_\mathbf{v}$: Consider, for instance, the tensor product $i\otimes j=k$. This is equivalent to the following vertex in the physical model:
		\begin{equation}
			\begin{tikzpicture}[baseline=(current bounding box.center),scale=0.6]
			\draw (330:1cm) to node[below,near start] {\small $j$} (0,0);
			\draw (210:1cm) to node[below,near start] {\small $i$} (0,0);
			\draw (90:1cm) to node[right,near start] {\small $k$} (0,0);
			\end{tikzpicture}
		\end{equation}
		Hence, if this tensor product exists in the fusion category, the corresponding vertex is an allowed configuration in the lattice model.
		
		Furthermore, we can define the so-called $F$-symbols for a fusion category (they sometimes also appear under the name $6j$-symbols, especially in physics \footnote{The $6j$ symbols are defined slightly different than the definition we give for the $F$-symbols here, but there is an easy relation between the two, see Appendix~\ref{app:tetrahedral}}). In a fusion category, this is a family of maps given by
		\begin{equation}
			F_i^{jkl}:\mathrm{Hom}(i,(j\otimes k)\otimes l)\rightarrow \mathrm{Hom}(i,j\otimes(k\otimes l)),
		\end{equation}
		where $\mathrm{Hom}(X,Y)$ denotes the space of morphisms from the object $X$ to the object $Y$. The matrix representation of these maps is given by
		\begin{equation}
			\label{eq:Fsymbol}
			\begin{tikzpicture}[baseline=(current bounding box.center),scale=0.8]
			\draw (0,0.5) -- (0,1);
			\draw (0,1) -- (-0.5,1.5);
			\draw (-0.5,1.5) -- (-1,2);
			\draw (-0.5,1.5) -- (0,2);
			\draw (0,1) -- (0.5,1.5);
			\draw (0.5,1.5) -- (1,2);
			\node at (0,0.25) {\small $i$};
			\node at (-1,2.25) {\small $j$};
			\node at (0,2.25) {\small $k$};
			\node at (1,2.25) {\small $l$};
			\node at (-0.5,1.1) {\small $m$};
			\end{tikzpicture}=\sum_n\left(F_{i}^{jkl}\right)_{mn}
			\begin{tikzpicture}[baseline=(current bounding box.center),scale=0.8]
			\draw (0,0.5) -- (0,1);
			\draw (0,1) -- (-0.5,1.5);
			\draw (-0.5,1.5) -- (-1,2);
			\draw (0.5,1.5) -- (0,2);
			\draw (0,1) -- (0.5,1.5);
			\draw (0.5,1.5) -- (1,2);
			\node at (0,0.25) {\small $i$};
			\node at (-1,2.25) {\small $j$};
			\node at (0,2.25) {\small $k$};
			\node at (1,2.25) {\small $l$};
			\node at (0.5,1.1) {\small $n$};
			\end{tikzpicture}.
		\end{equation}
		These $F$-symbols have to fulfil the \emph{pentagon equation} (which also appeared in \eqref{eq:pentagon}):
		\begin{equation}
			\left(F_m^{nkl}\right)_{pl}
			\left(F_m^{ijs}\right)_{nr}
			=
			\sum_q
			\left(F_{p}^{ijk}\right)_{nq}
			\left(F_m^{iql}\right)_{pr}
			\left(F_{r}^{jkl}\right)_{qs}.
		\end{equation}
		The fusion category is called \emph{unitary} if its $F$-symbols are unitary. In a unitary fusion category, the inverse $F$-symbol is given by the Hermitian conjugate:
		\begin{equation}
			\left(F_i^{jkl}\right)_{mn}^{-1}=\left(F_i^{jkl}\right)_{mn}^\dagger=\overline{\left(F_i^{jkl}\right)}_{nm},
		\end{equation}
		where the bar above the $F$-symbols denotes the complex conjugate. In terms of string diagrams, this is 
		\begin{equation}
			\label{eq:hermit}
			\begin{tikzpicture}[baseline=(current bounding box.center),scale=0.8]
			\draw (0,0.5) -- (0,1);
			\draw (0,1) -- (-0.5,1.5);
			\draw (-0.5,1.5) -- (-1,2);
			\draw (0.5,1.5) -- (0,2);
			\draw (0,1) -- (0.5,1.5);
			\draw (0.5,1.5) -- (1,2);
			\node at (0,0.25) {\small $i$};
			\node at (-1,2.25) {\small $j$};
			\node at (0,2.25) {\small $k$};
			\node at (1,2.25) {\small $l$};
			\node at (0.5,1.1) {\small $n$};
			\end{tikzpicture}=\sum_m\left(F_{i}^{jkl}\right)^\dagger_{nm}
			\begin{tikzpicture}[baseline=(current bounding box.center),scale=0.8]
			\draw (0,0.5) -- (0,1);
			\draw (0,1) -- (-0.5,1.5);
			\draw (-0.5,1.5) -- (-1,2);
			\draw (-0.5,1.5) -- (0,2);
			\draw (0,1) -- (0.5,1.5);
			\draw (0.5,1.5) -- (1,2);
			\node at (0,0.25) {\small $i$};
			\node at (-1,2.25) {\small $j$};
			\node at (0,2.25) {\small $k$};
			\node at (1,2.25) {\small $l$};
			\node at (-0.5,1.1) {\small $m$};
			\end{tikzpicture}.
		\end{equation}
		Another consequence of unitarity is that the category is spherical (see \cite[Proposition 8.23]{Etingof2002}), which yields a \emph{mirror symmetry} (see \cite{Hong09}), i.e., we can make statements about the diagram that is horizontally mirrored in terms of the previously defined $F$-symbols:
		\begin{align}
			\label{eq:mirror}
			\begin{tikzpicture}[baseline=(current bounding box.center),xscale=0.8,yscale=-0.8]
			\draw (0,0.5) -- (0,1);
			\draw (0,1) -- (-0.5,1.5);
			\draw (-0.5,1.5) -- (-1,2);
			\draw (-0.5,1.5) -- (0,2);
			\draw (0,1) -- (0.5,1.5);
			\draw (0.5,1.5) -- (1,2);
			\node at (0,0.25) {\small $i$};
			\node at (-1,2.25) {\small $j$};
			\node at (0,2.25) {\small $k$};
			\node at (1,2.25) {\small $l$};
			\node at (-0.5,1.1) {\small $m$};
			\end{tikzpicture}&=\sum_n\overline{\left(F_{i}^{jkl}\right)}_{mn}
			\begin{tikzpicture}[baseline=(current bounding box.center),xscale=0.8,yscale=-0.8]
			\draw (0,0.5) -- (0,1);
			\draw (0,1) -- (-0.5,1.5);
			\draw (-0.5,1.5) -- (-1,2);
			\draw (0.5,1.5) -- (0,2);
			\draw (0,1) -- (0.5,1.5);
			\draw (0.5,1.5) -- (1,2);
			\node at (0,0.25) {\small $i$};
			\node at (-1,2.25) {\small $j$};
			\node at (0,2.25) {\small $k$};
			\node at (1,2.25) {\small $l$};
			\node at (0.5,1.1) {\small $n$};
			\end{tikzpicture}\\
			\label{eq:mirror2}
			\begin{tikzpicture}[baseline=(current bounding box.center),xscale=0.8,yscale=-0.8]
			\draw (0,0.5) -- (0,1);
			\draw (0,1) -- (-0.5,1.5);
			\draw (-0.5,1.5) -- (-1,2);
			\draw (0.5,1.5) -- (0,2);
			\draw (0,1) -- (0.5,1.5);
			\draw (0.5,1.5) -- (1,2);
			\node at (0,0.25) {\small $i$};
			\node at (-1,2.25) {\small $j$};
			\node at (0,2.25) {\small $k$};
			\node at (1,2.25) {\small $l$};
			\node at (0.5,1.1) {\small $n$};
			\end{tikzpicture}&=\sum_m\left(F_{i}^{jkl}\right)^\mathsf{T}_{nm}
			\begin{tikzpicture}[baseline=(current bounding box.center),xscale=0.8,yscale=-0.8]
			\draw (0,0.5) -- (0,1);
			\draw (0,1) -- (-0.5,1.5);
			\draw (-0.5,1.5) -- (-1,2);
			\draw (-0.5,1.5) -- (0,2);
			\draw (0,1) -- (0.5,1.5);
			\draw (0.5,1.5) -- (1,2);
			\node at (0,0.25) {\small $i$};
			\node at (-1,2.25) {\small $j$};
			\node at (0,2.25) {\small $k$};
			\node at (1,2.25) {\small $l$};
			\node at (-0.5,1.1) {\small $m$};
			\end{tikzpicture}.
		\end{align}
		The last equation is a combination of \eqref{eq:hermit} and \eqref{eq:mirror}.
		
		In order to compute the matrix elements of the Hamiltonian we need three more relations within the graphical calculus that basically all follow from the fact that we use a specific normalization for trivalent vertices, namely
		\begin{equation}
			\label{eq:norm}
			\left(\frac{d_k}{d_i d_j}\right)^\frac{1}{4}
			\begin{tikzpicture}[baseline=(current bounding box.center),scale=0.6]
			\draw (330:1cm) to node[below,near start] {\small $j$} (0,0);
			\draw (210:1cm) to node[below,near start] {\small $i$} (0,0);
			\draw (90:1cm) to node[right,near start] {\small $k$} (0,0);
			\end{tikzpicture}.
		\end{equation}
		We sometimes use the notation $V_{ij}^k$ for trivalent vertices of this form. Using this normalization, we can conclude the value of the loop stated in \eqref{eq:groundstate1}:
		\begin{equation}
			\label{eq:loop}
			\begin{tikzpicture}[baseline=(current bounding box.center),scale=0.7]
			\draw[->-] (0,0) circle (0.5cm);
			\node at (0.75,0) {\small $i$};
			\end{tikzpicture}=
			\begin{tikzpicture}[baseline=(current bounding box.center),scale=1.6]
			\draw[dotted] (0,0.25) to node[right, near start] {\small $0$} (0,0.5);
			\draw[->-] (0,0.5) to [bend right=90] node [right] {\small $i$} (0,1);
			\draw[->-] (0,0.5) to [bend left=90] node [left] {\small $i^*$} (0,1);
			\draw[dotted] (0,1) to node[right,near end] {\small $0$} (0,1.25);
			\end{tikzpicture}=d_i\ 
			\begin{tikzpicture}[baseline=(current bounding box.center),scale=1.6]
			\draw[dotted] (0,0.25) to node [right] {$0$} (0,1.25);
			\end{tikzpicture}.
		\end{equation}
		This can be generalized to the \emph{bigon relation}:
		\begin{equation}
			\label{eq:UFCbigon}
			\begin{tikzpicture}[baseline=(current bounding box.center),scale=1.6]
			\draw[->-] (0,0.25) to node[right, near start] {\small $k$} (0,0.5);
			\draw[->-] (0,0.5) to [bend right=90] node [right] {\small $j$} (0,1);
			\draw[->-] (0,0.5) to [bend left=90] node [left] {\small $i$} (0,1);
			\draw[->-] (0,1) to node[right,near end] {\small $k'$} (0,1.25);
			\end{tikzpicture}=\sqrt{\frac{d_i d_j}{d_k}}\delta_{k,k'}\ 
			\begin{tikzpicture}[baseline=(current bounding box.center),scale=1.6]
			\draw[->-] (0,0.25) to node [right] {$k$} (0,1.25);
			\end{tikzpicture}.
		\end{equation}
		This fulfils the local constraint stated in \eqref{eq:groundstate2}. The last relation we need is the \emph{completeness relation}:
		\begin{equation}
			\label{eq:completeness}
			\begin{tikzpicture}[baseline={([yshift=-0.1cm]current bounding box.center)},scale=0.7]
			\draw[->-] (-0.5,-0.5) to node[left] {$i$} (-0.5,1.25);
			\draw[->-] (0.5,-0.5) to node[right] {$j$} (0.5,1.25);
			\end{tikzpicture}
			=\sum_k \sqrt{\frac{d_k}{d_i d_j}}\ 
			\begin{tikzpicture}[baseline={([yshift=-0.1cm]current bounding box.center)},scale=0.7]
			\draw[->-] (-0.5,-0.5) -- (0,0);
			\draw[->-] (0.5,-0.5) -- (0,0);
			\draw[->-] (0,0) to node[right] {$k$} (0,0.75);
			\draw[->-] (0,0.75) -- (-0.5,1.25);
			\draw[->-] (0,0.75) -- (0.5,1.25);
			\node at (-0.5,-0.75) {$i$};
			\node at (-0.5,1.5) {$i$};
			\node at (0.5,-0.75) {$j$};
			\node at (0.5,1.5) {$j$};
			\end{tikzpicture}.
		\end{equation}
		Note that the set of relations \eqref{eq:Fsymbol}, \eqref{eq:loop}, \eqref{eq:UFCbigon} and \eqref{eq:completeness} is equivalent to the set of local constraints on the ground state given in the main paper. Here, the constraint~\eqref{eq:groundstate1} is implicitly contained via the normalization of trivalent vertices \eqref{eq:norm} (see \cite{Bonderson2007} for more details).

		\section{Calculation of the operators $G$ and $H$}
		\label{app:GHcalc}
		
		The local constraints \eqref{eq:groundstate4} and \eqref{eq:groundstate5} replace a single condition from the original model, namely
		\begin{equation}
			\Phi\left(
			\begin{tikzpicture}[baseline={([yshift=-3pt]current bounding box.center)}, decoration={markings,mark=at position .6 with {\arrow[>=stealth]{>}}},scale=0.5]
			\draw[{postaction=decorate}] (0.8,0.5) to node[below] {\small $j$} (1.5,1);
			\draw[{postaction=decorate}] (0.8,1.5) to node[above] {\small $i$} (1.5,1);
			\draw[{postaction=decorate}] (3.2,0.5) to node[below] {\small $k$} (2.5,1);
			\draw[{postaction=decorate}] (3.2,1.5) to node[above] {\small $l$} (2.5,1);
			\draw[{postaction=decorate}] (2.5,1) to node[above] {\small $m$} (1.5,1);
			\draw[fill=LightGray,rounded corners,LightGray] (-0.2,0) rectangle (0.8,2);
			\draw[fill=LightGray,rounded corners,LightGray] (3.2,0) rectangle (4.2,2);
			\end{tikzpicture}\right) 
			=\sum_n \left(F_k^{j^*i^*l^*}\right)_{mn}\ \Phi\left(
			\begin{tikzpicture}[baseline={([yshift=-3pt]current bounding box.center)}, decoration={markings,mark=at position .6 with {\arrow[>=stealth]{>}}},scale=0.5]
			\draw[{postaction=decorate}] (0.8,0.5) to node[below] {\small $j$} (2,0.5);
			\draw[{postaction=decorate}] (3.2,0.5) to node[below] {\small $k$} (2,0.5);
			\draw[{postaction=decorate}] (0.8,1.5) to node[above] {\small $i$} (2,1.5);
			\draw[{postaction=decorate}] (3.2,1.5) to node[above] {\small $l$} (2,1.5);
			\draw[{postaction=decorate}] (2,0.5) to node[right] {\small $n$} (2,1.5);
			\draw[fill=LightGray,rounded corners,LightGray] (-0.2,0) rectangle (0.8,2);
			\draw[fill=LightGray,rounded corners,LightGray] (3.2,0) rectangle (4.2,2);
			\end{tikzpicture}
			\right).
		\end{equation}
		To translate these diagrams into ones that do not contain horizontal lines, we have to distinguish two different cases (this is also the reason for rotating the diagrams in the following), namely 
		\begin{align}
			\Phi\left(
			\begin{tikzpicture}[baseline={([yshift=-3pt]current bounding box.center)}, decoration={markings,mark=at position .6 with {\arrow[>=stealth]{>}}},scale=0.5]
			\draw[fill=LightGray,rounded corners,LightGray] (0,-0.2) rectangle (2,0.8);
			\draw[fill=LightGray,rounded corners,LightGray] (0,3.2) rectangle (2,4.2);
			\draw[{postaction=decorate}] (0.5,0.8) to node[left] {\small $i$} (0.5,2.5);
			\draw[{postaction=decorate}] (0.5,2.5) to node[left] {\small $k$} (0.5,3.2);
			\draw[{postaction=decorate}] (1.5,0.8) to node[right] {\small $j$} (1.5,1.5);
			\draw[{postaction=decorate}] (1.5,1.5) to node[right] {\small $l$} (1.5,3.2);
			\draw[{postaction=decorate}] (1.5,1.5) to node[above] {\small $m$} (0.5,2.5);
			\end{tikzpicture}\right) &=\sum_n \left(G_{ij}^{kl}\right)_{mn}\ \Phi\left(
			\begin{tikzpicture}[baseline={([yshift=-3pt]current bounding box.center)}, decoration={markings,mark=at position .6 with {\arrow[>=stealth]{>}}},scale=0.5]
			\draw[fill=LightGray,rounded corners,LightGray] (0,-0.2) rectangle (2,0.8);
			\draw[fill=LightGray,rounded corners,LightGray] (0,3.2) rectangle (2,4.2);
			\draw[{postaction=decorate}] (0.5,0.8) to node[left] {\small $i$} (1,1.5);
			\draw[{postaction=decorate}] (1.5,0.8) to node[right] {\small $j\ $} (1,1.5);
			\draw[{postaction=decorate}] (1,1.5) to node[right] {\small $n$} (1,2.5);
			\draw[{postaction=decorate}] (1,2.5) to node[left] {\small $\ k$} (0.5,3.2);
			\draw[{postaction=decorate}] (1,2.5) to node[right] {\small $l$} (1.5,3.2);
			\end{tikzpicture}
			\right)\\
			\Phi\left(
			\begin{tikzpicture}[baseline={([yshift=-3pt]current bounding box.center)}, decoration={markings,mark=at position .6 with {\arrow[>=stealth]{>}}},scale=0.5]
			\draw[fill=LightGray,rounded corners,LightGray] (0,-0.2) rectangle (2,0.8);
			\draw[fill=LightGray,rounded corners,LightGray] (0,3.2) rectangle (2,4.2);
			\draw[{postaction=decorate}] (0.5,0.8) to node[left] {\small $i$} (0.5,1.5);
			\draw[{postaction=decorate}] (0.5,1.5) to node[left] {\small $k$} (0.5,3.2);
			\draw[{postaction=decorate}] (1.5,0.8) to node[right] {\small $j$} (1.5,2.5);
			\draw[{postaction=decorate}] (1.5,2.5) to node[right] {\small $l$} (1.5,3.2);
			\draw[{postaction=decorate}] (0.5,1.5) to node[above] {\small $m$} (1.5,2.5);
			\end{tikzpicture}\right) &=\sum_n \left(H_{ij}^{kl}\right)_{mn}\ \Phi\left(
			\begin{tikzpicture}[baseline={([yshift=-3pt]current bounding box.center)}, decoration={markings,mark=at position .6 with {\arrow[>=stealth]{>}}},scale=0.5]
			\draw[fill=LightGray,rounded corners,LightGray] (0,-0.2) rectangle (2,0.8);
			\draw[fill=LightGray,rounded corners,LightGray] (0,3.2) rectangle (2,4.2);
			\draw[{postaction=decorate}] (0.5,0.8) to node[left] {\small $i$} (1,1.5);
			\draw[{postaction=decorate}] (1.5,0.8) to node[right] {\small $j\ $} (1,1.5);
			\draw[{postaction=decorate}] (1,1.5) to node[right] {\small $n$} (1,2.5);
			\draw[{postaction=decorate}] (1,2.5) to node[left] {\small $\ k$} (0.5,3.2);
			\draw[{postaction=decorate}] (1,2.5) to node[right] {\small $l$} (1.5,3.2);
			\end{tikzpicture}
			\right)
		\end{align}
		The operators $H$ and $G$ are expressed in terms of $F$-symbols and quantum dimensions:
		\begin{align}
			\left(G_{ij}^{kl}\right)_{mn}&=\sqrt{\frac{d_m d_n}{d_j d_k}} \overline{\left(F_n^{iml}\right)}_{kj}\\
			\left(H_{ij}^{kl}\right)_{mn}&=\sqrt{\frac{d_m d_n}{d_i d_l}} \left(F_n^{kmj}\right)_{il}.
		\end{align}
		In the following, we show explicitly how the operator $H$ is computed. $G$ can then be determined analogously.
		\begin{align}
			\begin{tikzpicture}[scale=0.7,decoration={markings,mark=at position .55 with {\arrow[>=stealth]{>}}},baseline=(current bounding box.center)]
			\node (a) at (0,-0.25) {\small $i$};
			\node (c) at (0,2.25) {\small $k$};
			\node (b) at (1,-0.25) {\small $j$};
			\node (d) at (1,2.25) {\small $l$};
			\draw[{postaction=decorate}] (0,0) -- (0,0.6);
			\draw[{postaction=decorate}] (0,0.6) -- (0,2);
			\draw[{postaction=decorate}] (1,0) -- (1,1.4);
			\draw[{postaction=decorate}] (1,1.4) -- (1,2);
			\draw[{postaction=decorate}] (0,0.6) -- node[above] {$m$} (1,1.4);
			\end{tikzpicture}
			&= \sum_n \sqrt{\frac{d_n}{d_i d_j}}\ 
			\begin{tikzpicture}[scale=0.7,decoration={markings,mark=at position .55 with {\arrow[>=stealth]{>}}},baseline=(current bounding box.center)]
			\node (a) at (-0.25,0.1) {\small $i$};
			\node (c) at (0,2) {\small $k$};
			\node (b) at (1.25,0.1) {\small $j$};
			\node (d) at (1,2) {\small $l$};
			\draw[{postaction=decorate}] (0,0.6) -- (0,1.75);
			\draw[{postaction=decorate}] (0.5,-0.25) -- (1,0.25);
			\draw (1,0.25) -- (1,1.4);
			\draw[{postaction=decorate}] (1,1.4) -- (1,1.75);
			\draw[{postaction=decorate}] (0,0.6) -- node[above] {\small $m$} (1,1.4);
			\draw[{postaction=decorate}] (0.5,-0.25) -- (0,0.25);
			\draw (0,0.25) -- (0,0.6);
			\draw[{postaction=decorate}] (0.5,-0.75) to node[right] {\small $n$} (0.5,-0.25);
			\draw[{postaction=decorate}] (0,-1.25) -- (0.5,-0.75);
			\draw[{postaction=decorate}] (1,-1.25) -- (0.5,-0.75);
			\node at (-0.25,-1.25) {\small $i$};
			\node at (1.25,-1.25) {\small $j$};
			\end{tikzpicture}\\
			&=\sum_{n,\alpha} \sqrt{\frac{d_n}{d_i d_j}} \left(F_n^{kmj}\right)_{i\alpha}
			\begin{tikzpicture}[scale=0.7,decoration={markings,mark=at position .61 with {\arrow[>=stealth]{>}}},baseline=(current bounding box.center)]
			\node (a) at (-0.25,0.1) {\small $i$};
			\node (c) at (0,2) {\small $k$};
			\node (b) at (1.25,0.1) {\small $\alpha$};
			\node (d) at (1,2) {\small $l$};
			\draw[{postaction=decorate}] (0,0.6) -- (0,1.75);
			\draw[{postaction=decorate}] (0.5,-0.25) -- (1,0.25);
			\draw (1,0.25) -- (1,0.6);
			\draw[{postaction=decorate}] (1,1.4) -- (1,1.75);
			\draw[{postaction=decorate}] (1,1.4) -- (1,1.75);
			\draw[{postaction=decorate}] (1,0.6) -- (0.6,1);
			\draw (0.6,1) -- (1,1.4);
			\draw[{postaction=decorate}] (1,0.6) -- (1.4,1);
			\draw (1.4,1) -- (1,1.4);
			\draw[{postaction=decorate}] (0.5,-0.25) -- (0,0.25);
			\draw (0,0.25) -- (0,0.6);
			\draw[{postaction=decorate}] (0.5,-0.75) to node[right] {\small $n$} (0.5,-0.25);
			\draw[{postaction=decorate}] (0,-1.25) -- (0.5,-0.75);
			\draw[{postaction=decorate}] (1,-1.25) -- (0.5,-0.75);
			\node at (-0.25,-1.25) {\small $i$};
			\node at (1.25,-1.25) {\small $j$};
			\node at (1.6,1) {\small $j$};
			\node at (0.4,1) {\small $m$};
			\end{tikzpicture}\\
			&=\sum_{n,\alpha} \sqrt{\frac{d_n}{d_i d_j}} \left(F_n^{kmj}\right)_{i\alpha}\sqrt{\frac{d_m d_j}{d_l}} \delta_{l,\alpha}
			\begin{tikzpicture}[scale=0.7,decoration={markings,mark=at position .5 with {\arrow[>=stealth]{>}}},baseline=(current bounding box.center)]
			\draw[{postaction=decorate}] (0.5,-0.25) -- (0,0.25);
			\draw[{postaction=decorate}] (0.5,-0.25) -- (1,0.25);
			\draw[{postaction=decorate}] (0.5,-0.75) to node[right] {\small $n$} (0.5,-0.25);
			\draw[{postaction=decorate}] (0,-1.25) -- (0.5,-0.75);
			\draw[{postaction=decorate}] (1,-1.25) -- (0.5,-0.75);
			\node at (0,-1.5) {\small $i$};
			\node at (1,-1.5) {\small $j$};
			\node at (0,0.5) {\small $k$};
			\node at (1,0.5) {\small $l$};
			\end{tikzpicture}\\
			&=\sum_n \sqrt{\frac{d_m d_n}{d_i d_l}} \left(F_n^{kmj}\right)_{il}
			\begin{tikzpicture}[scale=0.7,decoration={markings,mark=at position .5 with {\arrow[>=stealth]{>}}},baseline=(current bounding box.center)]
			\draw[{postaction=decorate}] (0.5,-0.25) -- (0,0.25);
			\draw[{postaction=decorate}] (0.5,-0.25) -- (1,0.25);
			\draw[{postaction=decorate}] (0.5,-0.75) to node[right] {\small $n$} (0.5,-0.25);
			\draw[{postaction=decorate}] (0,-1.25) -- (0.5,-0.75);
			\draw[{postaction=decorate}] (1,-1.25) -- (0.5,-0.75);
			\node at (0,-1.5) {\small $i$};
			\node at (1,-1.5) {\small $j$};
			\node at (0,0.5) {\small $k$};
			\node at (1,0.5) {\small $l$};
			\end{tikzpicture}
		\end{align}
		Here, we have used the completeness relation in the fist step and an $F$-move and the bigon relation afterwards. Note that imposing tetrahedral symmetry on the $F$-symbols implies $(G_{ij}^{kl})_{mn}=(H_{ij}^{kl})_{mn}$ \cite{Hong09}.

		\section{Properties of the Hamiltonian}
		\label{app:Hamprops}
		
		The Hamiltonian we have constructed in the main paper has a variety of properties which are necessary for it to be exactly solvable: It is a Hermitian operator which is projector-valued and, furthermore, it commutes when applied to different vertices and plaquettes. Here, we prove each of these properties.
		
		\subsection{Hermicity}
		\label{sec:Hermicity}
		
		Being Hermitian is a crucial property of a Hamiltonian to describe an actual physical system. In this Subsection we show that the Hamiltonian constructed in this paper possess this attribute.
		
		Obviously, the operator $Q_\mathbf{v}$ is Hermitian. Therefore, it remains to show that $B_\mathbf{p}$ is a Hermitian operator, too. For this, recall that $B_\mathbf{p}$ maps an initial string-net configuration $\vec{p}=\{g,h,i,j,k,l\}$ to the final configuration $\vec{p}\,'=\{g',h',i',j',k',l'\}$ via
		\begin{equation}
			\begin{tikzpicture}[scale=0.6,baseline={([yshift=-0.1cm]current bounding box.center)}]
			\draw[->-] (90:1) -- (150:1);
			\draw[->-] (150:1) -- (210:1);
			\draw[->-] (210:1) -- (270:1);
			\draw[->-] (270:1) -- (330:1);
			\draw[->-] (330:1) -- (30:1);
			\draw[->-] (30:1) -- (90:1);
			\foreach \a in {30,90,150,210,270,330} \draw[-<-] (\a:1) -- (\a:1.5);
			\node at (0,0) {$\vec{p}$};
			\end{tikzpicture}\mapsto\sum_{\vec{p}\,'}C(\vec{p},\vec{p}\,')
			\begin{tikzpicture}[scale=0.6,baseline={([yshift=-0.1cm]current bounding box.center)}]
			\draw[->-] (90:1) -- (150:1);
			\draw[->-] (150:1) -- (210:1);
			\draw[->-] (210:1) -- (270:1);
			\draw[->-] (270:1) -- (330:1);
			\draw[->-] (330:1) -- (30:1);
			\draw[->-] (30:1) -- (90:1);
			\foreach \a in {30,90,150,210,270,330} \draw[-<-] (\a:1) -- (\a:1.5);
			\node at (0,0) {$\vec{p}\,'$};
			\end{tikzpicture},
		\end{equation}
		where the coefficients $C(\vec{p},\vec{p}\,')$ are the respective matrix elements \eqref{Bps} of the transformation performed by $B_\mathbf{p}$. In order to show that the operator $B_\mathbf{p}$ is Hermitian we need to show that
		\begin{equation}
			C(\vec{p},\vec{p}\,')=\overline{C(\vec{p}\,',\vec{p})}\label{eq:Toshow}.
		\end{equation}
		This can be verified using the graphical calculus of UFCs (see Appendix~\ref{app:UFCgraphcalc}). Before we start proving Eq.~\eqref{eq:Toshow}, we would like to give an intuition on how to graphically compute the matrix elements of a natural transformation by reviewing the example of the $F$-symbol. In terms of its matrix elements, the action of the transformation $F$ is given by
		\begin{equation}
			\begin{tikzpicture}[baseline=(current bounding box.center),scale=0.6]
			\draw (0,0.5) -- (0,1);
			\draw (0,1) -- (-0.5,1.5);
			\draw (-0.5,1.5) -- (-1,2);
			\draw (-0.5,1.5) -- (0,2);
			\draw (0,1) -- (0.5,1.5);
			\draw (0.5,1.5) -- (1,2);
			\node at (0,0.25) {\small $i$};
			\node at (-1,2.25) {\small $j$};
			\node at (0,2.25) {\small $k$};
			\node at (1,2.25) {\small $l$};
			\node at (-0.5,1.1) {\small $m$};
			\end{tikzpicture}=\sum_n\left(F_{i}^{jkl}\right)_{mn}
			\begin{tikzpicture}[baseline=(current bounding box.center),scale=0.6]
			\draw (0,0.5) -- (0,1);
			\draw (0,1) -- (-0.5,1.5);
			\draw (-0.5,1.5) -- (-1,2);
			\draw (0.5,1.5) -- (0,2);
			\draw (0,1) -- (0.5,1.5);
			\draw (0.5,1.5) -- (1,2);
			\node at (0,0.25) {\small $i$};
			\node at (-1,2.25) {\small $j$};
			\node at (0,2.25) {\small $k$};
			\node at (1,2.25) {\small $l$};
			\node at (0.5,1.1) {\small $n$};
			\end{tikzpicture}.
		\end{equation}
		If we want to compute a fixed matrix element $\left(F_{i}^{jkl}\right)_{mn}$ of $F$ we can do this by calculating the scalar product of the initial string diagram with the final string diagram, which is defined by composing (i.e., vertically stacking) the string-diagrams and tracing over all uncontracted labels. For this calculation, it is necessary to have an orthonormal basis of tree diagrams, i.e.,
		\begin{equation}\left\{\frac{1}{\left(d_i d_j d_k d_l\right)^\frac{1}{4}}
			\begin{tikzpicture}[baseline=(current bounding box.center),scale=0.6]
			\draw (0,0.5) -- (0,1);
			\draw (0,1) -- (-0.5,1.5);
			\draw (-0.5,1.5) -- (-1,2);
			\draw (0.5,1.5) -- (0,2);
			\draw (0,1) -- (0.5,1.5);
			\draw (0.5,1.5) -- (1,2);
			\node at (0,0.25) {\small $i$};
			\node at (-1,2.25) {\small $j$};
			\node at (0,2.25) {\small $k$};
			\node at (1,2.25) {\small $l$};
			\node at (0.5,1.1) {\small $n$};
			\end{tikzpicture}\right\}.
		\end{equation}
		Note that the normalization factor $\frac{1}{\left(d_i d_j d_k d_l\right)^\frac{1}{4}}$ comes from the normalization of vertices in \eqref{eq:norm} with an additional factor of $1/\sqrt{d_i}$ in order to make the diagram normalized with respect to the trace. The calculation of the matrix element $\left(F_{i}^{jkl}\right)_{mn}$ then works as follows: Due to the mirror symmetry of the $F$-symbols we receive
		\begin{align}\frac{1}{\sqrt{d_i d_j d_k d_l}}
			\Bigg\langle
			\begin{tikzpicture}[baseline=(current bounding box.center),scale=0.6]
			\draw (0,0.5) -- (0,1);
			\draw (0,1) -- (-0.5,1.5);
			\draw (-0.5,1.5) -- (-1,2);
			\draw (0.5,1.5) -- (0,2);
			\draw (0,1) -- (0.5,1.5);
			\draw (0.5,1.5) -- (1,2);
			\node at (0,0.25) {\small $i$};
			\node at (-1,2.25) {\small $j$};
			\node at (0,2.25) {\small $k$};
			\node at (1,2.25) {\small $l$};
			\node at (0.5,1.1) {\small $n$};
			\end{tikzpicture}
			\Bigg|
			\begin{tikzpicture}[baseline=(current bounding box.center),scale=0.6]
			\draw (0,0.5) -- (0,1);
			\draw (0,1) -- (-0.5,1.5);
			\draw (-0.5,1.5) -- (-1,2);
			\draw (-0.5,1.5) -- (0,2);
			\draw (0,1) -- (0.5,1.5);
			\draw (0.5,1.5) -- (1,2);
			\node at (0,0.25) {\small $i$};
			\node at (-1,2.25) {\small $j$};
			\node at (0,2.25) {\small $k$};
			\node at (1,2.25) {\small $l$};
			\node at (-0.5,1.1) {\small $m$};
			\end{tikzpicture}
			\Bigg\rangle
			&=\frac{1}{\sqrt{d_i d_j d_k d_l}}
			\begin{tikzpicture}[baseline=(j.base),scale=0.6]
			\clip(-1.75,-1) rectangle (2,4.5);
			\draw (0,0.5) -- (0,1);
			\draw (0,1) -- (-0.5,1.5);
			\coordinate (j) at (-1,2);
			\draw (-0.5,1.5) -- (j);
			\draw (-0.5,1.5) -- node[above] {\small $k$} (0,2);
			\draw (0,1) -- (0.5,1.5);
			\draw (0.5,1.5) -- (1,2);
			\node at (-0.5,1.1) {\small $m$};
			\node at (0.5,1.1) {\small $l$};
			\draw (0,0.5) arc (-180:0:0.75cm);
			\rotatebox[]{180}{
				\begin{scope}[yshift=-4cm]
				\draw (0,0.5) -- (0,1);
				\draw (0,1) -- (-0.5,1.5);
				\draw (-0.5,1.5) -- (-1,2);
				\draw (-0.5,1.5) -- (0,2);
				\draw (0,1) -- (0.5,1.5);
				\draw (0.5,1.5) -- (1,2);
				\node at (-0.5,1.1) {\rotatebox[]{180}{\small $n$}};
				\node at (0.5,1.1) {\rotatebox[]{180}{\small $j$}};
				\draw (0,0.5) arc (0:-180:0.75cm);
				\end{scope}
			}
			\draw (1.5,0.5) -- (1.5,3.5);
			\node at (-0.3,0.3) {\small $i$};
			\node at (-0.3,3.7) {\small $i$};
			\end{tikzpicture}\\
			&=
			\frac{1}{\sqrt{d_i d_j d_k d_l}} \sum_{m'}\left(F_{i}^{jkl}\right)_{mm'}
			\begin{tikzpicture}[baseline=(j.base),scale=0.6]
			\clip(-1.75,-1) rectangle (2,4.5);
			\draw (0,0.5) -- (0,1);
			\draw (0,1) -- (-0.5,1.5);
			\draw (-0.5,1.5) -- node[left] {\small $j$} (-1,2);
			\draw (0.5,1.5) -- node[left] {\small $k$} (0,2);
			\draw (0,1) -- (0.5,1.5);
			\draw (0.5,1.5) -- node[right] {\small $l$} (1,2);
			\node at (0.5,1.1) {\small $m'$};
			\draw (0,0.5) arc (-180:0:0.75cm);
			\rotatebox[]{180}{
				\begin{scope}[yshift=-4cm]
				\draw (0,0.5) -- (0,1);
				\draw (0,1) -- (-0.5,1.5);
				\draw (-0.5,1.5) -- (-1,2);
				\draw (-0.5,1.5) -- (0,2);
				\draw (0,1) -- (0.5,1.5);
				\draw (0.5,1.5) -- (1,2);
				\node at (-0.5,1.1) {\rotatebox[]{180}{\small $n$}};
				\draw (0,0.5) arc (0:-180:0.75cm);
				\end{scope}
			}
			\draw (1.5,0.5) -- (1.5,3.5);
			\node at (-0.3,0.3) {\small $i$};
			\node at (-0.3,3.7) {\small $i$};
			\end{tikzpicture}\\
			&=
			\frac{1}{\sqrt{d_i d_j d_k d_l}} \sum_{m'}\left(F_{i}^{jkl}\right)_{mm'} \sqrt{\frac{d_k d_l}{d_n}}\delta_{m',n}
			\begin{tikzpicture}[baseline=(j.base),scale=0.6]
			\clip(-1.75,-1) rectangle (2,4.5);
			\draw (0,0.5) -- (0,1);
			\draw (0,1) -- (-0.5,1.5);
			\draw (-0.5,1.5) -- node[left] {\small $j$} (-1,2);
			\draw (0,1) -- (0.5,1.5);
			\draw (0.5,1.5) -- node[right] {\small $n$} (1,2);
			\draw (0,0.5) arc (-180:0:0.75cm);
			\rotatebox[]{180}{
				\begin{scope}[yshift=-4cm]
				\draw (0,0.5) -- (0,1);
				\draw (0,1) -- (-0.5,1.5);
				\draw (-0.5,1.5) -- (-1,2);
				\draw (0,1) -- (0.5,1.5);
				\draw (0.5,1.5) -- (1,2);
				\draw (0,0.5) arc (0:-180:0.75cm);
				\end{scope}
			}
			\draw (1.5,0.5) -- (1.5,3.5);
			\node at (-0.3,0.3) {\small $i$};
			\node at (-0.3,3.7) {\small $i$};
			\end{tikzpicture}\\
			&=
			\frac{1}{\sqrt{d_i d_j d_k d_l}} \left(F_{i}^{jkl}\right)_{mn} \sqrt{\frac{d_k d_l}{d_n}} \sqrt{\frac{d_j d_n}{d_i}}
			\,\,\begin{tikzpicture}[baseline=(current bounding box.center)]
			\draw (0,0) circle (0.5cm);
			\node (i) at (-0.7,0) {\small $i$};
			\end{tikzpicture}\\
			&=\frac{1}{\sqrt{d_i d_j d_k d_l}}  \left(F_{i}^{jkl}\right)_{mn} \sqrt{\frac{d_k d_l}{d_n}} \sqrt{\frac{d_j d_n}{d_i}} d_i\\
			&=\left(F_{i}^{jkl}\right)_{mn}.
		\end{align}
		Note that we have taken the right trace for the $i$-string. Due to sphericality of the UFC, left and right trace coincide.
		
		The same calculation can be performed to receive the matrix element $C(\vec{p},\vec{p}\,')$ of the operator $B_\mathbf{p}$. This gives
		\begin{equation}
			C(\vec{p},\vec{p}\,')=
			\begin{tikzpicture}[scale=0.6,baseline={([yshift=0.25cm]center.base)}]
			\clip(-2,-3) rectangle (3.5,6);
			\coordinate (center) at (90:1);
			\draw (center) -- (150:1);
			\draw (150:1) -- (210:1);
			\draw (210:1) -- (270:1);
			\draw (270:1) -- (330:1);
			\draw (330:1) -- (30:1);
			\draw (30:1) -- (center);
			\foreach \a in {30,90,150,210,270,330} \draw (\a:1) -- (\a:1.5);
			\node at (0,0) {$\vec{p}$};
			\rotatebox[]{180}{
				\begin{scope}[yshift=-3cm]
				\draw (90:1) -- (150:1);
				\draw (150:1) -- (210:1);
				\draw (210:1) -- (270:1);
				\draw (270:1) -- (330:1);
				\draw (330:1) -- (30:1);
				\draw (30:1) -- (90:1);
				\foreach \a in {30,90,150,210,270,330} \draw (\a:1) -- (\a:1.5);
				\node at (0,0) {\rotatebox[]{180}{$\vec{p}\,'$}};
				\end{scope}
			}
			\begin{scope}[xshift=1.29904cm,yshift=1.5cm]
			\draw (0,-0.75) to [bend right] (0,0.75);
			\end{scope}
			\begin{scope}[xshift=-1.29904cm,yshift=1.5cm]
			\draw (0,-0.75) to [bend left] (0,0.75);
			\end{scope}
			\begin{scope}[xshift=1.29904cm,yshift=-1.5cm]
			\draw (0,0.75) to [bend right] (0,5.25);
			\end{scope}
			\begin{scope}[xshift=-1.29904cm,yshift=-1.5cm]
			\draw (0,0.75) to [bend left] (0,5.25);
			\end{scope}
			\draw (3,-1.5) arc (0:-180:1.5cm);
			\draw (3,4.5) arc (0:180:1.5cm);
			\draw (3,-1.5) to (3,4.5);
			\end{tikzpicture}\label{eq:Cpps}.
		\end{equation}
		Due to the graphical calculus, there is an obvious way to compute the matrix elements $C(\vec{p}\,',\vec{p})$ of the opposite operation which transforms the plaquette $\vec{p}\,'$ into $\vec{p}$. This is, just interchanging the plaquettes
		\begin{equation}
			C(\vec{p}\,',\vec{p})=
			\begin{tikzpicture}[scale=0.6,baseline={([yshift=0.25cm]center.base)}]
			\clip(-2,-3) rectangle (3.5,6);
			\coordinate (center) at (90:1);
			\draw (center) -- (150:1);
			\draw (150:1) -- (210:1);
			\draw (210:1) -- (270:1);
			\draw (270:1) -- (330:1);
			\draw (330:1) -- (30:1);
			\draw (30:1) -- (center);
			\foreach \a in {30,90,150,210,270,330} \draw (\a:1) -- (\a:1.5);
			\node at (0,0) {$\vec{p}\,'$};
			\rotatebox[]{180}{
				\begin{scope}[yshift=-3cm]
				\draw (90:1) -- (150:1);
				\draw (150:1) -- (210:1);
				\draw (210:1) -- (270:1);
				\draw (270:1) -- (330:1);
				\draw (330:1) -- (30:1);
				\draw (30:1) -- (90:1);
				\foreach \a in {30,90,150,210,270,330} \draw (\a:1) -- (\a:1.5);
				\node at (0,0) {\rotatebox[]{180}{$\vec{p}$}};
				\end{scope}
			}
			\begin{scope}[xshift=1.29904cm,yshift=1.5cm]
			\draw (0,-0.75) to [bend right] (0,0.75);
			\end{scope}
			\begin{scope}[xshift=-1.29904cm,yshift=1.5cm]
			\draw (0,-0.75) to [bend left] (0,0.75);
			\end{scope}
			\begin{scope}[xshift=1.29904cm,yshift=-1.5cm]
			\draw (0,0.75) to [bend right] (0,5.25);
			\end{scope}
			\begin{scope}[xshift=-1.29904cm,yshift=-1.5cm]
			\draw (0,0.75) to [bend left] (0,5.25);
			\end{scope}
			\draw (3,-1.5) arc (0:-180:1.5cm);
			\draw (3,4.5) arc (0:180:1.5cm);
			\draw (3,-1.5) to (3,4.5);
			\end{tikzpicture}.
		\end{equation}
		As a consequence of mirror symmetry \eqref{eq:mirror}, complex conjugation of the $C(\vec{p}\,',\vec{p})$ is then given by horizontal reflection of the entire string-diagram:
		\begin{equation}
			\overline{C(\vec{p}\,',\vec{p})}=
			\begin{tikzpicture}[scale=0.6,baseline={([yshift=0.25cm]center.base)}]
			\clip(-2,-3) rectangle (3.5,6);
			\coordinate (center) at (90:1);
			\draw (center) -- (150:1);
			\draw (150:1) -- (210:1);
			\draw (210:1) -- (270:1);
			\draw (270:1) -- (330:1);
			\draw (330:1) -- (30:1);
			\draw (30:1) -- (center);
			\foreach \a in {30,90,150,210,270,330} \draw (\a:1) -- (\a:1.5);
			\node at (0,0) {$\vec{p}$};
			\rotatebox[]{180}{
				\begin{scope}[yshift=-3cm]
				\draw (90:1) -- (150:1);
				\draw (150:1) -- (210:1);
				\draw (210:1) -- (270:1);
				\draw (270:1) -- (330:1);
				\draw (330:1) -- (30:1);
				\draw (30:1) -- (90:1);
				\foreach \a in {30,90,150,210,270,330} \draw (\a:1) -- (\a:1.5);
				\node at (0,0) {\rotatebox[]{180}{$\vec{p}\,'$}};
				\end{scope}
			}
			\begin{scope}[xshift=1.29904cm,yshift=1.5cm]
			\draw (0,-0.75) to [bend right] (0,0.75);
			\end{scope}
			\begin{scope}[xshift=-1.29904cm,yshift=1.5cm]
			\draw (0,-0.75) to [bend left] (0,0.75);
			\end{scope}
			\begin{scope}[xshift=1.29904cm,yshift=-1.5cm]
			\draw (0,0.75) to [bend right] (0,5.25);
			\end{scope}
			\begin{scope}[xshift=-1.29904cm,yshift=-1.5cm]
			\draw (0,0.75) to [bend left] (0,5.25);
			\end{scope}
			\draw (3,-1.5) arc (0:-180:1.5cm);
			\draw (3,4.5) arc (0:180:1.5cm);
			\draw (3,-1.5) to (3,4.5);
			\end{tikzpicture}\label{eq:Cppsbar}.
		\end{equation}
		Comparing Eq.~\eqref{eq:Cpps} with Eq.~\eqref{eq:Cppsbar} gives the desired expression in Eq.~\eqref{eq:Toshow}. Hence, the operator $B_\mathbf{p}$ is indeed Hermitian.
		
		\subsection{Projector}
		\label{sec:Projector}
		
		In the original Levin-Wen model, the $B_\mathbf{p}$ and $Q_\mathbf{v}$ are projectors (see \cite{LW05}, Appendix~C). Here, we show that the same properties also hold for our construction. Obviously, $Q_\mathbf{v}$ is projector-valued. For $B_\mathbf{p}$, we simultaneously act with the operators $B_\mathbf{p}=\sum_t a_t B_\mathbf{p}^t$ and $B_\mathbf{p}=\sum_s a_s B_\mathbf{p}^s$ on the plaquette $\mathbf{p}$. The corresponding joint operator is $B_\mathbf{p}^2=\sum_{s,t} a_s a_t B_\mathbf{p}^s B_\mathbf{p}^t$. Graphically, $B_\mathbf{p}^2$ acts by adding type $s$ and type $t$ loops to the plaquette and summing over $s,t$. For the sake of clarity, we omit drawing the plaquette itself in the following calculation and only fuse these two loops together. However, we implicitly mean the $B_\mathbf{p}^2$ action on a plaquette (see Fig.~\ref{fig:proj}).
		
		\begin{align}
			&B_\mathbf{p}^2
			=
			\sum_{s,t} \frac{d_s d_t}{D^4}
			\begin{tikzpicture} [rotate=90, scale=1.3,baseline=(current bounding box.center), decoration={markings,mark=at position .5 with {\arrow[>=stealth]{<}}}, scale=1]
			\draw[{postaction=decorate}] (0,2.73205) -- node[above] {\small{$s$}} (0.5,1.86603); 
			\draw[{postaction=decorate}] (-1,2.73205) -- node[left] {\small{$s$}} (0,2.73205); 
			\draw[{postaction=decorate}] (-1.5,1.86603) -- node[below] {\small{$s$}} (-1,2.73205); 
			\draw[{postaction=decorate}] (-1,1) -- node[below] {\small{$s$}} (-1.5,1.86603); 
			\draw[{postaction=decorate}] (0,1) -- node[right] {\small{$s$}} (-1,1); 
			\draw[{postaction=decorate}] (0.5,1.86603) -- node[above] {\small{$s$}} (0,1); 
			\draw[{postaction=decorate}] (-0.8,2.43205) -- node[right] {\small{$t$}} (-0.2,2.43205); 
			\draw[{postaction=decorate}] (-0.2,1.3) -- node[left] {\small{$t$}} (-0.8,1.3); 
			\draw[{postaction=decorate}] (0.13,1.86603) -- node[below] {\small{$t$}} (-0.2,1.3); 
			\draw[{postaction=decorate}] (-0.2,2.43205) -- node[below] {\small{$t$}} (0.13,1.86603); 
			\draw[{postaction=decorate}] (-1.13,1.86603) -- node[above] {\small{$t$}} (-0.8,2.43205); 
			\draw[{postaction=decorate}] (-0.8,1.3) -- node[above] {\small{$t$}} (-1.13,1.86603); 
			\draw[dotted] (0.13,1.86603) -- (0.30492,2.16603);
			\draw[dotted] (-1.13,1.86603) -- (-1.30492,1.56603);
			\end{tikzpicture}\\
			&=
			\sum_{\substack{s,t\\ \alpha,\beta}}\frac{d_s d_t}{D^4} \sqrt{\frac{d_\alpha}{d_s d_t}}\sqrt{\frac{d_{\beta^*}}{d_{s^*} d_{t^*}}} N_{st}^\alpha N_{s^*t^*}^{\beta^*}
			\begin{tikzpicture} [rotate=90, scale=1.3,baseline=(current bounding box.center), decoration={markings,mark=at position .5 with {\arrow[>=stealth]{<}}}]
			\draw[{postaction=decorate}] (0,2.73205) -- node[above] {\small{$s$}} (0.5,1.86603); 
			\draw[{postaction=decorate}] (-1.5,1.86603) -- node[below] {\small{$s$}} (-1,2.73205); 
			\draw[{postaction=decorate}] (-1,1) -- node[below] {\small{$s$}} (-1.5,1.86603); 
			\draw[{postaction=decorate}] (0.5,1.86603) -- node[above] {\small{$s$}} (0,1); 
			\draw[{postaction=decorate}] (0.13,1.86603) -- node[below] {\small{$t$}} (-0.2,1.3); 
			\draw[{postaction=decorate}] (-0.2,2.43205) -- node[below] {\small{$t$}} (0.13,1.86603); 
			\draw[{postaction=decorate}] (-1.13,1.86603) -- node[above] {\small{$t$}} (-0.8,2.43205); 
			\draw[{postaction=decorate}] (-0.8,1.3) -- node[above] {\small{$t$}} (-1.13,1.86603); 
			\draw[dotted] (0.13,1.86603) -- (0.30492,2.16603);
			\draw[dotted] (-1.13,1.86603) -- (-1.30492,1.56603);
			\coordinate (lsup) at (-0.6,2.58205);
			\coordinate (lsdown) at (-0.4,2.58205);
			\draw[{postaction=decorate}] (lsup) -- node[left] {\small{$\beta$}} (lsdown);
			\coordinate (slcdown) at (-0.7,2.43205);
			\coordinate (lcdown) at (-0.7,2.73205);
			\coordinate (slcup) at (-0.3,2.43205);
			\coordinate (lcup) at (-0.3,2.73205);
			\draw (slcdown) -- (lsup);
			\draw (lcdown) -- (lsup);
			\draw (lsdown) -- (slcup);
			\draw (lsdown) -- (lcup);
			\draw[{postaction=decorate}] (lcup) -- (0,2.73205);
			\draw[{postaction=decorate}] (-1,2.73205) -- (lcdown);
			\draw (slcdown) -- (-0.8,2.43205);
			\draw (slcup) -- (-0.2,2.43205);
			\coordinate (isup) at (-0.6,1.15);
			\coordinate (isdown) at (-0.4,1.15);
			\draw[{postaction=decorate}] (isdown) -- node[right] {\small{$\alpha$}} (isup);
			\coordinate (sicdown) at (-0.7,1);
			\coordinate (icdown) at (-0.7,1.3);
			\coordinate (sicup) at (-0.3,1);
			\coordinate (icup) at (-0.3,1.3);
			\draw (isup) -- (sicdown);
			\draw (isup) -- (icdown);
			\draw (sicup) -- (isdown);
			\draw (icup) -- (isdown);
			\draw[{postaction=decorate}] (0,1) -- (sicup);
			\draw[{postaction=decorate}] (sicdown) -- (-1,1);	
			\draw (icup) -- (-0.2,1.3);
			\draw (icdown) -- (-0.8,1.3);
			\end{tikzpicture}\label{P1}
			\\\begin{split}&=
				\sum_{\substack{s,t\\ \alpha,\beta}}\frac{\sqrt{d_\alpha d_{\beta^*}}}{D^4} N_{st}^\alpha N_{s^*t^*}^{\beta^*} \delta_{\alpha,\beta} \sqrt{\frac{d_{s^*}d_{t^*}}{d_{\beta^*}}} \sqrt{\frac{d_t d_s}{d_\alpha}} \left(F_{s^*}^{s^*t^*t}\right)_{\beta^*1} \overline{\left(F_1^{\beta^*ts}\right)}_{s^*\alpha}\\ 
				&\hspace{40pt}\sqrt{\frac{d_t d_s}{d_\alpha}} \sqrt{\frac{d_{s^*} d_{t^*}}{d_{\beta^*}}} \left(F_s^{t^*ts}\right)_{1\alpha} \overline{\left(F_1^{s^*t^*\alpha}\right)}_{\beta^*s}
				\ \begin{tikzpicture}[rotate=90, scale=0.6,baseline=(current bounding box.center), decoration={markings,mark=at position .5 with {\arrow[>=stealth]{<}}}]
				\draw[{postaction=decorate}] (-0.8,2.43205) -- (-0.2,2.43205); 
				\draw[{postaction=decorate}] (-0.2,1.3) -- (-0.8,1.3); 
				\draw[{postaction=decorate}] (0.13,1.86603) -- (-0.2,1.3); 
				\draw[{postaction=decorate}] (-0.2,2.43205) -- (0.13,1.86603); 
				\draw[{postaction=decorate}] (-1.13,1.86603) -- (-0.8,2.43205); 
				\draw[{postaction=decorate}] (-0.8,1.3) -- (-1.13,1.86603); 
				\node (alpha) at (-0.5,1.86603) {\small{$\alpha$}};
				\end{tikzpicture}\label{P2}\end{split}
			\\&=
			\sum_{s,t,\alpha} \frac{\sqrt{d_{s^*}d_{t^*} d_s d_t}}{D^4} \left(N_{st}^\alpha\right)^2 \left|\left(F_s^{t^*ts}\right)_{1\alpha}\right|^2\ \left|\left(F_1^{s^*t^*\alpha}\right)_{\alpha^*s}\right|^2
			\ \begin{tikzpicture}[rotate=90, scale=0.6,baseline=(current bounding box.center), decoration={markings,mark=at position .5 with {\arrow[>=stealth]{<}}}]
			\draw[{postaction=decorate}] (-0.8,2.43205) -- (-0.2,2.43205); 
			\draw[{postaction=decorate}] (-0.2,1.3) -- (-0.8,1.3); 
			\draw[{postaction=decorate}] (0.13,1.86603) -- (-0.2,1.3); 
			\draw[{postaction=decorate}] (-0.2,2.43205) -- (0.13,1.86603); 
			\draw[{postaction=decorate}] (-1.13,1.86603) -- (-0.8,2.43205); 
			\draw[{postaction=decorate}] (-0.8,1.3) -- (-1.13,1.86603); 
			\node (alpha) at (-0.5,1.86603) {\small{$\alpha$}};
			\end{tikzpicture}\label{P3}
			\\&=
			\sum_{s,t,\alpha} \frac{\sqrt{d_{s^*}d_{t^*} d_s d_t}}{D^4} N_{st}^\alpha \frac{d_\alpha}{d_{t^*} d_s}
			\ \begin{tikzpicture}[rotate=90, scale=0.6,baseline=(current bounding box.center), decoration={markings,mark=at position .5 with {\arrow[>=stealth]{<}}}]
			\draw[{postaction=decorate}] (-0.8,2.43205) -- (-0.2,2.43205); 
			\draw[{postaction=decorate}] (-0.2,1.3) -- (-0.8,1.3); 
			\draw[{postaction=decorate}] (0.13,1.86603) -- (-0.2,1.3); 
			\draw[{postaction=decorate}] (-0.2,2.43205) -- (0.13,1.86603); 
			\draw[{postaction=decorate}] (-1.13,1.86603) -- (-0.8,2.43205); 
			\draw[{postaction=decorate}] (-0.8,1.3) -- (-1.13,1.86603); 
			\node (alpha) at (-0.5,1.86603) {\small{$\alpha$}};
			\end{tikzpicture}\label{P4}
			\\&=
			\sum_{s,t,\alpha} \frac{1}{D^4} d_s d_t N_{t\alpha^*}^{s^*}
			\ \begin{tikzpicture}[rotate=90, scale=0.6,baseline=(current bounding box.center), decoration={markings,mark=at position .5 with {\arrow[>=stealth]{<}}}]
			\draw[{postaction=decorate}] (-0.8,2.43205) -- (-0.2,2.43205); 
			\draw[{postaction=decorate}] (-0.2,1.3) -- (-0.8,1.3); 
			\draw[{postaction=decorate}] (0.13,1.86603) -- (-0.2,1.3); 
			\draw[{postaction=decorate}] (-0.2,2.43205) -- (0.13,1.86603); 
			\draw[{postaction=decorate}] (-1.13,1.86603) -- (-0.8,2.43205); 
			\draw[{postaction=decorate}] (-0.8,1.3) -- (-1.13,1.86603); 
			\node (alpha) at (-0.5,1.86603) {\small{$\alpha$}};
			\end{tikzpicture}\label{P5}
			\\&=
			\frac{1}{D^4} \sum_{t,\alpha} d_t \left(\sum_{s^*} d_{s^*} N_{t\alpha^*}^{s^*}\right)
			\ \begin{tikzpicture}[rotate=90, scale=0.6,baseline=(current bounding box.center), decoration={markings,mark=at position .5 with {\arrow[>=stealth]{<}}}]
			\draw[{postaction=decorate}] (-0.8,2.43205) -- (-0.2,2.43205); 
			\draw[{postaction=decorate}] (-0.2,1.3) -- (-0.8,1.3); 
			\draw[{postaction=decorate}] (0.13,1.86603) -- (-0.2,1.3); 
			\draw[{postaction=decorate}] (-0.2,2.43205) -- (0.13,1.86603); 
			\draw[{postaction=decorate}] (-1.13,1.86603) -- (-0.8,2.43205); 
			\draw[{postaction=decorate}] (-0.8,1.3) -- (-1.13,1.86603); 
			\node (alpha) at (-0.5,1.86603) {\small{$\alpha$}};
			\end{tikzpicture}\label{P6}
			\\&=
			\frac{1}{D^4} \sum_{t,\alpha} d_t d_t d_{\alpha^*}
			\ \begin{tikzpicture}[rotate=90, scale=0.6,baseline=(current bounding box.center), decoration={markings,mark=at position .5 with {\arrow[>=stealth]{<}}}]
			\draw[{postaction=decorate}] (-0.8,2.43205) -- (-0.2,2.43205); 
			\draw[{postaction=decorate}] (-0.2,1.3) -- (-0.8,1.3); 
			\draw[{postaction=decorate}] (0.13,1.86603) -- (-0.2,1.3); 
			\draw[{postaction=decorate}] (-0.2,2.43205) -- (0.13,1.86603); 
			\draw[{postaction=decorate}] (-1.13,1.86603) -- (-0.8,2.43205); 
			\draw[{postaction=decorate}] (-0.8,1.3) -- (-1.13,1.86603); 
			\node (alpha) at (-0.5,1.86603) {\small{$\alpha$}};
			\end{tikzpicture}\label{P7}
			\\&=
			\frac{D^2}{D^4} \sum_{\alpha} d_{\alpha^*}
			\ \begin{tikzpicture}[rotate=90, scale=0.6,baseline=(current bounding box.center), decoration={markings,mark=at position .5 with {\arrow[>=stealth]{<}}}]
			\draw[{postaction=decorate}] (-0.8,2.43205) -- (-0.2,2.43205); 
			\draw[{postaction=decorate}] (-0.2,1.3) -- (-0.8,1.3); 
			\draw[{postaction=decorate}] (0.13,1.86603) -- (-0.2,1.3); 
			\draw[{postaction=decorate}] (-0.2,2.43205) -- (0.13,1.86603); 
			\draw[{postaction=decorate}] (-1.13,1.86603) -- (-0.8,2.43205); 
			\draw[{postaction=decorate}] (-0.8,1.3) -- (-1.13,1.86603); 
			\node (alpha) at (-0.5,1.86603) {\small{$\alpha$}};
			\end{tikzpicture}\label{P8}
			\\&=
			\sum_\alpha \frac{d_\alpha}{D^2}
			\ \begin{tikzpicture}[rotate=90, scale=0.6,baseline=(current bounding box.center), decoration={markings,mark=at position .5 with {\arrow[>=stealth]{<}}}]
			\draw[{postaction=decorate}] (-0.8,2.43205) -- (-0.2,2.43205); 
			\draw[{postaction=decorate}] (-0.2,1.3) -- (-0.8,1.3); 
			\draw[{postaction=decorate}] (0.13,1.86603) -- (-0.2,1.3); 
			\draw[{postaction=decorate}] (-0.2,2.43205) -- (0.13,1.86603); 
			\draw[{postaction=decorate}] (-1.13,1.86603) -- (-0.8,2.43205); 
			\draw[{postaction=decorate}] (-0.8,1.3) -- (-1.13,1.86603); 
			\node (alpha) at (-0.5,1.86603) {\small{$\alpha$}};
			\end{tikzpicture}\ 
			=B_\mathbf{p}\label{P9}
		\end{align}
		\noindent
		Let us explain each step of this somewhat cumbersome calculation in more detail.
		\begin{itemize}
			\item[\eqref{P1}] In the first step, we just insert the definition of $B_\mathbf{p}^2$ and apply two completeness relations \eqref{eq:completeness}. $N_{ij}^k$ denotes the dimension of the fusion vertex $V_{ij}^k$. In the case of multiplicity-free fusion rules, $N_{ij}^k=\delta_{ij}^k$, see \eqref{eq:Nijk}. The factor $N_{st}^\alpha N_{s^*t^*}^{\beta^*}$ is necessary to ensure that the sum only involves valid configurations.
			\item [\eqref{P2}] In order to come up with this equality, we evaluate the diagram from \eqref{P1} vertex-wise in the same way as we did in the computation of the $B_\mathbf{p}^s$ matrix elements by using the relations from \itemone\, and \itemfour. The Kronecker delta comes from the fact that the only possible way to evaluate this diagram is the case where $\alpha=\beta$.
			\item[\eqref{P3}] In this step, we carry out the Kronecker delta, simplify the factor and use the fact that $\left(F_{s^*}^{s^*t^*t}\right)_{\beta^*0}\overline{\left(F_0^{\beta^*ts}\right)}_{s^*\alpha}=\overline{\left(F_s^{t^*ts}\right)_{0\alpha}} \left(F_0^{s^*t^*\alpha}\right)_{\beta^*s}$. This is simply a consequence from the pentagon identity. Another way to check this is by redoing the calculation from \itemone\,\, whilst fusing the unit object to the other side.
			\item[\eqref{P4}] Since we only consider the multiplicity-free case $N_{st}^\alpha\in\left\{0,1\right\}$, we have $\left(N_{st}^\alpha\right)^2=N_{st}^\alpha$. Furthermore, the $F$-symbols involved in \eqref{P3} are all one-dimensional. Hence, we receive $\left|\left(F_s^{t^*ts}\right)_{0\alpha}\right|^2=\frac{d_\alpha}{d_{t^*}d_s}$ and due to unitarity of the fusion category $\left|\left(F_0^{s^*t^*\alpha}\right)_{\alpha^*s}\right|^2=1$. For further details, see \cite{Bonderson2007}.
			\item[\eqref{P5}] Here, we use that $N_{st}^\alpha=N_{t\alpha^*}^{s^*}$. This can be verified by looking at the respective fusion vertices $V_{st}^\alpha$ and $V_{t\alpha^*}^{s^*}$. $V_{st}^\alpha$ can be transformed into $V_{t\alpha^*}^{s^*}$ by bending the legs up/down accordingly
			\begin{equation}
				\begin{tikzpicture}[baseline=(current bounding box.center), decoration={markings,mark=at position .5 with {\arrow[>=stealth]{>}}},scale=0.8]
				\draw[{postaction=decorate}] (330:1cm) to node[below] {\small $t$} (0,0);
				\draw[{postaction=decorate}] (210:1cm) to [bend left=-45] node[below] {\small $s$} (0,0);
				\draw[{postaction=decorate}] (0,0) to [bend right=-45] node[left] {\small $\alpha$} (1,1);
				\draw[{postaction=decorate}] (-2,1) to [bend right=45] node[left] {\small $s$} (210:1cm);
				\draw[{postaction=decorate}] (1,1) to [bend left=45] node[right] {\small $\alpha$} (2,-0.5);
				\end{tikzpicture}
				=A_{st}^\alpha B_t^{s^*\alpha}
				\begin{tikzpicture}[baseline=(current bounding box.center), decoration={markings,mark=at position .5 with {\arrow[>=stealth]{>}}},scale=0.8]
				\draw[{postaction=decorate}] (0,0)  to node[below] {\small $\alpha$} (330:1cm);
				\draw[{postaction=decorate}] (210:1cm) to node[below] {\small $t$} (0,0);
				\draw[{postaction=decorate}] (90:1cm) to node[right] {\small $s$} (0,0);
				\end{tikzpicture},
			\end{equation}
			where $A_{ij}^k=\sqrt{\frac{d_{i^*} d_k}{d_j}} \left(F_k^{ii^*k}\right)_{0j}$ and $B_{ij}^k=\sqrt{\frac{d_{j^*} d_k}{d_i}}\, \overline{\left(F_k^{kj^*j}\right)_{i0}}$  \cite{Bonderson2007}. Hence, the operation $A_{st}^\alpha B_t^{s^*\alpha}\neq 0$ is just a rescaling of the trivalent vertex by a scalar factor. Therefore, it does not change the dimension of the corresponding fusion spaces. 
			\item[\eqref{P6}] This is just a rearrangement of the sums. In addition, we use $d_i=d_{i^*}$, which holds due to sphericality of the UFC \cite[Definition 4.7.14]{Etingof2015}.
			\item[\eqref{P7}] Using the fact $d_i d_j=\sum_k N_{ij}^k d_k$ (see, e.g., \cite[Eq.~2.10]{Bonderson2008}) gives the desired equation.
			\item[\eqref{P8}] Since the string diagram does not depend on $t$ anymore, we can evaluate the sum over $t$ separately. We then use the definition of the total quantum dimension $D=\sqrt{\sum_{i=0}^{N}d_i^2}$.
			\item[\eqref{P9}] Again, because of the sphericality of the UFC we have $d_i=d_{i^*}$.
		\end{itemize}
		
		\subsection{Commutativity}\label{sec:Commutativity}
		
		To build an exactly solvable Hamiltonian, it is necessary that the individual operators $Q_\mathbf{v}$ and $B_\mathbf{p}$ all commute with each other. Here, we explicitly do the computation for the configuration that is depicted in Fig.~\ref{fig:commut}. It turns out that the only factors that differ between the different orders of application are those where internal indices of both plaqettes appear. Hence, to keep the calculation as clear as possible, we will only write out these factors and denote the others by $(\dots)$.
		
		\begin{align}
			B_\mathbf{p_2}&\Bigg(B_\mathbf{p_1}\Bigg\vert
			\begin{tikzpicture}[scale=0.6,baseline={([yshift=-0.1cm]current bounding box.center)}]
			\draw[->-] (90:1) to node[above] {\scriptsize$g_1$} (150:1);
			\draw[->-] (150:1) to node[left] {\scriptsize$l_1$} (210:1);
			\draw[->-] (210:1) to node[below] {\scriptsize$k_1$} (270:1);
			\draw[->-] (270:1) to node[below,pos=0.6] {\scriptsize$j_1$} (330:1);
			\draw[->-] (330:1) to node[right] {\scriptsize$i_1$} (30:1);
			\draw[->-] (30:1) to node[above,pos=0.4] {\scriptsize$h_1$} (90:1);
			\foreach \a in {90,150,210,270} \draw[-<-] (\a:1) -- (\a:1.5);
			\node at (90:1.75) {\scriptsize$\beta$};
			\node at (150:1.75) {\scriptsize$\alpha$};
			\node at (210:1.75) {\scriptsize$\nu$};
			\node at (270:1.75) {\scriptsize$\mu$};
			\begin{scope}[xshift=1.74cm]
			\draw[-<-] (210:1) to node[below] {\scriptsize$k_2$} (270:1);
			\draw[-<-] (270:1) to node[below,pos=0.6] {\scriptsize$j_2$} (330:1);
			\draw[-<-] (330:1) to node[right] {\scriptsize$i_2$} (30:1);
			\draw[-<-] (30:1) to node[above,pos=0.4] {\scriptsize$h_2$} (90:1);
			\draw[-<-] (90:1) to node[above] {\scriptsize$g_2$} (150:1);
			\foreach \a in {90,30,330,270} \draw[-<-] (\a:1) -- (\a:1.5);
			\node at (90:1.75) {\scriptsize$\gamma$};
			\node at (30:1.75) {\scriptsize$\delta$};
			\node at (330:1.75) {\scriptsize$\epsilon$};
			\node at (270:1.75) {\scriptsize$\eta$};
			\end{scope}
			\end{tikzpicture}
			\Bigg\rangle\Bigg)\\
			&=B_\mathbf{p_2}\Bigg(\sum_s\Bigg\vert
			\begin{tikzpicture}[scale=0.6,baseline={([yshift=-0.1cm]current bounding box.center)}]
			\draw[->-] (90:1) to node[above] {\scriptsize$g_1$} (150:1);
			\draw[->-] (150:1) to node[left] {\scriptsize$l_1$} (210:1);
			\draw[->-] (210:1) to node[below] {\scriptsize$k_1$} (270:1);
			\draw[->-] (270:1) to node[below,pos=0.6] {\scriptsize$j_1$} (330:1);
			\draw[->-] (330:1) to node[right] {\scriptsize$i_1$} (30:1);
			\draw[->-] (30:1) to node[above,pos=0.4] {\scriptsize$h_1$} (90:1);
			\draw[->-,red] (90:0.6) -- (150:0.6);
			\draw[->-,red] (150:0.6) -- (210:0.6);
			\draw[->-,red] (210:0.6) -- (270:0.6);
			\draw[->-,red] (270:0.6) -- (330:0.6);
			\draw[->-,red] (330:0.6) -- (30:0.6);
			\draw[->-,red] (30:0.6) -- (90:0.6);
			\node[red] at (0,0) {\scriptsize$s$};
			\foreach \a in {90,150,210,270} \draw[-<-] (\a:1) -- (\a:1.5);
			\node at (90:1.75) {\scriptsize$\beta$};
			\node at (150:1.75) {\scriptsize$\alpha$};
			\node at (210:1.75) {\scriptsize$\nu$};
			\node at (270:1.75) {\scriptsize$\mu$};
			\begin{scope}[xshift=1.74cm]
			\draw[-<-] (210:1) to node[below] {\scriptsize$k_2$} (270:1);
			\draw[-<-] (270:1) to node[below,pos=0.6] {\scriptsize$j_2$} (330:1);
			\draw[-<-] (330:1) to node[right] {\scriptsize$i_2$} (30:1);
			\draw[-<-] (30:1) to node[above,pos=0.4] {\scriptsize$h_2$} (90:1);
			\draw[-<-] (90:1) to node[above] {\scriptsize$g_2$} (150:1);
			\foreach \a in {90,30,330,270} \draw[-<-] (\a:1) -- (\a:1.5);
			\node at (90:1.75) {\scriptsize$\gamma$};
			\node at (30:1.75) {\scriptsize$\delta$};
			\node at (330:1.75) {\scriptsize$\epsilon$};
			\node at (270:1.75) {\scriptsize$\eta$};
			\end{scope}
			\end{tikzpicture}
			\Bigg\rangle\Bigg)\\
			&=B_\mathbf{p_2}\Bigg(\sum_s\sum_{\substack{g_1',h_1',i_1',\\j_1',k_1',l_1'}}\overline{\left(F_{i_1'}^{s h_1 g_2}\right)}_{h_1' i_1} \left(F_{i_1'}^{s j_1 k_2}\right)_{j_1'i_1}(\dots)\Bigg\vert
			\begin{tikzpicture}[scale=0.6,baseline={([yshift=-0.1cm]current bounding box.center)}]
			\draw[->-] (90:1) to node[above] {\scriptsize$g_1'$} (150:1);
			\draw[->-] (150:1) to node[left] {\scriptsize$l_1'$} (210:1);
			\draw[->-] (210:1) to node[below] {\scriptsize$k_1'$} (270:1);
			\draw[->-] (270:1) to node[below,pos=0.6] {\scriptsize$j_1'$} (330:1);
			\draw[->-] (330:1) to node[right] {\scriptsize$i_1'$} (30:1);
			\draw[->-] (30:1) to node[above,pos=0.4] {\scriptsize$h_1'$} (90:1);
			\foreach \a in {90,150,210,270} \draw[-<-] (\a:1) -- (\a:1.5);
			\node at (90:1.75) {\scriptsize$\beta$};
			\node at (150:1.75) {\scriptsize$\alpha$};
			\node at (210:1.75) {\scriptsize$\nu$};
			\node at (270:1.75) {\scriptsize$\mu$};
			\begin{scope}[xshift=1.74cm]
			\draw[-<-] (210:1) to node[below] {\scriptsize$k_2$} (270:1);
			\draw[-<-] (270:1) to node[below,pos=0.6] {\scriptsize$j_2$} (330:1);
			\draw[-<-] (330:1) to node[right] {\scriptsize$i_2$} (30:1);
			\draw[-<-] (30:1) to node[above,pos=0.4] {\scriptsize$h_2$} (90:1);
			\draw[-<-] (90:1) to node[above] {\scriptsize$g_2$} (150:1);
			\foreach \a in {90,30,330,270} \draw[-<-] (\a:1) -- (\a:1.5);
			\node at (90:1.75) {\scriptsize$\gamma$};
			\node at (30:1.75) {\scriptsize$\delta$};
			\node at (330:1.75) {\scriptsize$\epsilon$};
			\node at (270:1.75) {\scriptsize$\eta$};
			\end{scope}
			\end{tikzpicture}
			\Bigg\rangle\Bigg)\\
			&=\sum_{s,t}\sum_{\substack{g_1',h_1',i_1',\\j_1',k_1',l_1'}}\overline{\left(F_{i_1'}^{s h_1 g_2}\right)}_{h_1' i_1} \left(F_{i_1'}^{s j_1 k_2}\right)_{j_1'i_1}(\dots)\Bigg\vert
			\begin{tikzpicture}[scale=0.6,baseline={([yshift=-0.1cm]current bounding box.center)}]
			\draw[->-] (90:1) to node[above] {\scriptsize$g_1'$} (150:1);
			\draw[->-] (150:1) to node[left] {\scriptsize$l_1'$} (210:1);
			\draw[->-] (210:1) to node[below] {\scriptsize$k_1'$} (270:1);
			\draw[->-] (270:1) to node[below,pos=0.6] {\scriptsize$j_1'$} (330:1);
			\draw[->-] (330:1) to node[left] {\scriptsize$i_1'$} (30:1);
			\draw[->-] (30:1) to node[above,pos=0.4] {\scriptsize$h_1'$} (90:1);
			\foreach \a in {90,150,210,270} \draw[-<-] (\a:1) -- (\a:1.5);
			\node at (90:1.75) {\scriptsize$\beta$};
			\node at (150:1.75) {\scriptsize$\alpha$};
			\node at (210:1.75) {\scriptsize$\nu$};
			\node at (270:1.75) {\scriptsize$\mu$};
			\begin{scope}[xshift=1.74cm]
			\draw[-<-] (210:1) to node[below] {\scriptsize$k_2$} (270:1);
			\draw[-<-] (270:1) to node[below,pos=0.6] {\scriptsize$j_2$} (330:1);
			\draw[-<-] (330:1) to node[right] {\scriptsize$i_2$} (30:1);
			\draw[-<-] (30:1) to node[above,pos=0.4] {\scriptsize$h_2$} (90:1);
			\draw[-<-] (90:1) to node[above] {\scriptsize$g_2$} (150:1);
			\foreach \a in {90,30,330,270} \draw[-<-] (\a:1) -- (\a:1.5);
			\draw[-<-,red] (90:0.6) -- (150:0.6);
			\draw[-<-,red] (150:0.6) -- (210:0.6);
			\draw[-<-,red] (210:0.6) -- (270:0.6);
			\draw[-<-,red] (270:0.6) -- (330:0.6);
			\draw[-<-,red] (330:0.6) -- (30:0.6);
			\draw[-<-,red] (30:0.6) -- (90:0.6);
			\node[red] at (0,0) {\scriptsize$t$};
			\node at (90:1.75) {\scriptsize$\gamma$};
			\node at (30:1.75) {\scriptsize$\delta$};
			\node at (330:1.75) {\scriptsize$\epsilon$};
			\node at (270:1.75) {\scriptsize$\eta$};
			\end{scope}
			\end{tikzpicture}
			\Bigg\rangle\\
			&=\sum_{s,t}\sum_{\substack{g_1',h_1',i_1',\\j_1',k_1',l_1'}}\sum_{\substack{g_2',h_2',i_1'',\\j_2',k_2',l_2'}}\overline{\left(F_{i_1'}^{s h_1 g_2}\right)}_{h_1' i_1} \left(F_{i_1'}^{s j_1 k_2}\right)_{j_1'i_1}\left(F_{i_1''}^{h_1' g_2 t}\right)_{i_1' g_2'}\overline{\left(F_{i_1''}^{j_1'k_2t}\right)}_{i_1'k_2'}(\dots)\Bigg\vert
			\begin{tikzpicture}[scale=0.6,baseline={([yshift=-0.1cm]current bounding box.center)}]
			\draw[->-] (90:1) to node[above] {\scriptsize$g_1'$} (150:1);
			\draw[->-] (150:1) to node[left] {\scriptsize$l_1'$} (210:1);
			\draw[->-] (210:1) to node[below] {\scriptsize$k_1'$} (270:1);
			\draw[->-] (270:1) to node[below,pos=0.6] {\scriptsize$j_1'$} (330:1);
			\draw[->-] (330:1) to node[right] {\scriptsize$i_1''$} (30:1);
			\draw[->-] (30:1) to node[above,pos=0.4] {\scriptsize$h_1'$} (90:1);
			\foreach \a in {90,150,210,270} \draw[-<-] (\a:1) -- (\a:1.5);
			\node at (90:1.75) {\scriptsize$\beta$};
			\node at (150:1.75) {\scriptsize$\alpha$};
			\node at (210:1.75) {\scriptsize$\nu$};
			\node at (270:1.75) {\scriptsize$\mu$};
			\begin{scope}[xshift=1.74cm]
			\draw[-<-] (210:1) to node[below] {\scriptsize$k_2'$} (270:1);
			\draw[-<-] (270:1) to node[below,pos=0.6] {\scriptsize$j_2'$} (330:1);
			\draw[-<-] (330:1) to node[right] {\scriptsize$i_2'$} (30:1);
			\draw[-<-] (30:1) to node[above,pos=0.4] {\scriptsize$h_2'$} (90:1);
			\draw[-<-] (90:1) to node[above] {\scriptsize$g_2'$} (150:1);
			\foreach \a in {90,30,330,270} \draw[-<-] (\a:1) -- (\a:1.5);
			\node at (90:1.75) {\scriptsize$\gamma$};
			\node at (30:1.75) {\scriptsize$\delta$};
			\node at (330:1.75) {\scriptsize$\epsilon$};
			\node at (270:1.75) {\scriptsize$\eta$};
			\end{scope}
			\end{tikzpicture}
			\Bigg\rangle.
		\end{align}

		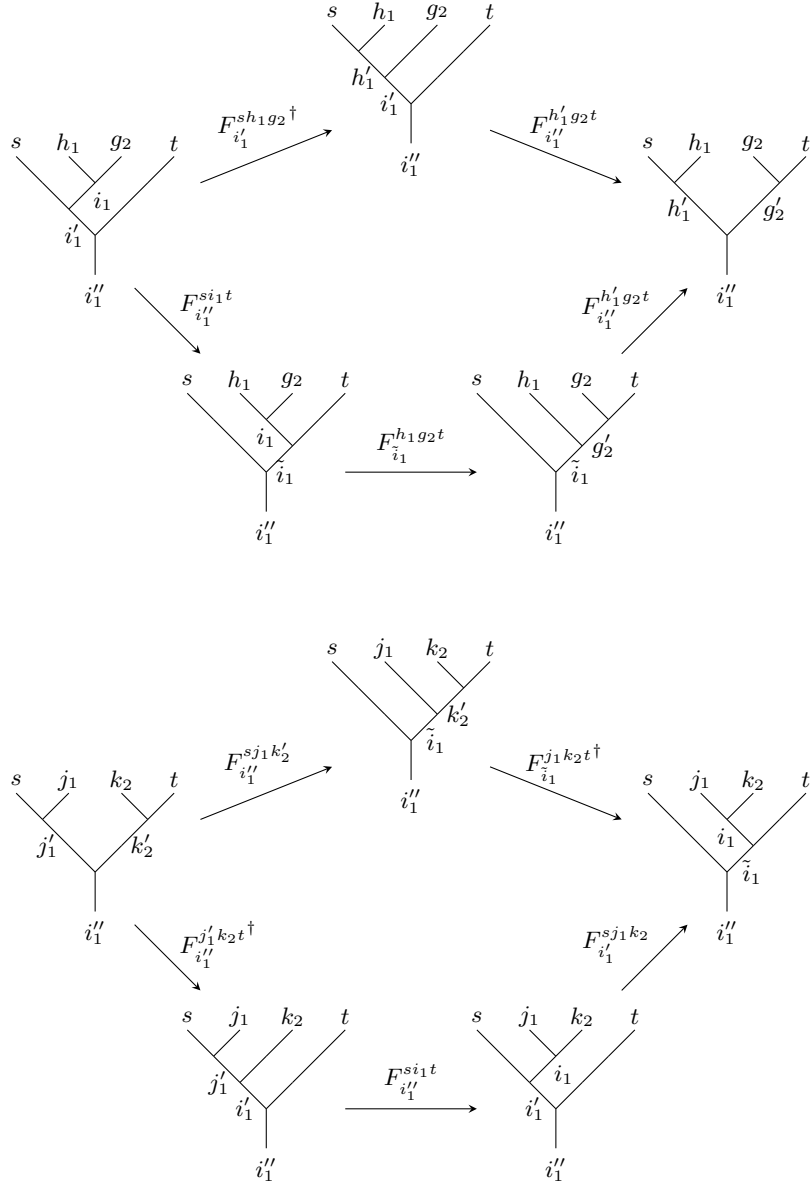
\begin{figure}[t]
			\centering
			\begin{tikzpicture}[scale=0.7]
			\draw (0,0.25) -- (0,1);
			\draw (0,1) -- (-0.5,1.5);
			\draw (0,1) -- (0.5,1.5);
			\draw (0.5,1.5) -- (1,2);
			\draw (1,2) -- (1.5,2.5);
			\draw (-0.5,1.5) -- (-1,2);
			\draw (-1,2) -- (-1.5,2.5);
			\draw (-0.5,1.5) -- (0,2);
			\draw (0,2) -- (0.5,2.5);
			\draw (0,2) -- (-0.5,2.5);
			\node at (0,-0.15) {\small $i_1''$};
			\node at (-1.5,2.75) {\small $s$};
			\node at (-0.5,2.75) {\small $h_1$};
			\node at (0.5,2.75) {\small $g_2$};
			\node at (1.5,2.75) {\small $t$};
			\node at (0.15,1.65) {\small $i_1$};
			\node at (-0.4,1) {\small $i_1'$};
			\draw[->,>=stealth] (2,2) to node[above] {\small ${F_{i_1'}^{sh_1g_2}}^\dagger$\ \ \ } (4.5,3);
			\draw[->,>=stealth] (7.5,3) to node[above] {\small\ \ $F_{i_1''}^{h_1'g_2t}$} (10,2);
			\draw[->,>=stealth] (0.75,0) to node[right,pos=0.3] {\small \ \ $F_{i_1''}^{si_1t}$} (2,-1.25);
			\draw[->,>=stealth] (10,-1.25) to node[left,pos=0.7] {\small $F_{i_1''}^{h_1'g_2t}$\ \ } (11.25,0);
			\draw[->,>=stealth] (4.75,-3.5) to node[above] {\small ${F_{\tilde{i}_1}^{h_1g_2t}}$} (7.25,-3.5);
			\begin{scope}[xshift=6cm,yshift=2.5cm]
			\draw (0,0.25) -- (0,1);
			\draw (0,1) -- (-0.5,1.5);
			\draw (0,1) -- (0.5,1.5);
			\draw (0.5,1.5) -- (1,2);
			\draw (1,2) -- (1.5,2.5);
			\draw (-0.5,1.5) -- (-1,2);
			\draw (-1,2) -- (-1.5,2.5);
			\draw (-0.5,1.5) -- (0,2);
			\draw (0,2) -- (0.5,2.5);
			\draw (-1,2) -- (-0.5,2.5);
			\node at (0,-0.15) {\small $i_1''$};
			\node at (-1.5,2.75) {\small $s$};
			\node at (-0.5,2.75) {\small $h_1$};
			\node at (0.5,2.75) {\small $g_2$};
			\node at (1.5,2.75) {\small $t$};
			\node at (-0.9,1.5) {\small $h_1'$};
			\node at (-0.4,1) {\small $i_1'$};
			\end{scope}
			\begin{scope}[xshift=12cm]
			\draw (0,0.25) -- (0,1);
			\draw (0,1) -- (-0.5,1.5);
			\draw (0,1) -- (0.5,1.5);
			\draw (0.5,1.5) -- (1,2);
			\draw (1,2) -- (1.5,2.5);
			\draw (-0.5,1.5) -- (-1,2);
			\draw (-1,2) -- (-1.5,2.5);
			\draw (1,2) -- (0.5,2.5);
			\draw (-1,2) -- (-0.5,2.5);
			\node at (0,-0.15) {\small $i_1''$};
			\node at (-1.5,2.75) {\small $s$};
			\node at (-0.5,2.75) {\small $h_1$};
			\node at (0.5,2.75) {\small $g_2$};
			\node at (1.5,2.75) {\small $t$};
			\node at (-0.9,1.5) {\small $h_1'$};
			\node at (0.9,1.5) {\small $g_2'$};
			\end{scope}
			\begin{scope}[xshift=3.25cm,yshift=-4.5cm]
			\draw (0,0.25) -- (0,1);
			\draw (0,1) -- (-0.5,1.5);
			\draw (0,1) -- (0.5,1.5);
			\draw (0.5,1.5) -- (1,2);
			\draw (1,2) -- (1.5,2.5);
			\draw (-0.5,1.5) -- (-1,2);
			\draw (-1,2) -- (-1.5,2.5);
			\draw (0.5,1.5) -- (0,2);
			\draw (0,2) -- (0.5,2.5);
			\draw (0,2) -- (-0.5,2.5);
			\node at (0,-0.15) {\small $i_1''$};
			\node at (-1.5,2.75) {\small $s$};
			\node at (-0.5,2.75) {\small $h_1$};
			\node at (0.5,2.75) {\small $g_2$};
			\node at (1.5,2.75) {\small $t$};
			\node at (0,1.65) {\small $i_1$};
			\node at (0.45,1) {\small $\tilde{i}_1\ $};
			\end{scope}
			\begin{scope}[xshift=8.75cm,yshift=-4.5cm]
			\draw (0,0.25) -- (0,1);
			\draw (0,1) -- (-0.5,1.5);
			\draw (0,1) -- (0.5,1.5);
			\draw (0.5,1.5) -- (1,2);
			\draw (1,2) -- (1.5,2.5);
			\draw (-0.5,1.5) -- (-1,2);
			\draw (-1,2) -- (-1.5,2.5);
			\draw (1,2) -- (0.5,2.5);
			\draw (0.5,1.5) -- (0,2);
			\draw (0,2) -- (-0.5,2.5);
			\node at (0,-0.15) {\small $i_1''$};
			\node at (-1.5,2.75) {\small $s$};
			\node at (-0.5,2.75) {\small $h_1$};
			\node at (0.5,2.75) {\small $g_2$};
			\node at (1.5,2.75) {\small $t$};
			\node at (0.4,1) {\small \ $\tilde{i}_1$};
			\node at (0.9,1.5) {\small $g_2'$};
			\end{scope}
			\end{tikzpicture}
			\\[1cm]
			\begin{tikzpicture}[scale=0.7]
			\draw (0,0.25) -- (0,1);
			\draw (0,1) -- (-0.5,1.5);
			\draw (0,1) -- (0.5,1.5);
			\draw (0.5,1.5) -- (1,2);
			\draw (1,2) -- (1.5,2.5);
			\draw (-0.5,1.5) -- (-1,2);
			\draw (-1,2) -- (-1.5,2.5);
			\draw (1,2) -- (0.5,2.5);
			\draw (-1,2) -- (-0.5,2.5);
			\node at (0,-0.15) {\small $i_1''$};
			\node at (-1.5,2.75) {\small $s$};
			\node at (-0.5,2.75) {\small $j_1$};
			\node at (0.5,2.75) {\small $k_2$};
			\node at (1.5,2.75) {\small $t$};
			\node at (-0.9,1.5) {\small $j_1'$};
			\node at (0.9,1.5) {\small $k_2'$};
			\draw[->,>=stealth] (2,2) to node[above] {\small $F_{i_1''}^{sj_1k_2'}$\ \ \ } (4.5,3);
			\draw[->,>=stealth] (7.5,3) to node[above] {\small\ \ ${F_{\tilde{i}_1}^{j_1k_2t}}^\dagger$} (10,2);
			\draw[->,>=stealth] (0.75,0) to node[right,pos=0.3] {\small \ \ ${F_{i_1''}^{j_1'k_2t}}^\dagger$} (2,-1.25);
			\draw[->,>=stealth] (10,-1.25) to node[left,pos=0.7] {\small $F_{i_1'}^{sj_1k_2}$\ \ } (11.25,0);
			\draw[->,>=stealth] (4.75,-3.5) to node[above] {\small ${F_{i_1''}^{si_1t}}$} (7.25,-3.5);
			\begin{scope}[xshift=6cm,yshift=2.5cm]
			\draw (0,0.25) -- (0,1);
			\draw (0,1) -- (-0.5,1.5);
			\draw (0,1) -- (0.5,1.5);
			\draw (0.5,1.5) -- (1,2);
			\draw (1,2) -- (1.5,2.5);
			\draw (-0.5,1.5) -- (-1,2);
			\draw (-1,2) -- (-1.5,2.5);
			\draw (1,2) -- (0.5,2.5);
			\draw (0.5,1.5) -- (0,2);
			\draw (0,2) -- (-0.5,2.5);
			\node at (0,-0.15) {\small $i_1''$};
			\node at (-1.5,2.75) {\small $s$};
			\node at (-0.5,2.75) {\small $j_1$};
			\node at (0.5,2.75) {\small $k_2$};
			\node at (1.5,2.75) {\small $t$};
			\node at (0.4,1) {\small \ $\tilde{i}_1$};
			\node at (0.9,1.5) {\small $k_2'$};
			\end{scope}
			\begin{scope}[xshift=12cm]
			\draw (0,0.25) -- (0,1);
			\draw (0,1) -- (-0.5,1.5);
			\draw (0,1) -- (0.5,1.5);
			\draw (0.5,1.5) -- (1,2);
			\draw (1,2) -- (1.5,2.5);
			\draw (-0.5,1.5) -- (-1,2);
			\draw (-1,2) -- (-1.5,2.5);
			\draw (0.5,1.5) -- (0,2);
			\draw (0,2) -- (0.5,2.5);
			\draw (0,2) -- (-0.5,2.5);
			\node at (0,-0.15) {\small $i_1''$};
			\node at (-1.5,2.75) {\small $s$};
			\node at (-0.5,2.75) {\small $j_1$};
			\node at (0.5,2.75) {\small $k_2$};
			\node at (1.5,2.75) {\small $t$};
			\node at (0,1.65) {\small $i_1$};
			\node at (0.4,1) {\small \ \ $\tilde{i}_1\ $};
			\end{scope}
			\begin{scope}[xshift=3.25cm,yshift=-4.5cm]
			\draw (0,0.25) -- (0,1);
			\draw (0,1) -- (-0.5,1.5);
			\draw (0,1) -- (0.5,1.5);
			\draw (0.5,1.5) -- (1,2);
			\draw (1,2) -- (1.5,2.5);
			\draw (-0.5,1.5) -- (-1,2);
			\draw (-1,2) -- (-1.5,2.5);
			\draw (-0.5,1.5) -- (0,2);
			\draw (0,2) -- (0.5,2.5);
			\draw (-1,2) -- (-0.5,2.5);
			\node at (0,-0.15) {\small $i_1''$};
			\node at (-1.5,2.75) {\small $s$};
			\node at (-0.5,2.75) {\small $j_1$};
			\node at (0.5,2.75) {\small $k_2$};
			\node at (1.5,2.75) {\small $t$};
			\node at (-0.9,1.5) {\small $j_1'$};
			\node at (-0.4,1) {\small $i_1'$};
			\end{scope}
			\begin{scope}[xshift=8.75cm,yshift=-4.5cm]
			\draw (0,0.25) -- (0,1);
			\draw (0,1) -- (-0.5,1.5);
			\draw (0,1) -- (0.5,1.5);
			\draw (0.5,1.5) -- (1,2);
			\draw (1,2) -- (1.5,2.5);
			\draw (-0.5,1.5) -- (-1,2);
			\draw (-1,2) -- (-1.5,2.5);
			\draw (-0.5,1.5) -- (0,2);
			\draw (0,2) -- (0.5,2.5);
			\draw (0,2) -- (-0.5,2.5);
			\node at (0,-0.15) {\small $i_1''$};
			\node at (-1.5,2.75) {\small $s$};
			\node at (-0.5,2.75) {\small $j_1$};
			\node at (0.5,2.75) {\small $k_2$};
			\node at (1.5,2.75) {\small $t$};
			\node at (0.15,1.65) {\small $i_1$};
			\node at (-0.4,1) {\small $i_1'$};
			\end{scope}
			\end{tikzpicture}
			\caption{\label{fig:penta}\textbf{Commuting diagrams.} The fact that these two diagrams commute yields the equations required to prove the commutativity of the magnetic flux operator.}
		\end{figure}

		Note that the above expression is a linear combination of string-net configurations, where in each coefficient we have a sum over the loop values $s$ and $t$ and also over the intermediate label $i_1'$. Performing the same calculation for the other order of operators yields
		\begin{align}
			\begin{split}
				\label{eq:commut}
				B_\mathbf{p_1}&\Bigg(B_\mathbf{p_2}\Bigg\vert
				\begin{tikzpicture}[scale=0.6,baseline={([yshift=-0.1cm]current bounding box.center)}]
				\draw[->-] (90:1) to node[above] {\scriptsize$g_1$} (150:1);
				\draw[->-] (150:1) to node[left] {\scriptsize$l_1$} (210:1);
				\draw[->-] (210:1) to node[below] {\scriptsize$k_1$} (270:1);
				\draw[->-] (270:1) to node[below,pos=0.6] {\scriptsize$j_1$} (330:1);
				\draw[->-] (330:1) to node[right] {\scriptsize$i_1$} (30:1);
				\draw[->-] (30:1) to node[above,pos=0.4] {\scriptsize$h_1$} (90:1);
				\foreach \a in {90,150,210,270} \draw[-<-] (\a:1) -- (\a:1.5);
				\node at (90:1.75) {\scriptsize$\beta$};
				\node at (150:1.75) {\scriptsize$\alpha$};
				\node at (210:1.75) {\scriptsize$\nu$};
				\node at (270:1.75) {\scriptsize$\mu$};
				\begin{scope}[xshift=1.74cm]
				\draw[-<-] (210:1) to node[below] {\scriptsize$k_2$} (270:1);
				\draw[-<-] (270:1) to node[below,pos=0.6] {\scriptsize$j_2$} (330:1);
				\draw[-<-] (330:1) to node[right] {\scriptsize$i_2$} (30:1);
				\draw[-<-] (30:1) to node[above,pos=0.4] {\scriptsize$h_2$} (90:1);
				\draw[-<-] (90:1) to node[above] {\scriptsize$g_2$} (150:1);
				\foreach \a in {90,30,330,270} \draw[-<-] (\a:1) -- (\a:1.5);
				\node at (90:1.75) {\scriptsize$\gamma$};
				\node at (30:1.75) {\scriptsize$\delta$};
				\node at (330:1.75) {\scriptsize$\epsilon$};
				\node at (270:1.75) {\scriptsize$\eta$};
				\end{scope}
				\end{tikzpicture}
				\Bigg\rangle\Bigg)\\
				&=\sum_{s,t}\sum_{\substack{g_1',h_1',\tilde{i}_1,\\j_1',k_1',l_1'}}\sum_{\substack{g_2',h_2',i_1'',\\j_2',k_2',l_2'}}\left(F_{\tilde{i}_1}^{h_1 g_2 t}\right)_{i_1 g_2'}\overline{\left(F_{\tilde{i}_1}^{j_1k_2t}\right)}_{i_1k_2'}\overline{\left(F_{i_1''}^{s h_1 g_2'}\right)}_{h_1' \tilde{i}_1} \left(F_{i_1''}^{s j_1 k_2'}\right)_{j_1'\tilde{i}_1}(\dots)\Bigg\vert
				\begin{tikzpicture}[scale=0.6,baseline={([yshift=-0.1cm]current bounding box.center)}]
				\draw[->-] (90:1) to node[above] {\scriptsize$g_1'$} (150:1);
				\draw[->-] (150:1) to node[left] {\scriptsize$l_1'$} (210:1);
				\draw[->-] (210:1) to node[below] {\scriptsize$k_1'$} (270:1);
				\draw[->-] (270:1) to node[below,pos=0.6] {\scriptsize$j_1'$} (330:1);
				\draw[->-] (330:1) to node[right] {\scriptsize$i_1''$} (30:1);
				\draw[->-] (30:1) to node[above,pos=0.4] {\scriptsize$h_1'$} (90:1);
				\foreach \a in {90,150,210,270} \draw[-<-] (\a:1) -- (\a:1.5);
				\node at (90:1.75) {\scriptsize$\beta$};
				\node at (150:1.75) {\scriptsize$\alpha$};
				\node at (210:1.75) {\scriptsize$\nu$};
				\node at (270:1.75) {\scriptsize$\mu$};
				\begin{scope}[xshift=1.74cm]
				\draw[-<-] (210:1) to node[below] {\scriptsize$k_2'$} (270:1);
				\draw[-<-] (270:1) to node[below,pos=0.6] {\scriptsize$j_2'$} (330:1);
				\draw[-<-] (330:1) to node[right] {\scriptsize$i_2'$} (30:1);
				\draw[-<-] (30:1) to node[above,pos=0.4] {\scriptsize$h_2'$} (90:1);
				\draw[-<-] (90:1) to node[above] {\scriptsize$g_2'$} (150:1);
				\foreach \a in {90,30,330,270} \draw[-<-] (\a:1) -- (\a:1.5);
				\node at (90:1.75) {\scriptsize$\gamma$};
				\node at (30:1.75) {\scriptsize$\delta$};
				\node at (330:1.75) {\scriptsize$\epsilon$};
				\node at (270:1.75) {\scriptsize$\eta$};
				\end{scope}
				\end{tikzpicture}
				\Bigg\rangle.
			\end{split}
		\end{align}
		
		Hence, the equality that we have to show reduces to
		\begin{equation}
			\begin{split}
				\sum_{s,t,i_1'}&\overline{\left(F_{i_1'}^{s h_1 g_2}\right)}_{h_1' i_1} \left(F_{i_1'}^{s j_1 k_2}\right)_{j_1'i_1}\left(F_{i_1''}^{h_1' g_2 t}\right)_{i_1' g_2'}\overline{\left(F_{i_1''}^{j_1'k_2t}\right)}_{i_1'k_2'}\\
				&=\sum_{s,t,\tilde{i}_1}\left(F_{\tilde{i}_1}^{h_1 g_2 t}\right)_{i_1 g_2'}\overline{\left(F_{\tilde{i}_1}^{j_1k_2t}\right)}_{i_1k_2'}\overline{\left(F_{i_1''}^{s h_1 g_2'}\right)}_{h_1' \tilde{i}_1} \left(F_{i_1''}^{s j_1 k_2'}\right)_{j_1'\tilde{i}_1}.
			\end{split}
		\end{equation}
		
		To show that this equality holds, we use the fact that the diagrams in Fig.~\ref{fig:penta} commute (which is due to Mac Lane's coherence theorem, see \cite{MacLane1998}). From these diagrams, we get the following identities:
		\begin{align}
			\left(F_{i_1'}^{s h_1 g_2}\right)^\dagger_{i_1h_1'}\left(F_{i_1''}^{h_1' g_2 t}\right)_{i_1' g_2'}&=\sum_{\tilde{i}_1}\left(F_{i_1''}^{si_1 t}\right)_{i_1'\tilde{i}_1} \left(F_{\tilde{i}_1}^{h_1g_2t}\right)_{i_1g_2'}\left(F_{i_1''}^{sh_1g_2'}\right)^\dagger_{\tilde{i}_1h_1'}\\
			\left(F_{i_1''}^{sj_1k_2'}\right)_{j_1'\tilde{i}_1}\left(F_{\tilde{i}_1}^{j_1k_2t}\right)^\dagger_{k_2'i_1}&=\sum_{i_1'} \left(F_{i_1''}^{j_1'k_2t}\right)^\dagger_{k_2'i_1'}\left(F_{i_1'}^{sj_1k_2}\right)_{j_1'i_1}\left(F_{i_1''}^{si_1t}\right)_{i_1'\tilde{i}_1}.
		\end{align}
		Inserting these identities into \eqref{eq:commut} (and using the fact that $\left(F_u^{xyz}\right)^\dagger_{\alpha\beta}=\overline{\left(F_u^{xyz}\right)}_{\beta\alpha}$) yields
		\begin{equation}
			\begin{split}
				\sum_{s,t,i_1',\tilde{i}_1}&\left(F_{i_1'}^{s j_1 k_2}\right)_{j_1'i_1}\overline{\left(F_{i_1''}^{j_1'k_2t}\right)}_{i_1'k_2'}\left(F_{i_1''}^{si_1 t}\right)_{i_1'\tilde{i}_1} \left(F_{\tilde{i}_1}^{h_1g_2t}\right)_{i_1g_2'}\overline{\left(F_{i_1''}^{sh_1g_2'}\right)}_{h_1'\tilde{i}_1}\\
				&=\sum_{s,t,\tilde{i}_1,i_1'}\overline{\left(F_{i_1''}^{j_1'k_2t}\right)}_{i_1'k_2'}\left(F_{i_1'}^{sj_1k_2}\right)_{j_1'i_1}\left(F_{i_1''}^{si_1t}\right)_{i_1'\tilde{i}_1}\left(F_{\tilde{i}_1}^{h_1 g_2 t}\right)_{i_1 g_2'}\overline{\left(F_{i_1''}^{s h_1 g_2'}\right)}_{h_1' \tilde{i}_1},
			\end{split}
		\end{equation}
		therefore
		\begin{equation}
			B_\mathbf{p_2}B_\mathbf{p_1}=B_\mathbf{p_1}B_\mathbf{p_2}.
		\end{equation}
		This calculation can analogously be done for any configuration of neighboring plaquettes. One always has to consider two commuting diagrams like those depicted in Fig.~\ref{fig:penta}.

		\section{Self-consistency constraints and the original Hamiltonian}
		\label{app:tetrahedral}
		This appendix is dedicated to the self-consistency constraints which are imposed by Levin and Wen in their original Hamiltonian (\cite[Eq.~(8) and Eq.~(14)]{LW05}). We show that our model translates into the original one when adding these additional symmetries. A special focus will be on the tetrahedral symmetry condition since it is the most restrictive constraint. We will also briefly explain why it is needed in the original model. In order to differ the Levin-Wen and the general UFC construction we use curly letters for items occurring in the original Levin-Wen Hamiltonian.
		
		In contrast to the UFC definition, the $\mathcal{F}$-symbols used in the original Levin-Wen model \cite{LW05} are defined in a slightly different fashion
		\begin{equation}
			\begin{tikzpicture}[decoration={markings,mark=at position .5 with {\arrow[>=stealth]{>}}},baseline=(current bounding box.center)]
			\draw[{postaction=decorate}] (0.5,0.9) to node[above] {\scriptsize{$i$}} (1,0.5);
			\draw[{postaction=decorate}] (0.5,0.1) to node[below] {\scriptsize{$j$}} (1,0.5);
			\draw[{postaction=decorate}] (1.5,0.5) to node[above] {\scriptsize{$m$}} (1,0.5);
			\draw[{postaction=decorate}] (2,0.9) to node[above] {\scriptsize{$k$}} (1.5,0.5);
			\draw[{postaction=decorate}] (2,0.1) to node[below] {\scriptsize{$l$}} (1.5,0.5);
			\end{tikzpicture}
			=
			\sum_f \mathcal{F}^{ijm}_{kln}
			\begin{tikzpicture}[decoration={markings,mark=at position .5 with {\arrow[>=stealth]{>}}},baseline=(current bounding box.center)]
			\draw[{postaction=decorate}] (0.5,0.9) to node[above] {\scriptsize{$i$}} (1.25,0.9);
			\draw[{postaction=decorate}] (2,0.9) to node[above] {\scriptsize{$k$}} (1.25,0.9);
			\draw[{postaction=decorate}] (0.5,0.1) to node[below] {\scriptsize{$j$}} (1.25,0.1);
			\draw[{postaction=decorate}] (2,0.1) to node[below] {\scriptsize{$l$}} (1.25,0.1);
			\draw[{postaction=decorate}] (1.25,0.1) to node[right] {\scriptsize{$n$}} (1.25,0.9);
			\end{tikzpicture}.\label{LWFsymbol}
		\end{equation}
		\noindent
		This action is not defined in a wide class of fusion categories since it involves horizontal lines. In order to illustrate the meaning of those let us give a brief physical outline. Physically, the wave function
		\begin{equation*}
			\Phi\left(
			\begin{tikzpicture}[baseline={([yshift=-3pt]current bounding box.center)},scale=0.5]
			\draw[fill=LightGray,rounded corners,LightGray] (0,0) rectangle (1,2);
			\end{tikzpicture}
			\right)
		\end{equation*}
		gives the probability amplitude of a physical process corresponding to the string diagram inside the grey box. For example, this may be a scattering process of anyons. As usual in physics literature, we use the convention that time goes upwards \footnote{Alternatively, we could impose time to go left or right. This would go along with a rotation of all string diagrams by 90 degrees. However, this modification of convention does not change the physical process.}. In this picture, a horizontal line denotes a physical process, which happens without passing of time. Hence, it should have a probability amplitude of zero. Whereas, if
		\begin{equation}
			\begin{tikzpicture}[baseline={([yshift=-3pt]current bounding box.center)}, decoration={markings,mark=at position .6 with {\arrow[>=stealth]{>}}},scale=0.5]
			\draw[{postaction=decorate}] (0.5,0.8) to node[left] {\small $i$} (0.5,2.5);
			\draw[{postaction=decorate}] (0.5,2.5) to node[left] {\small $k$} (0.5,3.2);
			\draw[{postaction=decorate}] (1.5,0.8) to node[right] {\small $j$} (1.5,1.5);
			\draw[{postaction=decorate}] (1.5,1.5) to node[right] {\small $l$} (1.5,3.2);
			\draw[{postaction=decorate}] (1.5,1.5) to node[above] {\small $m$} (0.5,2.5);
			\end{tikzpicture}
			=
			\begin{tikzpicture}[baseline={([yshift=-3pt]current bounding box.center)}, decoration={markings,mark=at position .6 with {\arrow[>=stealth]{>}}},scale=0.5]
			\draw[{postaction=decorate}] (0.5,0.8) to node[left] {\small $i$} (0.5,1.5);
			\draw[{postaction=decorate}] (0.5,1.5) to node[left] {\small $k$} (0.5,3.2);
			\draw[{postaction=decorate}] (1.5,0.8) to node[right] {\small $j$} (1.5,2.5);
			\draw[{postaction=decorate}] (1.5,2.5) to node[right] {\small $l$} (1.5,3.2);
			\draw[{postaction=decorate}] (0.5,1.5) to node[above] {\small $m$} (1.5,2.5);
			\end{tikzpicture}
			=:
			\begin{tikzpicture}[baseline={([yshift=-3pt]current bounding box.center)}, decoration={markings,mark=at position .6 with {\arrow[>=stealth]{>}}},scale=0.5]
			\draw[{postaction=decorate}] (0.5,0.8) to node[left] {\small $i$} (0.5,2);
			\draw[{postaction=decorate}] (0.5,2) to node[left] {\small $k$} (0.5,3.2);
			\draw[{postaction=decorate}] (1.5,0.8) to node[right] {\small $j$} (1.5,2);
			\draw[{postaction=decorate}] (1.5,2) to node[right] {\small $l$} (1.5,3.2);
			\draw[{postaction=decorate}] (0.5,2) to node[above] {\small $m$} (1.5,2);
			\end{tikzpicture}
			\label{eq:G=H}
		\end{equation}
		\noindent
		the horizontal line can be defined as a reasonable physical process by means of Eq.~\eqref{eq:G=H}. It only holds if we require an additional symmetry of the Levin-Wen $\mathcal{F}$-symbols, namely tetrahedral symmetry \cite{Hong09}
		\begin{equation}
			\mathcal{F}_{kln}^{ijm}=\mathcal{F}_{jin}^{lkm^*}=\mathcal{F}_{lkn^*}^{jim}=\sqrt{\frac{d_m d_n}{d_j d_l}}\mathcal{F}_{k^*nl}^{imj}.\label{tetrahedralSymmetry}
		\end{equation}
		\noindent
		In terms of the UFC $F$-symbols, Eq.~\eqref{tetrahedralSymmetry} can be written as \cite{Soejima2020}
		\begin{equation}
			\left(F^{jkl}_{i}\right)_{mn}=\left(F^{kji^*}_{l^*}\right)_{mn^*}=\left(F^{i^*lk}_{j^*}\right)_{m^*n}=\sqrt{\frac{d_m d_n}{d_j d_l}}\left(F_{i^*}^{m^*kn^*}\right)_{j^*l^*}.
		\end{equation}
		\noindent
		The fact that tetrahedral symmetry \eqref{tetrahedralSymmetry} is mandatory for Eq.~\eqref{LWFsymbol} to be a defined operation makes it a necessary condition for a standard Levin-Wen construction. However, this restriction is too strong \cite{HW12} and does not hold in any UFC. We show a particular counterexample in Appendix~\ref{app:H3}.
		
		The remaining symmetry conditions for the original Levin-Wen model are
		\begin{align}
			\mathcal{F}_{j^*i^*0}^{ijk}&=\sqrt{\frac{d_k}{d_i d_j}}\delta_{ijk}\label{normalizationConvention}\\
			\overline{\mathcal{F}_{kln}^{ijm}}&=\mathcal{F}_{k^*l^*n^*}^{i^*j^*m^*},\label{unitarityCondition}\\
			\mathcal{F}_{q^*kr^*}^{jip} \mathcal{F}_{mls^*}^{riq^*}&=\sum_{n=0}^{N} \mathcal{F}_{kp^*n}^{mlq} \mathcal{F}_{mns^*}^{jip} \mathcal{F}_{lkr^*}^{js^*n}\label{pentagonIdentity}
		\end{align}
		\noindent
		Eq.~\eqref{normalizationConvention} is a normalization convention for the $\mathcal{F}$-symbols. It ensures that a gauge transformed $\mathcal{F}$-symbol belongs to the same local rules (\cite[Eq.~(4) -- (7)]{LW05}) as the untransformed operation. The unitarity condition \eqref{unitarityCondition} ensures Hermicity of the original Levin-Wen model. The last constraint \eqref{pentagonIdentity} is just the pentagon equation \eqref{eq:pentagon} for the Levin-Wen $\mathcal{F}$-symbols. In contrast to the other conditions this is not an additional symmetry since it is also required in our construction \footnote{In fact, the pentagon equation is a coherence condition for the associator in any monoidal category, see \cite{Etingof2015}.}.
		
		We now show that our model segues into the original Levin-Wen model when claiming tetrahedral symmetry \eqref{tetrahedralSymmetry}, the normalization convention \eqref{normalizationConvention} and the unitarity condition \eqref{unitarityCondition} as additional symmetries. First, we observe that the only difference between the two Hamiltonians is in the matrix elements of the magnetic flux operator. In the original Levin-Wen model, they are given by
		\begin{equation}
			\mathcal{B}_{\mathbf{p},ghijkl}^{s,g'h'i'j'k'l'}(abcdef)=\mathcal{F}^{al^*g}_{s^*g'l'^*}\mathcal{F}^{bg^*h}_{s^*h'g'^*}\mathcal{F}^{ch^*i}_{s^*i',h'^*}\mathcal{F}^{di^*j}_{s^*j'i'^*}\mathcal{F}^{ej^*k}_{s^*k'j'^*}\mathcal{F}^{fk^*l}_{s^*l'k'^*}\label{eq:LWmagneticFlux}.
		\end{equation}
		\noindent
		We start by translating the UFC $F$-symbols in $B_\mathbf{p}^s$ to Levin-Wen $\mathcal{F}$-symbols \eqref{LWFsymbol} as in $\mathcal{B}_\mathbf{p}^s$. They are related by \cite{Soejima2020}
		\begin{equation}
			\left(F_l^{ijk}\right)_{mn}=\mathcal{F}^{j^*i^*m}_{lk^*n}.
		\end{equation}
		\noindent
		\begin{equation}\begin{array}[b]{r@{}c@{}l} 
		B_{\mathbf{p},ghijkl}^{s,g'h'i'j'k'l'}(abcdef)&{}={}&\sqrt{d_s d_{s^*}}\sqrt{\frac{d_{g^*}d_h d_{i'} d_j d_{k^*} d_{l'^*}}{d_{g'^*}d_{h'} d_{i} d_{j'} d_{k'^*} d_{l^*}}}
		\mathcal{F}^{sgg'^*}_{g^*s^*0} 
		\overline{\mathcal{F}_{b^*h^*h'}^{s^*g'g^*}}
		\overline{\mathcal{F}_{i'ci}^{h^*s^*h'}}
		\mathcal{F}_{i'd^*i}^{j^*s^*j'} 
		\mathcal{F}_{jj^*j'}^{s^*s0} 
		\overline{\mathcal{F}_{ej'^*j}^{skk'^*}}
		\mathcal{F}_{l'^*sg'^*}^{gal^*}
		\overline{\mathcal{F}_{l'^*sk'^*}^{kf^*l^*}}\\
		&{}\overset{\eqref{unitarityCondition}}{=}{}&
		\sqrt{d_s d_{s^*}}
		\sqrt{\frac{d_{g^*}d_h d_{i'} d_j d_{k^*} d_{l'^*}}{d_{g'^*}d_{h'} d_{i} d_{j'} d_{k'^*} d_{l^*}}}
		\mathcal{F}^{sgg'^*}_{g^*s^*0} 
		\mathcal{F}_{jj^*j'}^{s^*s0} 
		\mathcal{F}_{bhh'^*}^{sg'^*g} 
		\mathcal{F}_{i'^*c^*i^*}^{hsh'^*} 
		\mathcal{F}_{i'd^*i}^{j^*s^*j'} 
		\mathcal{F}_{e^*j'j^*}^{s^*k^*k'} 
		\mathcal{F}_{l's^*k'}^{k^*fl} 
		\mathcal{F}_{l'^*sg'^*}^{gal^*}.
		\label{eq:BpsTetra}
		\end{array}\end{equation}
		\noindent
		We now rewrite each $\mathcal{F}$-symbol in Eq.~\eqref{eq:BpsTetra} individually by using the normalization convention \eqref{normalizationConvention} and the tetrahedral symmetry condition \eqref{tetrahedralSymmetry}. For better transparency of this calculation, we numerate the different $\mathcal{F}$-symbols occurring in Eq.~\eqref{tetrahedralSymmetry}
		\begin{equation}
		\underbrace{\mathcal{F}_{kln}^{ijm}}_{(1)}=
		\underbrace{\mathcal{F}_{jin}^{lkm^*}}_{(2)}=
		\underbrace{\mathcal{F}_{lkn^*}^{jim}}_{(3)}=
		\underbrace{\sqrt{\frac{d_m d_n}{d_j d_l}}\mathcal{F}_{k^*nl}^{imj}}_{(4)}
		\end{equation}
		and point out which equality we use by $(x)\rightarrow (y)$. This gives
		\begin{equation}\begin{array}[c]{r@{}c@{}l} 
		\mathcal{F}^{sgg'^*}_{g^*s^*0}&{}\overset{\eqref{normalizationConvention}}{=}{}&\sqrt{\frac{d_{g'^*}}{d_s d_g}}\label{eq:tetra1}\\[1.5em]
		\mathcal{F}_{jj^*j'}^{s^*s0} &{}\overset{(4)\rightarrow (2)}{=}{}&
		\sqrt{\frac{d_{j'}}{d_s d_{j^*}}} \mathcal{F}_{0s^*j^*}^{j'j^*s^*}
		\overset{(3)\rightarrow (4)}{=}
		\mathcal{F}_{sj1}^{j^*s^*j'}
		\overset{\eqref{normalizationConvention}}{=}\sqrt{\frac{d_{j'}}{d_{s^*} d_{j^*}}}\\[1.5em]
		\mathcal{F}_{bhh'^*}^{sg'^*g} &{}\overset{(2)\rightarrow (3)}{=}{}& \mathcal{F}_{sg'^*h'}^{bhg^*}\overset{(1)\rightarrow (4)}{=} \sqrt{\frac{d_{g^*}d_{h'}}{d_h d_{g'^*}}} \mathcal{F}_{s^*h'g'^*}^{bg^*h}\\[1.5em]
		\mathcal{F}_{i'^*c^*i^*}^{hsh'^*} &{}\overset{(4)\rightarrow (2)}{=}{}&\sqrt{\frac{d_{h'^*} d_{i^*}}{d_s d_{c^*}}} \mathcal{F}_{h'^*hc^*}^{i^*i's^*} \overset{(3)\rightarrow (4)}{=} \mathcal{F}_{h^*ch'^*}^{i's^*i^*} \overset{(1)\rightarrow (2)}{=} \mathcal{F}_{s^*i'h'^*}^{ch^*i}\\[1.5em]
		\mathcal{F}_{i'd^*i}^{j^*s^*j'} &{}\overset{(1)\rightarrow (3)}{=}{}& \mathcal{F}_{d^*i'i^*}^{s^*j^*j'} \overset{(1)\rightarrow (4)}{=} \sqrt{\frac{d_{j'} d_{i^*}}{d_{j^*} d_{i'}}} \mathcal{F}_{di^*i'}^{s^*j'j^*} \overset{(2)\rightarrow (3)}{=} \sqrt{\frac{d_{j'} d_{i^*}}{d_{j^*} d_{i'}}} \mathcal{F}_{s^*j'i'^*}^{di^*j}\\[1.5em]
		\mathcal{F}_{e^*j'j^*}^{s^*k^*k'} &{}\overset{(1)\rightarrow (4)}{=}{}& \sqrt{\frac{d_{k'} d_{j^*}}{d_{k^*} d_{j'}}} \mathcal{F}_{ej^*j'}^{s^*k'k^*} \overset{(2)\rightarrow (3)}{=} \sqrt{\frac{d_{k'} d_{j^*}}{d_{k^*} d_{j'}}} \mathcal{F}_{s^*k'j'^*}^{ej^*k}\\[1.5em]
		\mathcal{F}_{l's^*k'}^{k^*fl} &{}\overset{(1)\rightarrow (3)}{=}{}& \mathcal{F}_{s^*l'k'^*}^{fk^*l}\\[1.5em]
		\mathcal{F}_{l'^*sg'^*}^{gal^*} &{}\overset{(3)\rightarrow (4)}{=}{}& \sqrt{\frac{d_{l^*} d_{g'}}{d_g d_{l'^*}}} \mathcal{F}_{s^*g'l'^*}^{al^*g}
		\end{array}\end{equation}
		\noindent
		Reinserting \eqref{eq:tetra1} into \eqref{eq:BpsTetra} yields
		\begin{equation}
			B_{\mathbf{p},ghijkl}^{s,g'h'i'j'k'l'}(abcdef)=\mathcal{F}^{al^*g}_{s^*g'l'^*}\mathcal{F}^{bg^*h}_{s^*h'g'^*}\mathcal{F}^{ch^*i}_{s^*i',h'^*}\mathcal{F}^{di^*j}_{s^*j'i'^*}\mathcal{F}^{ej^*k}_{s^*k'j'^*}\mathcal{F}^{fk^*l}_{s^*l'k'^*},
		\end{equation}
		which is exactly the original Levin-Wen magnetic flux operator $\mathcal{B}_\mathbf{p}^s$ \eqref{eq:LWmagneticFlux}. Therefore, our model is a true generalization of the original Levin-Wen Hamiltonian presented in \cite{LW05}.

		\section{The Haagerup $\mathcal{H}_3$ fusion category and tetrahedral symmetry}\label{app:H3}

		In this appendix, we examine the $F$-symbols of the Haagerup $\mathcal{H}_3$ fusion category and show that they do not fulfill the tetrahedral symmetry condition \eqref{tetrahedralSymmetry}. Therefore, the original Levin-Wen model cannot be built from $\mathcal{H}_3$. However, it can be used as an input for our generalized Levin-Wen construction since it has the structure of a unitary fusion category.
		
		The fusion category $\mathcal{H}_3$ was first found in \cite{GS12}. It has six simple objects, namely $\{1,\alpha,\alpha^*,\rho,\arho,\asrho\}$. Note that, in contrast to the nomenclature of the string types, here we use $1$ to denote the trivial object instead of $0$ which follows the usual convention in category theory. The quantum dimensions of the objects are specified in Table~\ref{HaagerupObjects} and the fusion rules are given in Table~\ref{HaagerupFusionRules}.

		The $F$-symbols of $\mathcal{H}_3$ have been explicitly computed in \cite{Osborne2019}. It is easy to see that these $F$-symbols do not fulfill the tetrahedral symmetry condition \eqref{tetrahedralSymmetry}. For instance, consider the map
		\begin{equation}
			\begin{tikzpicture}[scale=1,baseline=(current bounding box.center), decoration={markings,mark=at position .6 with {\arrow[>=stealth]{>}}}]
			\node (u1) at (0,-0.25) {$\rho$};
			\coordinate (v11) at (0,0.5);
			\coordinate (v12) at (-0.5,1);
			\node (x1) at (-1,1.5) {$1$};
			\node (y1) at (0,1.5) {$\alpha$};
			\node (z1) at (1,1.5) {$\asrho$};
			
			\draw[{postaction=decorate}] (u1) -- (v11);
			\draw[{postaction=decorate}] (v11) -- (z1);
			\draw[{postaction=decorate}] (v11) to node[below left] {$\alpha$} (v12);
			\draw[{postaction=decorate}] (v12) -- (y1);
			\draw[{postaction=decorate}] (v12) -- (x1);
			\end{tikzpicture}
			\xrightarrow{\left(F_{\rho}^{1\alpha\asrho}\right)_{\alpha\rho}}
			\begin{tikzpicture}[scale=1,baseline=(current bounding box.center), decoration={markings,mark=at position .6 with {\arrow[>=stealth]{>}}}]
			\node (u2) at (0,-0.25) {$\rho$};
			\coordinate (v21) at (0,0.5);
			\coordinate (v22) at (0.5,1);
			\node (x2) at (-1,1.5) {$1$};
			\node (y2) at (0,1.5) {$\alpha$};
			\node (z2) at (1,1.5) {$\asrho$};
			
			\draw[{postaction=decorate}] (u2) -- (v21);
			\draw[{postaction=decorate}] (v21) -- (x2);
			\draw[{postaction=decorate}] (v21) to node[below right] {$\rho$} (v22);
			\draw[{postaction=decorate}] (v22) -- (y2);
			\draw[{postaction=decorate}] (v22) -- (z2);
			\end{tikzpicture},
		\end{equation}
		\noindent
		where $\left(F_{\rho}^{1\alpha\asrho}\right)_{\alpha\rho} = 1$ is a valid transformation between string diagrams. The tetrahedral symmetry condition \eqref{tetrahedralSymmetryUFC} for this $F$-symbol then imposes the relations
		\begin{equation}
			\left(F_{\rho}^{1\alpha\asrho}\right)_{\alpha\rho}=\left(F_{\asrho}^{\alpha 1\rho}\right)_{\alpha\rho}=\left(F_{1}^{\rho\asrho\alpha}\right)_{\alpha^*\rho}.\label{TetraH3}
		\end{equation}
		\noindent
		However, both of the latter $F$-symbols belong to invalid transformations with respect to the fusion rules. Therefore, the corresponding fusion vector spaces are zero-dimensional and \eqref{TetraH3} is not fulfilled. Furthermore, even a gauge transformation
		\begin{equation}
			\left(F_i^{jkl}\right)'_{mn} = \frac{u_l^{in} u_n^{jk}}{u_m^{ij} u_l^{mk}} \left(F_i^{jkl}\right)_{mn}
		\end{equation}
		\noindent
		does not lead to tetrahedrally symmetric $F$-symbols, since the dimension of the corresponding fusion spaces cannot be changed by the choice of gauge.
		
		\begin{table}[t]
			\centering
			\begin{tabular}{c|c}
				\textbf{Object} $i$ & \textbf{Quantum dimension} $d_{i}$\\
				\hline
				1 & $1$\\
				$\alpha$ & $1$\\
				$\alpha^*$ & $1$\\
				$\rho$ & $\frac{3+\sqrt{13}}{2}$\\
				$\arho$ & $\frac{3+\sqrt{13}}{2}$\\
				$\asrho$ & $\frac{3+\sqrt{13}}{2}$
			\end{tabular}
			\caption{Simple objects of the Haagerup $\mathcal{H}_3$ fusion category and their corresponding quantum dimensions.}\label{HaagerupObjects}
		\end{table}
		
		\begin{table}[t]
			\centering
			\begin{tabular}{c||c|c|c|c|c|c}
				& $1$ & $\alpha$ & $\alpha^*$ & $\rho$ & $\arho$ & $\asrho$\\
				\hhline{==|=|=|=|=|=}
				$1$ & $1$ & $\alpha$ & $\alpha^*$ & $\rho$ & $\arho$ & $\asrho$\\
				\hline
				$\alpha$ & $\alpha$ & $\alpha^*$ & $1$ & $\arho$ & $\asrho$ & $\rho$\\
				\hline
				$\alpha^*$ & $\alpha^*$ & $1$ & $\alpha$ & $\asrho$ & $\rho$ & $\arho$\\
				\hline
				$\rho$ & $\rho$ & $\asrho$ & $\arho$ & $1+Z$ & $\alpha^*+Z$ & $\alpha+Z$\\
				\hline
				$\arho$ & $\arho$ & $\rho$ & $\asrho$ & $\alpha+Z$ & $1+Z$ & $\alpha^*+Z$\\
				\hline
				$\asrho$ & $\asrho$ & $\arho$ & $\rho$ & $\alpha^*+Z$ & $\alpha+Z$ & $1+Z$
			\end{tabular}
			\caption{Fusion rules $i\otimes j=\sum_{k} N_{ij}^k k$ of the Haagerup $\mathcal{H}_3$ fusion category using the abbreviation $Z=\rho+\arho+\asrho$.}\label{HaagerupFusionRules}
		\end{table}
		
	\end{appendix}
	
	\twocolumngrid
	
	\bibliography{Stringnetbib}

\begin{thebibliography}{72}%
\makeatletter
\providecommand \@ifxundefined [1]{%
 \@ifx{#1\undefined}
}%
\providecommand \@ifnum [1]{%
 \ifnum #1\expandafter \@firstoftwo
 \else \expandafter \@secondoftwo
 \fi
}%
\providecommand \@ifx [1]{%
 \ifx #1\expandafter \@firstoftwo
 \else \expandafter \@secondoftwo
 \fi
}%
\providecommand \natexlab [1]{#1}%
\providecommand \enquote  [1]{``#1''}%
\providecommand \bibnamefont  [1]{#1}%
\providecommand \bibfnamefont [1]{#1}%
\providecommand \citenamefont [1]{#1}%
\providecommand \href@noop [0]{\@secondoftwo}%
\providecommand \href [0]{\begingroup \@sanitize@url \@href}%
\providecommand \@href[1]{\@@startlink{#1}\@@href}%
\providecommand \@@href[1]{\endgroup#1\@@endlink}%
\providecommand \@sanitize@url [0]{\catcode `\\12\catcode `\$12\catcode
  `\&12\catcode `\#12\catcode `\^12\catcode `\_12\catcode `\%12\relax}%
\providecommand \@@startlink[1]{}%
\providecommand \@@endlink[0]{}%
\providecommand \url  [0]{\begingroup\@sanitize@url \@url }%
\providecommand \@url [1]{\endgroup\@href {#1}{\urlprefix }}%
\providecommand \urlprefix  [0]{URL }%
\providecommand \Eprint [0]{\href }%
\providecommand \doibase [0]{https://doi.org/}%
\providecommand \selectlanguage [0]{\@gobble}%
\providecommand \bibinfo  [0]{\@secondoftwo}%
\providecommand \bibfield  [0]{\@secondoftwo}%
\providecommand \translation [1]{[#1]}%
\providecommand \BibitemOpen [0]{}%
\providecommand \bibitemStop [0]{}%
\providecommand \bibitemNoStop [0]{.\EOS\space}%
\providecommand \EOS [0]{\spacefactor3000\relax}%
\providecommand \BibitemShut  [1]{\csname bibitem#1\endcsname}%
\let\auto@bib@innerbib\@empty
\bibitem [{\citenamefont {Landau}(1965)}]{Landau1965}%
  \BibitemOpen
  \bibfield  {author} {\bibinfo {author} {\bibfnamefont {L.~D.}\ \bibnamefont
  {Landau}},\ }\bibfield  {title} {\bibinfo {title} {On the theory of phase
  transitions},\ }in\ \href
  {https://doi.org/10.1016/b978-0-08-010586-4.50034-1} {\emph {\bibinfo
  {booktitle} {Collected Papers of L. D. Landau}}},\ \bibinfo {editor} {edited
  by\ \bibinfo {editor} {\bibfnamefont {D.}~\bibnamefont {ter Haar}}}\
  (\bibinfo  {publisher} {Pergamon Press},\ \bibinfo {address} {Oxford, New
  York},\ \bibinfo {year} {1965})\ pp.\ \bibinfo {pages} {193--216}\BibitemShut
  {NoStop}%
\bibitem [{\citenamefont {Tsui}\ \emph {et~al.}(1982)\citenamefont {Tsui},
  \citenamefont {Stormer},\ and\ \citenamefont {Gossard}}]{FQH1}%
  \BibitemOpen
  \bibfield  {author} {\bibinfo {author} {\bibfnamefont {D.~C.}\ \bibnamefont
  {Tsui}}, \bibinfo {author} {\bibfnamefont {H.~L.}\ \bibnamefont {Stormer}},\
  and\ \bibinfo {author} {\bibfnamefont {A.~C.}\ \bibnamefont {Gossard}},\
  }\bibfield  {title} {\bibinfo {title} {{Two-Dimensional Magnetotransport in
  the Extreme Quantum Limit}},\ }\href
  {https://doi.org/10.1103/physrevlett.48.1559} {\bibfield  {journal} {\bibinfo
   {journal} {Phys. Rev. Lett.}\ }\textbf {\bibinfo {volume} {48}},\ \bibinfo
  {pages} {1559} (\bibinfo {year} {1982})}\BibitemShut {NoStop}%
\bibitem [{\citenamefont {Laughlin}(1983)}]{FQH2}%
  \BibitemOpen
  \bibfield  {author} {\bibinfo {author} {\bibfnamefont {R.~B.}\ \bibnamefont
  {Laughlin}},\ }\bibfield  {title} {\bibinfo {title} {{Anomalous Quantum Hall
  Effect: An Incompressible Quantum Fluid with Fractionally Charged
  Excitations}},\ }\href {https://doi.org/10.1103/physrevlett.50.1395}
  {\bibfield  {journal} {\bibinfo  {journal} {Phys. Rev. Lett.}\ }\textbf
  {\bibinfo {volume} {50}},\ \bibinfo {pages} {1395} (\bibinfo {year}
  {1983})}\BibitemShut {NoStop}%
\bibitem [{\citenamefont {Wen}(1990)}]{Wen1990}%
  \BibitemOpen
  \bibfield  {author} {\bibinfo {author} {\bibfnamefont {X.~G.}\ \bibnamefont
  {Wen}},\ }\bibfield  {title} {\bibinfo {title} {{Topological} {Orders} in
  {Rigid} {States}},\ }\href {https://doi.org/10.1142/s0217979290000139}
  {\bibfield  {journal} {\bibinfo  {journal} {Int. J. Mod. Phys. B}\ }\textbf
  {\bibinfo {volume} {04}},\ \bibinfo {pages} {239} (\bibinfo {year}
  {1990})}\BibitemShut {NoStop}%
\bibitem [{\citenamefont {Wen}(1995)}]{Wen1995}%
  \BibitemOpen
  \bibfield  {author} {\bibinfo {author} {\bibfnamefont {X.-G.}\ \bibnamefont
  {Wen}},\ }\bibfield  {title} {\bibinfo {title} {{Topological orders and edge
  excitations in fractional quantum Hall states}},\ }\href
  {https://doi.org/10.1080/00018739500101566} {\bibfield  {journal} {\bibinfo
  {journal} {Adv. Phys.}\ }\textbf {\bibinfo {volume} {44}},\ \bibinfo {pages}
  {405} (\bibinfo {year} {1995})}\BibitemShut {NoStop}%
\bibitem [{\citenamefont {Haldane}(1983{\natexlab{a}})}]{Haldane1983}%
  \BibitemOpen
  \bibfield  {author} {\bibinfo {author} {\bibfnamefont {F.~D.~M.}\
  \bibnamefont {Haldane}},\ }\bibfield  {title} {\bibinfo {title} {Continuum
  dynamics of the $1$-{D Heisenberg} antiferromagnet: Identification with the
  $o(3)$ nonlinear sigma model},\ }\href
  {https://doi.org/10.1016/0375-9601(83)90631-x} {\bibfield  {journal}
  {\bibinfo  {journal} {Phys. Lett. A}\ }\textbf {\bibinfo {volume} {93}},\
  \bibinfo {pages} {464} (\bibinfo {year} {1983}{\natexlab{a}})}\BibitemShut
  {NoStop}%
\bibitem [{\citenamefont {Haldane}(1983{\natexlab{b}})}]{Haldane1983a}%
  \BibitemOpen
  \bibfield  {author} {\bibinfo {author} {\bibfnamefont {F.~D.~M.}\
  \bibnamefont {Haldane}},\ }\bibfield  {title} {\bibinfo {title} {{Nonlinear
  Field Theory of Large-Spin Heisenberg Antiferromagnets: Semiclassically
  Quantized Solitons of the One-Dimensional Easy-Axis N{\'{e}}el State}},\
  }\href {https://doi.org/10.1103/physrevlett.50.1153} {\bibfield  {journal}
  {\bibinfo  {journal} {Phys. Rev. Lett.}\ }\textbf {\bibinfo {volume} {50}},\
  \bibinfo {pages} {1153} (\bibinfo {year} {1983}{\natexlab{b}})}\BibitemShut
  {NoStop}%
\bibitem [{\citenamefont {Pollmann}\ \emph {et~al.}(2010)\citenamefont
  {Pollmann}, \citenamefont {Turner}, \citenamefont {Berg},\ and\ \citenamefont
  {Oshikawa}}]{PTBO10}%
  \BibitemOpen
  \bibfield  {author} {\bibinfo {author} {\bibfnamefont {F.}~\bibnamefont
  {Pollmann}}, \bibinfo {author} {\bibfnamefont {A.~M.}\ \bibnamefont
  {Turner}}, \bibinfo {author} {\bibfnamefont {E.}~\bibnamefont {Berg}},\ and\
  \bibinfo {author} {\bibfnamefont {M.}~\bibnamefont {Oshikawa}},\ }\bibfield
  {title} {\bibinfo {title} {Entanglement spectrum of a topological phase in
  one dimension},\ }\bibfield  {journal} {\bibinfo  {journal} {Phys. Rev. B}\
  }\textbf {\bibinfo {volume} {81}},\ \href
  {https://doi.org/10.1103/physrevb.81.064439} {10.1103/physrevb.81.064439}
  (\bibinfo {year} {2010})\BibitemShut {NoStop}%
\bibitem [{\citenamefont {Pollmann}\ \emph {et~al.}(2012)\citenamefont
  {Pollmann}, \citenamefont {Berg}, \citenamefont {Turner},\ and\ \citenamefont
  {Oshikawa}}]{PBTO12}%
  \BibitemOpen
  \bibfield  {author} {\bibinfo {author} {\bibfnamefont {F.}~\bibnamefont
  {Pollmann}}, \bibinfo {author} {\bibfnamefont {E.}~\bibnamefont {Berg}},
  \bibinfo {author} {\bibfnamefont {A.~M.}\ \bibnamefont {Turner}},\ and\
  \bibinfo {author} {\bibfnamefont {M.}~\bibnamefont {Oshikawa}},\ }\bibfield
  {title} {\bibinfo {title} {Symmetry protection of topological phases in
  one-dimensional quantum spin systems},\ }\bibfield  {journal} {\bibinfo
  {journal} {Phys. Rev. B}\ }\textbf {\bibinfo {volume} {85}},\ \href
  {https://doi.org/10.1103/physrevb.85.075125} {10.1103/physrevb.85.075125}
  (\bibinfo {year} {2012})\BibitemShut {NoStop}%
\bibitem [{\citenamefont {Wen}(2017)}]{Wen2017}%
  \BibitemOpen
  \bibfield  {author} {\bibinfo {author} {\bibfnamefont {X.-G.}\ \bibnamefont
  {Wen}},\ }\bibfield  {title} {\bibinfo {title} {Colloquium: {Zoo} of
  quantum-topological phases of matter},\ }\bibfield  {journal} {\bibinfo
  {journal} {Rev. Mod. Phys.}\ }\textbf {\bibinfo {volume} {89}},\ \href
  {https://doi.org/10.1103/revmodphys.89.041004} {10.1103/revmodphys.89.041004}
  (\bibinfo {year} {2017})\BibitemShut {NoStop}%
\bibitem [{\citenamefont {Wen}\ and\ \citenamefont {Niu}(1990)}]{WN90}%
  \BibitemOpen
  \bibfield  {author} {\bibinfo {author} {\bibfnamefont {X.~G.}\ \bibnamefont
  {Wen}}\ and\ \bibinfo {author} {\bibfnamefont {Q.}~\bibnamefont {Niu}},\
  }\bibfield  {title} {\bibinfo {title} {{Ground-state degeneracy of the
  fractional quantum Hall states in the presence of a random potential and on
  high-genus Riemann surfaces}},\ }\href
  {https://doi.org/10.1103/physrevb.41.9377} {\bibfield  {journal} {\bibinfo
  {journal} {Phys. Rev. B}\ }\textbf {\bibinfo {volume} {41}},\ \bibinfo
  {pages} {9377} (\bibinfo {year} {1990})}\BibitemShut {NoStop}%
\bibitem [{\citenamefont {Blok}\ and\ \citenamefont {Wen}(1990)}]{BW90}%
  \BibitemOpen
  \bibfield  {author} {\bibinfo {author} {\bibfnamefont {B.}~\bibnamefont
  {Blok}}\ and\ \bibinfo {author} {\bibfnamefont {X.~G.}\ \bibnamefont {Wen}},\
  }\bibfield  {title} {\bibinfo {title} {{Effective theories of the fractional
  quantum Hall effect: Hierarchy construction}},\ }\href
  {https://doi.org/10.1103/physrevb.42.8145} {\bibfield  {journal} {\bibinfo
  {journal} {Phys. Rev. B}\ }\textbf {\bibinfo {volume} {42}},\ \bibinfo
  {pages} {8145} (\bibinfo {year} {1990})}\BibitemShut {NoStop}%
\bibitem [{\citenamefont {Read}(1990)}]{Read1990}%
  \BibitemOpen
  \bibfield  {author} {\bibinfo {author} {\bibfnamefont {N.}~\bibnamefont
  {Read}},\ }\bibfield  {title} {\bibinfo {title} {{Excitation structure of the
  hierarchy scheme in the fractional quantum Hall effect}},\ }\href
  {https://doi.org/10.1103/physrevlett.65.1502} {\bibfield  {journal} {\bibinfo
   {journal} {Phys. Rev. Lett.}\ }\textbf {\bibinfo {volume} {65}},\ \bibinfo
  {pages} {1502} (\bibinfo {year} {1990})}\BibitemShut {NoStop}%
\bibitem [{\citenamefont {Fröhlich}\ and\ \citenamefont
  {Kerler}(1991)}]{FK91}%
  \BibitemOpen
  \bibfield  {author} {\bibinfo {author} {\bibfnamefont {J.}~\bibnamefont
  {Fröhlich}}\ and\ \bibinfo {author} {\bibfnamefont {T.}~\bibnamefont
  {Kerler}},\ }\bibfield  {title} {\bibinfo {title} {{Universality in quantum
  Hall systems}},\ }\href {https://doi.org/10.1016/0550-3213(91)90360-a}
  {\bibfield  {journal} {\bibinfo  {journal} {Nucl. Phys. B}\ }\textbf
  {\bibinfo {volume} {354}},\ \bibinfo {pages} {369} (\bibinfo {year}
  {1991})}\BibitemShut {NoStop}%
\bibitem [{\citenamefont {Zaletel}\ \emph {et~al.}(2013)\citenamefont
  {Zaletel}, \citenamefont {Mong},\ and\ \citenamefont {Pollmann}}]{ZMP13}%
  \BibitemOpen
  \bibfield  {author} {\bibinfo {author} {\bibfnamefont {M.~P.}\ \bibnamefont
  {Zaletel}}, \bibinfo {author} {\bibfnamefont {R.~S.~K.}\ \bibnamefont
  {Mong}},\ and\ \bibinfo {author} {\bibfnamefont {F.}~\bibnamefont
  {Pollmann}},\ }\bibfield  {title} {\bibinfo {title} {{Topological
  Characterization of Fractional Quantum Hall Ground States from Microscopic
  Hamiltonians}},\ }\bibfield  {journal} {\bibinfo  {journal} {Phys. Rev.
  Lett.}\ }\textbf {\bibinfo {volume} {110}},\ \href
  {https://doi.org/10.1103/physrevlett.110.236801}
  {10.1103/physrevlett.110.236801} (\bibinfo {year} {2013})\BibitemShut
  {NoStop}%
\bibitem [{\citenamefont {Wen}\ \emph {et~al.}(1989)\citenamefont {Wen},
  \citenamefont {Wilczek},\ and\ \citenamefont {Zee}}]{WWZ89}%
  \BibitemOpen
  \bibfield  {author} {\bibinfo {author} {\bibfnamefont {X.~G.}\ \bibnamefont
  {Wen}}, \bibinfo {author} {\bibfnamefont {F.}~\bibnamefont {Wilczek}},\ and\
  \bibinfo {author} {\bibfnamefont {A.}~\bibnamefont {Zee}},\ }\bibfield
  {title} {\bibinfo {title} {Chiral spin states and superconductivity},\ }\href
  {https://doi.org/10.1103/physrevb.39.11413} {\bibfield  {journal} {\bibinfo
  {journal} {Phys. Rev. B}\ }\textbf {\bibinfo {volume} {39}},\ \bibinfo
  {pages} {11413} (\bibinfo {year} {1989})}\BibitemShut {NoStop}%
\bibitem [{\citenamefont {Read}\ and\ \citenamefont {Sachdev}(1991)}]{RS91}%
  \BibitemOpen
  \bibfield  {author} {\bibinfo {author} {\bibfnamefont {N.}~\bibnamefont
  {Read}}\ and\ \bibinfo {author} {\bibfnamefont {S.}~\bibnamefont {Sachdev}},\
  }\bibfield  {title} {\bibinfo {title} {Large-{$N$} expansion for frustrated
  quantum antiferromagnets},\ }\href
  {https://doi.org/10.1103/physrevlett.66.1773} {\bibfield  {journal} {\bibinfo
   {journal} {Phys. Rev. Lett.}\ }\textbf {\bibinfo {volume} {66}},\ \bibinfo
  {pages} {1773} (\bibinfo {year} {1991})}\BibitemShut {NoStop}%
\bibitem [{\citenamefont {Wen}(1991)}]{Wen1991}%
  \BibitemOpen
  \bibfield  {author} {\bibinfo {author} {\bibfnamefont {X.~G.}\ \bibnamefont
  {Wen}},\ }\bibfield  {title} {\bibinfo {title} {Mean-field theory of
  spin-liquid states with finite energy gap and topological orders},\ }\href
  {https://doi.org/10.1103/physrevb.44.2664} {\bibfield  {journal} {\bibinfo
  {journal} {Phys. Rev. B}\ }\textbf {\bibinfo {volume} {44}},\ \bibinfo
  {pages} {2664} (\bibinfo {year} {1991})}\BibitemShut {NoStop}%
\bibitem [{\citenamefont {Senthil}\ and\ \citenamefont {Fisher}(2000)}]{SF00}%
  \BibitemOpen
  \bibfield  {author} {\bibinfo {author} {\bibfnamefont {T.}~\bibnamefont
  {Senthil}}\ and\ \bibinfo {author} {\bibfnamefont {M.~P.~A.}\ \bibnamefont
  {Fisher}},\ }\bibfield  {title} {\bibinfo {title} {$\mathbb{Z}_2$ gauge
  theory of electron fractionalization in strongly correlated systems},\ }\href
  {https://doi.org/10.1103/physrevb.62.7850} {\bibfield  {journal} {\bibinfo
  {journal} {Phys. Rev. B}\ }\textbf {\bibinfo {volume} {62}},\ \bibinfo
  {pages} {7850} (\bibinfo {year} {2000})}\BibitemShut {NoStop}%
\bibitem [{\citenamefont {Wen}(2002)}]{Wen2002}%
  \BibitemOpen
  \bibfield  {author} {\bibinfo {author} {\bibfnamefont {X.-G.}\ \bibnamefont
  {Wen}},\ }\bibfield  {title} {\bibinfo {title} {Quantum orders and symmetric
  spin liquids},\ }\bibfield  {journal} {\bibinfo  {journal} {Phys. Rev. B}\
  }\textbf {\bibinfo {volume} {65}},\ \href
  {https://doi.org/10.1103/physrevb.65.165113} {10.1103/physrevb.65.165113}
  (\bibinfo {year} {2002})\BibitemShut {NoStop}%
\bibitem [{\citenamefont {Sachdev}\ and\ \citenamefont {Park}(2002)}]{SP02}%
  \BibitemOpen
  \bibfield  {author} {\bibinfo {author} {\bibfnamefont {S.}~\bibnamefont
  {Sachdev}}\ and\ \bibinfo {author} {\bibfnamefont {K.}~\bibnamefont {Park}},\
  }\bibfield  {title} {\bibinfo {title} {{Ground States of Quantum
  Antiferromagnets in Two Dimensions}},\ }\href
  {https://doi.org/10.1006/aphy.2002.6232} {\bibfield  {journal} {\bibinfo
  {journal} {Ann. Phys.}\ }\textbf {\bibinfo {volume} {298}},\ \bibinfo {pages}
  {58} (\bibinfo {year} {2002})}\BibitemShut {NoStop}%
\bibitem [{\citenamefont {Ioffe}\ \emph {et~al.}(2002)\citenamefont {Ioffe},
  \citenamefont {Feigel'man}, \citenamefont {Ioselevich}, \citenamefont
  {Ivanov}, \citenamefont {Troyer},\ and\ \citenamefont {Blatter}}]{Ioffe2002}%
  \BibitemOpen
  \bibfield  {author} {\bibinfo {author} {\bibfnamefont {L.~B.}\ \bibnamefont
  {Ioffe}}, \bibinfo {author} {\bibfnamefont {M.~V.}\ \bibnamefont
  {Feigel'man}}, \bibinfo {author} {\bibfnamefont {A.}~\bibnamefont
  {Ioselevich}}, \bibinfo {author} {\bibfnamefont {D.}~\bibnamefont {Ivanov}},
  \bibinfo {author} {\bibfnamefont {M.}~\bibnamefont {Troyer}},\ and\ \bibinfo
  {author} {\bibfnamefont {G.}~\bibnamefont {Blatter}},\ }\bibfield  {title}
  {\bibinfo {title} {{Topologically protected quantum bits using Josephson
  junction arrays}},\ }\href {https://doi.org/10.1038/415503a} {\bibfield
  {journal} {\bibinfo  {journal} {Nature}\ }\textbf {\bibinfo {volume} {415}},\
  \bibinfo {pages} {503} (\bibinfo {year} {2002})}\BibitemShut {NoStop}%
\bibitem [{\citenamefont {Dennis}\ \emph {et~al.}(2002)\citenamefont {Dennis},
  \citenamefont {Kitaev}, \citenamefont {Landahl},\ and\ \citenamefont
  {Preskill}}]{DKLP02}%
  \BibitemOpen
  \bibfield  {author} {\bibinfo {author} {\bibfnamefont {E.}~\bibnamefont
  {Dennis}}, \bibinfo {author} {\bibfnamefont {A.}~\bibnamefont {Kitaev}},
  \bibinfo {author} {\bibfnamefont {A.}~\bibnamefont {Landahl}},\ and\ \bibinfo
  {author} {\bibfnamefont {J.}~\bibnamefont {Preskill}},\ }\bibfield  {title}
  {\bibinfo {title} {Topological quantum memory},\ }\href
  {https://doi.org/10.1063/1.1499754} {\bibfield  {journal} {\bibinfo
  {journal} {J. Math. Phys.}\ }\textbf {\bibinfo {volume} {43}},\ \bibinfo
  {pages} {4452} (\bibinfo {year} {2002})}\BibitemShut {NoStop}%
\bibitem [{\citenamefont {Kitaev}(2003)}]{Kitaev2003}%
  \BibitemOpen
  \bibfield  {author} {\bibinfo {author} {\bibfnamefont {A.~Y.}\ \bibnamefont
  {Kitaev}},\ }\bibfield  {title} {\bibinfo {title} {Fault-tolerant quantum
  computation by anyons},\ }\href
  {https://doi.org/10.1016/s0003-4916(02)00018-0} {\bibfield  {journal}
  {\bibinfo  {journal} {Ann. Phys.}\ }\textbf {\bibinfo {volume} {303}},\
  \bibinfo {pages} {2} (\bibinfo {year} {2003})}\BibitemShut {NoStop}%
\bibitem [{\citenamefont {Nayak}\ \emph {et~al.}(2008)\citenamefont {Nayak},
  \citenamefont {Simon}, \citenamefont {Stern}, \citenamefont {Freedman},\ and\
  \citenamefont {Sarma}}]{Nayak2008}%
  \BibitemOpen
  \bibfield  {author} {\bibinfo {author} {\bibfnamefont {C.}~\bibnamefont
  {Nayak}}, \bibinfo {author} {\bibfnamefont {S.~H.}\ \bibnamefont {Simon}},
  \bibinfo {author} {\bibfnamefont {A.}~\bibnamefont {Stern}}, \bibinfo
  {author} {\bibfnamefont {M.}~\bibnamefont {Freedman}},\ and\ \bibinfo
  {author} {\bibfnamefont {S.~D.}\ \bibnamefont {Sarma}},\ }\bibfield  {title}
  {\bibinfo {title} {Non-{Abelian} anyons and topological quantum
  computation},\ }\href {https://doi.org/10.1103/revmodphys.80.1083} {\bibfield
   {journal} {\bibinfo  {journal} {Rev. Mod. Phys.}\ }\textbf {\bibinfo
  {volume} {80}},\ \bibinfo {pages} {1083} (\bibinfo {year}
  {2008})}\BibitemShut {NoStop}%
\bibitem [{\citenamefont {Terhal}(2015)}]{Terhal2015}%
  \BibitemOpen
  \bibfield  {author} {\bibinfo {author} {\bibfnamefont {B.~M.}\ \bibnamefont
  {Terhal}},\ }\bibfield  {title} {\bibinfo {title} {Quantum error correction
  for quantum memories},\ }\href {https://doi.org/10.1103/revmodphys.87.307}
  {\bibfield  {journal} {\bibinfo  {journal} {Rev. Mod. Phys.}\ }\textbf
  {\bibinfo {volume} {87}},\ \bibinfo {pages} {307} (\bibinfo {year}
  {2015})}\BibitemShut {NoStop}%
\bibitem [{\citenamefont {Schollwöck}(2005)}]{Schollwoeck2005}%
  \BibitemOpen
  \bibfield  {author} {\bibinfo {author} {\bibfnamefont {U.}~\bibnamefont
  {Schollwöck}},\ }\bibfield  {title} {\bibinfo {title} {The density-matrix
  renormalization group},\ }\href {https://doi.org/10.1103/revmodphys.77.259}
  {\bibfield  {journal} {\bibinfo  {journal} {Rev. Mod. Phys.}\ }\textbf
  {\bibinfo {volume} {77}},\ \bibinfo {pages} {259} (\bibinfo {year}
  {2005})}\BibitemShut {NoStop}%
\bibitem [{\citenamefont {Schollwöck}(2011)}]{Schollwoeck2011}%
  \BibitemOpen
  \bibfield  {author} {\bibinfo {author} {\bibfnamefont {U.}~\bibnamefont
  {Schollwöck}},\ }\bibfield  {title} {\bibinfo {title} {The density-matrix
  renormalization group in the age of matrix product states},\ }\href
  {https://doi.org/10.1016/j.aop.2010.09.012} {\bibfield  {journal} {\bibinfo
  {journal} {Ann. Phys.}\ }\textbf {\bibinfo {volume} {326}},\ \bibinfo {pages}
  {96} (\bibinfo {year} {2011})}\BibitemShut {NoStop}%
\bibitem [{\citenamefont {Vidal}(2004)}]{Vidal2004}%
  \BibitemOpen
  \bibfield  {author} {\bibinfo {author} {\bibfnamefont {G.}~\bibnamefont
  {Vidal}},\ }\bibfield  {title} {\bibinfo {title} {{Efficient Simulation of
  One-Dimensional Quantum Many-Body Systems}},\ }\bibfield  {journal} {\bibinfo
   {journal} {Phys. Rev. Lett.}\ }\textbf {\bibinfo {volume} {93}},\ \href
  {https://doi.org/10.1103/physrevlett.93.040502}
  {10.1103/physrevlett.93.040502} (\bibinfo {year} {2004})\BibitemShut
  {NoStop}%
\bibitem [{\citenamefont {Verstraete}\ and\ \citenamefont
  {Cirac}(2004)}]{Verstraete2004}%
  \BibitemOpen
  \bibfield  {author} {\bibinfo {author} {\bibfnamefont {F.}~\bibnamefont
  {Verstraete}}\ and\ \bibinfo {author} {\bibfnamefont {J.~I.}\ \bibnamefont
  {Cirac}},\ }\bibfield  {title} {\bibinfo {title} {{Renormalization algorithms
  for Quantum-Many Body Systems in two and higher dimensions}},\ }\href@noop {}
  {\  (\bibinfo {year} {2004})},\ \Eprint
  {https://arxiv.org/abs/cond-mat/0407066v1} {cond-mat/0407066v1} \BibitemShut
  {NoStop}%
\bibitem [{\citenamefont {Trebst}\ \emph {et~al.}(2008)\citenamefont {Trebst},
  \citenamefont {Ardonne}, \citenamefont {Feiguin}, \citenamefont {Huse},
  \citenamefont {Ludwig},\ and\ \citenamefont {Troyer}}]{Trebst2008}%
  \BibitemOpen
  \bibfield  {author} {\bibinfo {author} {\bibfnamefont {S.}~\bibnamefont
  {Trebst}}, \bibinfo {author} {\bibfnamefont {E.}~\bibnamefont {Ardonne}},
  \bibinfo {author} {\bibfnamefont {A.}~\bibnamefont {Feiguin}}, \bibinfo
  {author} {\bibfnamefont {D.~A.}\ \bibnamefont {Huse}}, \bibinfo {author}
  {\bibfnamefont {A.~W.~W.}\ \bibnamefont {Ludwig}},\ and\ \bibinfo {author}
  {\bibfnamefont {M.}~\bibnamefont {Troyer}},\ }\bibfield  {title} {\bibinfo
  {title} {{Collective States of Interacting Fibonacci Anyons}},\ }\bibfield
  {journal} {\bibinfo  {journal} {Phys. Rev. Lett.}\ }\textbf {\bibinfo
  {volume} {101}},\ \href {https://doi.org/10.1103/physrevlett.101.050401}
  {10.1103/physrevlett.101.050401} (\bibinfo {year} {2008})\BibitemShut
  {NoStop}%
\bibitem [{\citenamefont {König}\ and\ \citenamefont
  {Bilgin}(2010)}]{Koenig2010}%
  \BibitemOpen
  \bibfield  {author} {\bibinfo {author} {\bibfnamefont {R.}~\bibnamefont
  {König}}\ and\ \bibinfo {author} {\bibfnamefont {E.}~\bibnamefont
  {Bilgin}},\ }\bibfield  {title} {\bibinfo {title} {Anyonic entanglement
  renormalization},\ }\bibfield  {journal} {\bibinfo  {journal} {Phys. Rev. B}\
  }\textbf {\bibinfo {volume} {82}},\ \href
  {https://doi.org/10.1103/physrevb.82.125118} {10.1103/physrevb.82.125118}
  (\bibinfo {year} {2010})\BibitemShut {NoStop}%
\bibitem [{\citenamefont {Haegeman}\ \emph {et~al.}(2011)\citenamefont
  {Haegeman}, \citenamefont {Cirac}, \citenamefont {Osborne}, \citenamefont
  {Pi{\v{z}}orn}, \citenamefont {Verschelde},\ and\ \citenamefont
  {Verstraete}}]{Haegeman2011}%
  \BibitemOpen
  \bibfield  {author} {\bibinfo {author} {\bibfnamefont {J.}~\bibnamefont
  {Haegeman}}, \bibinfo {author} {\bibfnamefont {J.~I.}\ \bibnamefont {Cirac}},
  \bibinfo {author} {\bibfnamefont {T.~J.}\ \bibnamefont {Osborne}}, \bibinfo
  {author} {\bibfnamefont {I.}~\bibnamefont {Pi{\v{z}}orn}}, \bibinfo {author}
  {\bibfnamefont {H.}~\bibnamefont {Verschelde}},\ and\ \bibinfo {author}
  {\bibfnamefont {F.}~\bibnamefont {Verstraete}},\ }\bibfield  {title}
  {\bibinfo {title} {{Time-Dependent Variational Principle for Quantum
  Lattices}},\ }\bibfield  {journal} {\bibinfo  {journal} {Phys. Rev. Lett.}\
  }\textbf {\bibinfo {volume} {107}},\ \href
  {https://doi.org/10.1103/physrevlett.107.070601}
  {10.1103/physrevlett.107.070601} (\bibinfo {year} {2011})\BibitemShut
  {NoStop}%
\bibitem [{\citenamefont {Haegeman}\ \emph {et~al.}(2012)\citenamefont
  {Haegeman}, \citenamefont {Pirvu}, \citenamefont {Weir}, \citenamefont
  {Cirac}, \citenamefont {Osborne}, \citenamefont {Verschelde},\ and\
  \citenamefont {Verstraete}}]{Haegeman2012}%
  \BibitemOpen
  \bibfield  {author} {\bibinfo {author} {\bibfnamefont {J.}~\bibnamefont
  {Haegeman}}, \bibinfo {author} {\bibfnamefont {B.}~\bibnamefont {Pirvu}},
  \bibinfo {author} {\bibfnamefont {D.~J.}\ \bibnamefont {Weir}}, \bibinfo
  {author} {\bibfnamefont {J.~I.}\ \bibnamefont {Cirac}}, \bibinfo {author}
  {\bibfnamefont {T.~J.}\ \bibnamefont {Osborne}}, \bibinfo {author}
  {\bibfnamefont {H.}~\bibnamefont {Verschelde}},\ and\ \bibinfo {author}
  {\bibfnamefont {F.}~\bibnamefont {Verstraete}},\ }\bibfield  {title}
  {\bibinfo {title} {Variational matrix product ansatz for dispersion
  relations},\ }\bibfield  {journal} {\bibinfo  {journal} {Phys. Rev. B}\
  }\textbf {\bibinfo {volume} {85}},\ \href
  {https://doi.org/10.1103/physrevb.85.100408} {10.1103/physrevb.85.100408}
  (\bibinfo {year} {2012})\BibitemShut {NoStop}%
\bibitem [{\citenamefont {Finch}\ \emph {et~al.}(2014)\citenamefont {Finch},
  \citenamefont {Frahm}, \citenamefont {Lewerenz}, \citenamefont {Milsted},\
  and\ \citenamefont {Osborne}}]{Finch2014}%
  \BibitemOpen
  \bibfield  {author} {\bibinfo {author} {\bibfnamefont {P.~E.}\ \bibnamefont
  {Finch}}, \bibinfo {author} {\bibfnamefont {H.}~\bibnamefont {Frahm}},
  \bibinfo {author} {\bibfnamefont {M.}~\bibnamefont {Lewerenz}}, \bibinfo
  {author} {\bibfnamefont {A.}~\bibnamefont {Milsted}},\ and\ \bibinfo {author}
  {\bibfnamefont {T.~J.}\ \bibnamefont {Osborne}},\ }\bibfield  {title}
  {\bibinfo {title} {Quantum phases of a chain of strongly interacting
  anyons},\ }\bibfield  {journal} {\bibinfo  {journal} {Phys. Rev. B}\ }\textbf
  {\bibinfo {volume} {90}},\ \href {https://doi.org/10.1103/physrevb.90.081111}
  {10.1103/physrevb.90.081111} (\bibinfo {year} {2014})\BibitemShut {NoStop}%
\bibitem [{\citenamefont {Singh}\ \emph {et~al.}(2014)\citenamefont {Singh},
  \citenamefont {Pfeifer}, \citenamefont {Vidal},\ and\ \citenamefont
  {Brennen}}]{Singh2014}%
  \BibitemOpen
  \bibfield  {author} {\bibinfo {author} {\bibfnamefont {S.}~\bibnamefont
  {Singh}}, \bibinfo {author} {\bibfnamefont {R.~N.~C.}\ \bibnamefont
  {Pfeifer}}, \bibinfo {author} {\bibfnamefont {G.}~\bibnamefont {Vidal}},\
  and\ \bibinfo {author} {\bibfnamefont {G.~K.}\ \bibnamefont {Brennen}},\
  }\bibfield  {title} {\bibinfo {title} {Matrix product states for anyonic
  systems and efficient simulation of dynamics},\ }\bibfield  {journal}
  {\bibinfo  {journal} {Phys. Rev. B}\ }\textbf {\bibinfo {volume} {89}},\
  \href {https://doi.org/10.1103/physrevb.89.075112}
  {10.1103/physrevb.89.075112} (\bibinfo {year} {2014})\BibitemShut {NoStop}%
\bibitem [{\citenamefont {Ginzburg}\ and\ \citenamefont {Landau}(1965)}]{GL50}%
  \BibitemOpen
  \bibfield  {author} {\bibinfo {author} {\bibfnamefont {V.~L.}\ \bibnamefont
  {Ginzburg}}\ and\ \bibinfo {author} {\bibfnamefont {L.~D.}\ \bibnamefont
  {Landau}},\ }\bibfield  {title} {\bibinfo {title} {On the theory of
  superconductivity},\ }in\ \href
  {https://doi.org/10.1016/b978-0-08-010586-4.50078-x} {\emph {\bibinfo
  {booktitle} {Collected Papers of L. D. Landau}}},\ \bibinfo {editor} {edited
  by\ \bibinfo {editor} {\bibfnamefont {D.}~\bibnamefont {ter Haar}}}\
  (\bibinfo  {publisher} {Pergamon Press},\ \bibinfo {address} {Oxford, New
  York},\ \bibinfo {year} {1965})\ pp.\ \bibinfo {pages} {546--568}\BibitemShut
  {NoStop}%
\bibitem [{\citenamefont {Witten}(1988)}]{Witten1988}%
  \BibitemOpen
  \bibfield  {author} {\bibinfo {author} {\bibfnamefont {E.}~\bibnamefont
  {Witten}},\ }\bibfield  {title} {\bibinfo {title} {Topological quantum field
  theory},\ }\href {https://doi.org/10.1007/bf01223371} {\bibfield  {journal}
  {\bibinfo  {journal} {Commun. Math. Phys.}\ }\textbf {\bibinfo {volume}
  {117}},\ \bibinfo {pages} {353} (\bibinfo {year} {1988})}\BibitemShut
  {NoStop}%
\bibitem [{\citenamefont {Witten}(1989)}]{Witten1989}%
  \BibitemOpen
  \bibfield  {author} {\bibinfo {author} {\bibfnamefont {E.}~\bibnamefont
  {Witten}},\ }\bibfield  {title} {\bibinfo {title} {{Quantum field theory and
  the Jones polynomial}},\ }\href {https://doi.org/10.1007/bf01217730}
  {\bibfield  {journal} {\bibinfo  {journal} {Commun. Math. Phys.}\ }\textbf
  {\bibinfo {volume} {121}},\ \bibinfo {pages} {351} (\bibinfo {year}
  {1989})}\BibitemShut {NoStop}%
\bibitem [{\citenamefont {Levin}\ and\ \citenamefont {Wen}(2005)}]{LW05}%
  \BibitemOpen
  \bibfield  {author} {\bibinfo {author} {\bibfnamefont {M.~A.}\ \bibnamefont
  {Levin}}\ and\ \bibinfo {author} {\bibfnamefont {X.-G.}\ \bibnamefont
  {Wen}},\ }\bibfield  {title} {\bibinfo {title} {String-net condensation: {A}
  physical mechanism for topological phases},\ }\href
  {https://doi.org/10.1103/PhysRevB.71.045110} {\bibfield  {journal} {\bibinfo
  {journal} {Phys. Rev. B}\ }\textbf {\bibinfo {volume} {71}},\ \bibinfo
  {pages} {045110} (\bibinfo {year} {2005})}\BibitemShut {NoStop}%
\bibitem [{\citenamefont {Kassel}(1995)}]{Kassel95}%
  \BibitemOpen
  \bibfield  {author} {\bibinfo {author} {\bibfnamefont {C.}~\bibnamefont
  {Kassel}},\ }\href@noop {} {\emph {\bibinfo {title} {Quantum {Groups}}}},\
  \bibinfo {series} {Graduate {Texts} in {Mathematics}}, Vol.\ \bibinfo
  {volume} {155}\ (\bibinfo  {publisher} {Springer},\ \bibinfo {year}
  {1995})\BibitemShut {NoStop}%
\bibitem [{\citenamefont {Etingof}\ \emph {et~al.}(2015)\citenamefont
  {Etingof}, \citenamefont {Gelaki}, \citenamefont {Nikshych},\ and\
  \citenamefont {Ostrik}}]{Etingof2015}%
  \BibitemOpen
  \bibfield  {author} {\bibinfo {author} {\bibfnamefont {P.}~\bibnamefont
  {Etingof}}, \bibinfo {author} {\bibfnamefont {S.}~\bibnamefont {Gelaki}},
  \bibinfo {author} {\bibfnamefont {D.}~\bibnamefont {Nikshych}},\ and\
  \bibinfo {author} {\bibfnamefont {V.}~\bibnamefont {Ostrik}},\ }\href@noop {}
  {\emph {\bibinfo {title} {Tensor {Categories}}}}\ (\bibinfo  {publisher}
  {American Math. Soc},\ \bibinfo {year} {2015})\BibitemShut {NoStop}%
\bibitem [{\citenamefont {Hong}(2009)}]{Hong09}%
  \BibitemOpen
  \bibfield  {author} {\bibinfo {author} {\bibfnamefont {S.-M.}\ \bibnamefont
  {Hong}},\ }\bibfield  {title} {\bibinfo {title} {On symmetrization of
  6j-symbols and {Levin}-{Wen} {Hamiltonian}},\ }\href@noop {} {\  (\bibinfo
  {year} {2009})},\ \Eprint {https://arxiv.org/abs/arXiv:0907.2204}
  {arXiv:0907.2204} \BibitemShut {NoStop}%
\bibitem [{\citenamefont {Osborne}\ \emph {et~al.}(2019)\citenamefont
  {Osborne}, \citenamefont {Stiegemann},\ and\ \citenamefont
  {Wolf}}]{Osborne2019}%
  \BibitemOpen
  \bibfield  {author} {\bibinfo {author} {\bibfnamefont {T.~J.}\ \bibnamefont
  {Osborne}}, \bibinfo {author} {\bibfnamefont {D.~E.}\ \bibnamefont
  {Stiegemann}},\ and\ \bibinfo {author} {\bibfnamefont {R.}~\bibnamefont
  {Wolf}},\ }\bibfield  {title} {\bibinfo {title} {{The $F$-Symbols for the
  $\mathcal{H}_3$ Fusion Category}},\ }\href@noop {} {\  (\bibinfo {year}
  {2019})},\ \Eprint {https://arxiv.org/abs/arXiv:1906.01322}
  {arXiv:1906.01322} \BibitemShut {NoStop}%
\bibitem [{\citenamefont {Hung}\ and\ \citenamefont {Wan}(2012)}]{HW12}%
  \BibitemOpen
  \bibfield  {author} {\bibinfo {author} {\bibfnamefont {L.-Y.}\ \bibnamefont
  {Hung}}\ and\ \bibinfo {author} {\bibfnamefont {Y.}~\bibnamefont {Wan}},\
  }\bibfield  {title} {\bibinfo {title} {String-net models with $\mathbb{Z}_n$
  fusion algebra},\ }\href {https://doi.org/10.1103/physrevb.86.235132}
  {\bibfield  {journal} {\bibinfo  {journal} {Phys. Rev. B}\ }\textbf {\bibinfo
  {volume} {86}},\ \bibinfo {pages} {235132} (\bibinfo {year}
  {2012})}\BibitemShut {NoStop}%
\bibitem [{\citenamefont {Lin}\ and\ \citenamefont {Levin}(2014)}]{LW14}%
  \BibitemOpen
  \bibfield  {author} {\bibinfo {author} {\bibfnamefont {C.-H.}\ \bibnamefont
  {Lin}}\ and\ \bibinfo {author} {\bibfnamefont {M.}~\bibnamefont {Levin}},\
  }\bibfield  {title} {\bibinfo {title} {Generalizations and limitations of
  string-net models},\ }\href {https://doi.org/10.1103/physrevb.89.195130}
  {\bibfield  {journal} {\bibinfo  {journal} {Physical Review B}\ }\textbf
  {\bibinfo {volume} {89}},\ \bibinfo {pages} {195130} (\bibinfo {year}
  {2014})}\BibitemShut {NoStop}%
\bibitem [{\citenamefont {K{\'{a}}d{\'{a}}r}\ \emph {et~al.}(2010)\citenamefont
  {K{\'{a}}d{\'{a}}r}, \citenamefont {Marzuoli},\ and\ \citenamefont
  {Rasetti}}]{KMR10}%
  \BibitemOpen
  \bibfield  {author} {\bibinfo {author} {\bibfnamefont {Z.}~\bibnamefont
  {K{\'{a}}d{\'{a}}r}}, \bibinfo {author} {\bibfnamefont {A.}~\bibnamefont
  {Marzuoli}},\ and\ \bibinfo {author} {\bibfnamefont {M.}~\bibnamefont
  {Rasetti}},\ }\bibfield  {title} {\bibinfo {title} {{Microscopic Description
  of $2D$ Topological Phases, Duality, and $3D$ State Sums}},\ }\href
  {https://doi.org/10.1155/2010/671039} {\bibfield  {journal} {\bibinfo
  {journal} {Advances in Mathematical Physics}\ }\textbf {\bibinfo {volume}
  {2010}},\ \bibinfo {pages} {1} (\bibinfo {year} {2010})}\BibitemShut
  {NoStop}%
\bibitem [{\citenamefont {K{\"o}nig}\ \emph {et~al.}(2010)\citenamefont
  {K{\"o}nig}, \citenamefont {Kuperberg},\ and\ \citenamefont
  {Reichardt}}]{KKR10}%
  \BibitemOpen
  \bibfield  {author} {\bibinfo {author} {\bibfnamefont {R.}~\bibnamefont
  {K{\"o}nig}}, \bibinfo {author} {\bibfnamefont {G.}~\bibnamefont
  {Kuperberg}},\ and\ \bibinfo {author} {\bibfnamefont {B.~W.}\ \bibnamefont
  {Reichardt}},\ }\bibfield  {title} {\bibinfo {title} {Quantum computation
  with {Turaev-Viro} codes},\ }\href
  {https://doi.org/10.1016/j.aop.2010.08.001} {\bibfield  {journal} {\bibinfo
  {journal} {Annals of Physics}\ }\textbf {\bibinfo {volume} {325}},\ \bibinfo
  {pages} {2707} (\bibinfo {year} {2010})}\BibitemShut {NoStop}%
\bibitem [{\citenamefont {{Kirillov Jr.}}(2011)}]{Ki11}%
  \BibitemOpen
  \bibfield  {author} {\bibinfo {author} {\bibfnamefont {A.}~\bibnamefont
  {{Kirillov Jr.}}},\ }\bibfield  {title} {\bibinfo {title} {String-net model
  of {Turaev-Viro} invariants},\ }\href@noop {} {\  (\bibinfo {year} {2011})},\
  \Eprint {https://arxiv.org/abs/1106.6033v1} {1106.6033v1} \BibitemShut
  {NoStop}%
\bibitem [{\citenamefont {{Kirillov Jr.}}\ and\ \citenamefont
  {Balsam}(2010)}]{KB10}%
  \BibitemOpen
  \bibfield  {author} {\bibinfo {author} {\bibfnamefont {A.}~\bibnamefont
  {{Kirillov Jr.}}}\ and\ \bibinfo {author} {\bibfnamefont {B.}~\bibnamefont
  {Balsam}},\ }\bibfield  {title} {\bibinfo {title} {{Turaev-Viro invariants as
  an extended TQFT}},\ }\href@noop {} {\  (\bibinfo {year} {2010})},\ \Eprint
  {https://arxiv.org/abs/1004.1533v3} {1004.1533v3} \BibitemShut {NoStop}%
\bibitem [{\citenamefont {Turaev}\ and\ \citenamefont
  {Virelizier}(2010)}]{TV10}%
  \BibitemOpen
  \bibfield  {author} {\bibinfo {author} {\bibfnamefont {V.}~\bibnamefont
  {Turaev}}\ and\ \bibinfo {author} {\bibfnamefont {A.}~\bibnamefont
  {Virelizier}},\ }\bibfield  {title} {\bibinfo {title} {On two approaches to
  $3$-dimensional {TQFTs}},\ }\href@noop {} {\  (\bibinfo {year} {2010})},\
  \Eprint {https://arxiv.org/abs/1006.3501v5} {1006.3501v5} \BibitemShut
  {NoStop}%
\bibitem [{\citenamefont {Balsam}(2010)}]{Balsam2010}%
  \BibitemOpen
  \bibfield  {author} {\bibinfo {author} {\bibfnamefont {B.}~\bibnamefont
  {Balsam}},\ }\bibfield  {title} {\bibinfo {title} {{Turaev-Viro invariants as
  an extended TQFT $III$}},\ }\href@noop {} {\  (\bibinfo {year} {2010})},\
  \Eprint {https://arxiv.org/abs/1012.0560v2} {1012.0560v2} \BibitemShut
  {NoStop}%
\bibitem [{\citenamefont {Fuchs}\ \emph {et~al.}(2002)\citenamefont {Fuchs},
  \citenamefont {Runkel},\ and\ \citenamefont {Schweigert}}]{FRS02}%
  \BibitemOpen
  \bibfield  {author} {\bibinfo {author} {\bibfnamefont {J.}~\bibnamefont
  {Fuchs}}, \bibinfo {author} {\bibfnamefont {I.}~\bibnamefont {Runkel}},\ and\
  \bibinfo {author} {\bibfnamefont {C.}~\bibnamefont {Schweigert}},\ }\bibfield
   {title} {\bibinfo {title} {{TFT} construction of {RCFT} correlators~{I}:
  partition functions},\ }\href {https://doi.org/10.1016/s0550-3213(02)00744-7}
  {\bibfield  {journal} {\bibinfo  {journal} {Nucl. Phys. B}\ }\textbf
  {\bibinfo {volume} {646}},\ \bibinfo {pages} {353} (\bibinfo {year}
  {2002})}\BibitemShut {NoStop}%
\bibitem [{\citenamefont {Fuchs}\ \emph
  {et~al.}(2004{\natexlab{a}})\citenamefont {Fuchs}, \citenamefont {Runkel},\
  and\ \citenamefont {Schweigert}}]{FRS04a}%
  \BibitemOpen
  \bibfield  {author} {\bibinfo {author} {\bibfnamefont {J.}~\bibnamefont
  {Fuchs}}, \bibinfo {author} {\bibfnamefont {I.}~\bibnamefont {Runkel}},\ and\
  \bibinfo {author} {\bibfnamefont {C.}~\bibnamefont {Schweigert}},\ }\bibfield
   {title} {\bibinfo {title} {{TFT} construction of {RCFT} correlators~{II}:
  unoriented world sheets},\ }\href
  {https://doi.org/10.1016/j.nuclphysb.2003.11.026} {\bibfield  {journal}
  {\bibinfo  {journal} {Nucl. Phys. B}\ }\textbf {\bibinfo {volume} {678}},\
  \bibinfo {pages} {511} (\bibinfo {year} {2004}{\natexlab{a}})}\BibitemShut
  {NoStop}%
\bibitem [{\citenamefont {Fuchs}\ \emph
  {et~al.}(2004{\natexlab{b}})\citenamefont {Fuchs}, \citenamefont {Runkel},\
  and\ \citenamefont {Schweigert}}]{FRS04b}%
  \BibitemOpen
  \bibfield  {author} {\bibinfo {author} {\bibfnamefont {J.}~\bibnamefont
  {Fuchs}}, \bibinfo {author} {\bibfnamefont {I.}~\bibnamefont {Runkel}},\ and\
  \bibinfo {author} {\bibfnamefont {C.}~\bibnamefont {Schweigert}},\ }\bibfield
   {title} {\bibinfo {title} {{TFT} construction of {RCFT} correlators~{III}:
  simple currents},\ }\href {https://doi.org/10.1016/j.nuclphysb.2004.05.014}
  {\bibfield  {journal} {\bibinfo  {journal} {Nucl. Phys. B}\ }\textbf
  {\bibinfo {volume} {694}},\ \bibinfo {pages} {277} (\bibinfo {year}
  {2004}{\natexlab{b}})}\BibitemShut {NoStop}%
\bibitem [{\citenamefont {Fuchs}\ \emph {et~al.}(2005)\citenamefont {Fuchs},
  \citenamefont {Runkel},\ and\ \citenamefont {Schweigert}}]{FRS05}%
  \BibitemOpen
  \bibfield  {author} {\bibinfo {author} {\bibfnamefont {J.}~\bibnamefont
  {Fuchs}}, \bibinfo {author} {\bibfnamefont {I.}~\bibnamefont {Runkel}},\ and\
  \bibinfo {author} {\bibfnamefont {C.}~\bibnamefont {Schweigert}},\ }\bibfield
   {title} {\bibinfo {title} {{TFT} construction of {RCFT} correlators~{IV}:
  Structure constants and correlation functions},\ }\href
  {https://doi.org/10.1016/j.nuclphysb.2005.03.018} {\bibfield  {journal}
  {\bibinfo  {journal} {Nucl. Phys. B}\ }\textbf {\bibinfo {volume} {715}},\
  \bibinfo {pages} {539} (\bibinfo {year} {2005})}\BibitemShut {NoStop}%
\bibitem [{\citenamefont {Fjelstad}\ \emph {et~al.}(2006)\citenamefont
  {Fjelstad}, \citenamefont {Fuchs}, \citenamefont {Runkel},\ and\
  \citenamefont {Schweigert}}]{FFRS06}%
  \BibitemOpen
  \bibfield  {author} {\bibinfo {author} {\bibfnamefont {J.}~\bibnamefont
  {Fjelstad}}, \bibinfo {author} {\bibfnamefont {J.}~\bibnamefont {Fuchs}},
  \bibinfo {author} {\bibfnamefont {I.}~\bibnamefont {Runkel}},\ and\ \bibinfo
  {author} {\bibfnamefont {C.}~\bibnamefont {Schweigert}},\ }\bibfield  {title}
  {\bibinfo {title} {{TFT} construction of {RCFT} correlators~{V}: Structure
  constants and correlation functions},\ }\href@noop {} {\bibfield  {journal}
  {\bibinfo  {journal} {Theor. Appl. Categor.}\ }\textbf {\bibinfo {volume}
  {16}},\ \bibinfo {pages} {342} (\bibinfo {year} {2006})}\BibitemShut
  {NoStop}%
\bibitem [{Note1()}]{Note1}%
  \BibitemOpen
  \bibinfo {note} {Note that we are always allowed to add and remove vacuum
  lines, since they do not change the meaning of the diagram. Mathematically,
  the loop added by the $B_\protect \mathbf {p}^s$ operator is a composition of
  an evaluation and a co-evaluation morphism. Therefore, a vacuum line at the
  top corner as well as at the bottom corner of the loop is already
  implicit.}\BibitemShut {Stop}%
\bibitem [{\citenamefont {Buerschaper}\ \emph {et~al.}(2009)\citenamefont
  {Buerschaper}, \citenamefont {Aguado},\ and\ \citenamefont
  {Vidal}}]{Buerschaper2009}%
  \BibitemOpen
  \bibfield  {author} {\bibinfo {author} {\bibfnamefont {O.}~\bibnamefont
  {Buerschaper}}, \bibinfo {author} {\bibfnamefont {M.}~\bibnamefont
  {Aguado}},\ and\ \bibinfo {author} {\bibfnamefont {G.}~\bibnamefont
  {Vidal}},\ }\bibfield  {title} {\bibinfo {title} {Explicit tensor network
  representation for the ground states of string-net models},\ }\href
  {https://doi.org/10.1103/physrevb.79.085119} {\bibfield  {journal} {\bibinfo
  {journal} {Phys. Rev. B}\ }\textbf {\bibinfo {volume} {79}},\ \bibinfo
  {pages} {085119} (\bibinfo {year} {2009})}\BibitemShut {NoStop}%
\bibitem [{\citenamefont {Şahinoğlu}\ \emph {et~al.}(2014)\citenamefont
  {Şahinoğlu}, \citenamefont {Williamson}, \citenamefont {Bultinck},
  \citenamefont {Mariën}, \citenamefont {Haegeman}, \citenamefont {Schuch},\
  and\ \citenamefont {Verstraete}}]{Sahinoglu2014}%
  \BibitemOpen
  \bibfield  {author} {\bibinfo {author} {\bibfnamefont {M.~B.}\ \bibnamefont
  {Şahinoğlu}}, \bibinfo {author} {\bibfnamefont {D.}~\bibnamefont
  {Williamson}}, \bibinfo {author} {\bibfnamefont {N.}~\bibnamefont
  {Bultinck}}, \bibinfo {author} {\bibfnamefont {M.}~\bibnamefont {Mariën}},
  \bibinfo {author} {\bibfnamefont {J.}~\bibnamefont {Haegeman}}, \bibinfo
  {author} {\bibfnamefont {N.}~\bibnamefont {Schuch}},\ and\ \bibinfo {author}
  {\bibfnamefont {F.}~\bibnamefont {Verstraete}},\ }\bibfield  {title}
  {\bibinfo {title} {Characterizing topological order with matrix product
  operators},\ }\href@noop {} {\  (\bibinfo {year} {2014})},\ \Eprint
  {https://arxiv.org/abs/arXiv:1409.2150} {arXiv:1409.2150} \BibitemShut
  {NoStop}%
\bibitem [{\citenamefont {Turaev}(1992)}]{TURAEV1992}%
  \BibitemOpen
  \bibfield  {author} {\bibinfo {author} {\bibfnamefont {V.~G.}\ \bibnamefont
  {Turaev}},\ }\bibfield  {title} {\bibinfo {title} {{Modular} {Categories}
  {and} $3$-{Manifold} {Invariants}},\ }\href
  {https://doi.org/10.1142/s0217979292000876} {\bibfield  {journal} {\bibinfo
  {journal} {Int. J. Mod. Phys. B}\ }\textbf {\bibinfo {volume} {06}},\
  \bibinfo {pages} {1807} (\bibinfo {year} {1992})}\BibitemShut {NoStop}%
\bibitem [{\citenamefont {Müger}(2003)}]{Mueger2003}%
  \BibitemOpen
  \bibfield  {author} {\bibinfo {author} {\bibfnamefont {M.}~\bibnamefont
  {Müger}},\ }\bibfield  {title} {\bibinfo {title} {From subfactors to
  categories and topology {II}: The quantum double of tensor categories and
  subfactors},\ }\href {https://doi.org/10.1016/s0022-4049(02)00248-7}
  {\bibfield  {journal} {\bibinfo  {journal} {J. Pure Appl. Algebra}\ }\textbf
  {\bibinfo {volume} {180}},\ \bibinfo {pages} {159} (\bibinfo {year}
  {2003})}\BibitemShut {NoStop}%
\bibitem [{Note2()}]{Note2}%
  \BibitemOpen
  \bibinfo {note} {The tensor product is defined not only for simple objects
  but for all objects of the category. We restrict ourselves to simple objects
  here because it makes it easier to draw the connection to the string types of
  the lattice model.}\BibitemShut {Stop}%
\bibitem [{Note3()}]{Note3}%
  \BibitemOpen
  \bibinfo {note} {The $6j$ symbols are defined slightly different than the
  definition we give for the $F$-symbols here, but there is an easy relation
  between the two, see Appendix~\ref {app:tetrahedral}}\BibitemShut {NoStop}%
\bibitem [{\citenamefont {Etingof}\ \emph {et~al.}(2005)\citenamefont
  {Etingof}, \citenamefont {Nikshych},\ and\ \citenamefont
  {Ostrik}}]{Etingof2002}%
  \BibitemOpen
  \bibfield  {author} {\bibinfo {author} {\bibfnamefont {P.}~\bibnamefont
  {Etingof}}, \bibinfo {author} {\bibfnamefont {D.}~\bibnamefont {Nikshych}},\
  and\ \bibinfo {author} {\bibfnamefont {V.}~\bibnamefont {Ostrik}},\
  }\bibfield  {title} {\bibinfo {title} {On fusion categories},\ }\href
  {https://doi.org/10.4007/annals.2005.162.581} {\bibfield  {journal} {\bibinfo
   {journal} {Ann. of Math.}\ }\textbf {\bibinfo {volume} {162}},\ \bibinfo
  {pages} {581} (\bibinfo {year} {2005})}\BibitemShut {NoStop}%
\bibitem [{\citenamefont {Bonderson}(2007)}]{Bonderson2007}%
  \BibitemOpen
  \bibfield  {author} {\bibinfo {author} {\bibfnamefont {P.}~\bibnamefont
  {Bonderson}},\ }\emph {\bibinfo {title} {Non-{Abelian} {Anyons} and
  {Interferometry}}},\ \href
  {https://thesis.library.caltech.edu/2447/2/thesis.pdf} {Ph.D. thesis},\
  \bibinfo  {school} {California Institute of Technology}, \bibinfo {address}
  {Pasadena, California} (\bibinfo {year} {2007})\BibitemShut {NoStop}%
\bibitem [{\citenamefont {Bonderson}\ \emph {et~al.}(2008)\citenamefont
  {Bonderson}, \citenamefont {Shtengel},\ and\ \citenamefont
  {Slingerland}}]{Bonderson2008}%
  \BibitemOpen
  \bibfield  {author} {\bibinfo {author} {\bibfnamefont {P.}~\bibnamefont
  {Bonderson}}, \bibinfo {author} {\bibfnamefont {K.}~\bibnamefont
  {Shtengel}},\ and\ \bibinfo {author} {\bibfnamefont {J.~K.}\ \bibnamefont
  {Slingerland}},\ }\bibfield  {title} {\bibinfo {title} {Interferometry of
  non-{Abelian} anyons},\ }\href {https://doi.org/10.1016/j.aop.2008.01.012}
  {\bibfield  {journal} {\bibinfo  {journal} {Ann. Phys.}\ }\textbf {\bibinfo
  {volume} {323}},\ \bibinfo {pages} {2709} (\bibinfo {year}
  {2008})}\BibitemShut {NoStop}%
\bibitem [{\citenamefont {{Mac Lane}}(1998)}]{MacLane1998}%
  \BibitemOpen
  \bibfield  {author} {\bibinfo {author} {\bibfnamefont {S.}~\bibnamefont {{Mac
  Lane}}},\ }\href
  {https://www.ebook.de/de/product/1381115/saunders_mac_lane_categories_for_the_working_mathematician.html}
  {\emph {\bibinfo {title} {Categories for the {Working} {Mathematician}}}}\
  (\bibinfo  {publisher} {Springer-Verlag New York Inc.},\ \bibinfo {year}
  {1998})\BibitemShut {NoStop}%
\bibitem [{Note4()}]{Note4}%
  \BibitemOpen
  \bibinfo {note} {Alternatively, we could impose time to go left or right.
  This would go along with a rotation of all string diagrams by 90 degrees.
  However, this modification of convention does not change the physical
  process.}\BibitemShut {Stop}%
\bibitem [{\citenamefont {Soejima}\ \emph {et~al.}(2020)\citenamefont
  {Soejima}, \citenamefont {Siva}, \citenamefont {Bultinck}, \citenamefont
  {Chatterjee}, \citenamefont {Pollmann},\ and\ \citenamefont
  {Zaletel}}]{Soejima2020}%
  \BibitemOpen
  \bibfield  {author} {\bibinfo {author} {\bibfnamefont {T.}~\bibnamefont
  {Soejima}}, \bibinfo {author} {\bibfnamefont {K.}~\bibnamefont {Siva}},
  \bibinfo {author} {\bibfnamefont {N.}~\bibnamefont {Bultinck}}, \bibinfo
  {author} {\bibfnamefont {S.}~\bibnamefont {Chatterjee}}, \bibinfo {author}
  {\bibfnamefont {F.}~\bibnamefont {Pollmann}},\ and\ \bibinfo {author}
  {\bibfnamefont {M.~P.}\ \bibnamefont {Zaletel}},\ }\bibfield  {title}
  {\bibinfo {title} {Isometric tensor network representation of string-net
  liquids},\ }\bibfield  {journal} {\bibinfo  {journal} {Phys. Rev. B}\
  }\textbf {\bibinfo {volume} {101}},\ \href
  {https://doi.org/10.1103/physrevb.101.085117} {10.1103/physrevb.101.085117}
  (\bibinfo {year} {2020})\BibitemShut {NoStop}%
\bibitem [{Note5()}]{Note5}%
  \BibitemOpen
  \bibinfo {note} {In fact, the pentagon equation is a coherence condition for
  the associator in any monoidal category, see \cite
  {Etingof2015}.}\BibitemShut {Stop}%
\bibitem [{\citenamefont {Grossman}\ and\ \citenamefont {Snyder}(2012)}]{GS12}%
  \BibitemOpen
  \bibfield  {author} {\bibinfo {author} {\bibfnamefont {P.}~\bibnamefont
  {Grossman}}\ and\ \bibinfo {author} {\bibfnamefont {N.}~\bibnamefont
  {Snyder}},\ }\bibfield  {title} {\bibinfo {title} {Quantum {Subgroups} of the
  {Haagerup} {Fusion} {Categories}},\ }\href
  {https://doi.org/10.1007/s00220-012-1427-x} {\bibfield  {journal} {\bibinfo
  {journal} {Commun. Math. Phys.}\ }\textbf {\bibinfo {volume} {311}},\
  \bibinfo {pages} {617} (\bibinfo {year} {2012})}\BibitemShut {NoStop}%
\end{thebibliography}%
	
\end{document}